\documentclass[prl,aps,twocolumn,showpacs,superscriptaddress,nofootinbib,preprintnumbers]{revtex4-2}

\usepackage{graphicx}
\usepackage{color}
\usepackage{amsmath,xfrac}
\usepackage{amssymb}
\usepackage{enumerate}
\usepackage{multirow}
\usepackage[colorlinks=true,citecolor=blue,urlcolor=blue]{hyperref}
\usepackage{subcaption}
\usepackage{ragged2e}
\usepackage{epsfig,psfrag,bm,amssymb,hyperref}
\usepackage{dcolumn}
\usepackage{bm}
\usepackage{soul}
\usepackage{cancel}
\usepackage{mathrsfs,amsfonts,color}
\usepackage[utf8]{inputenc}
\usepackage{hyperref}
\usepackage[normalem]{ulem}

\newcommand{\be}{\begin{equation}}
\newcommand{\ee}{\end{equation}}
\newcommand{\pd}{\partial}

\usepackage{tikz,xcolor,hyperref}

\definecolor{lime}{HTML}{A6CE39}
\DeclareRobustCommand{\orcidicon}{\hspace{-1mm}
	\begin{tikzpicture}
	\draw[lime, fill=lime] (0,0) 
	circle [radius=0.16] 
	node[white] {{\fontfamily{qag}\selectfont \tiny \,ID}};
	\draw[white, fill=white] (-0.0525,0.095) 
	circle [radius=0.007];
	\end{tikzpicture}
	\hspace{-3mm}
}

\foreach \x in {A, ..., Z}{\expandafter\xdef\csname orcid\x\endcsname{\noexpand\href{https://orcid.org/\csname orcidauthor\x\endcsname}
			{\noexpand\orcidicon}}
}

\begin{document}
% \preprint{UTWI-XX-2025}
\preprint{MITP-25-044}
\title{
Impostor Among {\LARGE{$\nu$}}s: Dark Radiation Masquerading as Self-Interacting Neutrinos
}

\begin{abstract}

Multiple cosmological observations hint at neutrino self-interactions beyond the Standard Model, yet such interactions face severe constraints from terrestrial experiments. We resolve this tension by introducing a model where active neutrinos resonantly convert to self-interacting dark radiation after BBN but before CMB epoch. This exploits the fact that cosmological observables cannot distinguish between neutrinos and dark radiation with the same abundance and free-streaming properties. Our mechanism, based on a simple type-I seesaw framework along with a keV-scale scalar mediator, achieves two objectives: (i) it produces strongly self-interacting dark radiation that imitates neutrino self-interactions favored by cosmological data, and (ii) it depletes the active neutrino energy density, relaxing cosmological neutrino mass bounds and easing the tension with neutrino oscillation data. The model naturally evades laboratory constraints through suppression of the neutrino-mediator coupling by the squared mass ratio of active and sterile neutrinos. We show that this scenario is favored over $\Lambda$CDM by the combined Planck and DESI data, while being consistent with all other constraints.
Our mechanism is 
testable in future laboratory probes of absolute neutrino mass and searches for sterile neutrinos. 
\end{abstract}

%\tableofcontents

\author{Anirban Das\orcidA{}}
\email{anirban.das@saha.ac.in}
\affiliation{Theory Division, Saha Institute of Nuclear Physics, 1/AF, Bidhannagar, Kolkata 700064, India}
\affiliation{Homi Bhabha National Institute, Training School Complex, Anushaktinagar, Mumbai
400094, India}

\author{P. S. Bhupal Dev\orcidB{}}
\email{bdev@wustl.edu}
\affiliation{Department of Physics and McDonnell Center for the Space Sciences, Washington University, St.~Louis, MO 63130, USA}
\affiliation{PRISMA$^+$ Cluster of Excellence \& Mainz Institute for Theoretical Physics, 
Johannes Gutenberg-Universit\"{a}t Mainz, 55099 Mainz, Germany}

\author{Christina Gao\orcidC{}}
\email{gaoy3@sustech.edu.cn}
\affiliation{Department of Physics, Southern University of Science and Technology, Shenzhen, 518055, China}

\author{Subhajit Ghosh\orcidD{}}
\email{sghosh@utexas.edu}
\affiliation{Texas Center for Cosmology and Astroparticle Physics, Weinberg Institute, Department of Physics, The University of Texas at Austin, Austin, TX 78712, USA}

\author{Taegyun Kim\orcidE{}}
\email{tkim12@alumni.nd.edu}
\affiliation{Uichang District Office, 468, Dogye-dong, Uichang-gu, Changwon, 51381, South Korea}
\affiliation{Department of Physics, Southern University of Science and Technology, Shenzhen, 518055, China}

\maketitle
%%%%%%%%%%%%%
\noindent
{\textbf {\textit{Introduction.--}}}
Neutrinos, as the most elusive Standard Model (SM) particles, may hold the key to discovering new physics. 
While the SM predicts neutrinos to be massless, neutrino oscillation experiments have precisely measured 
two non-zero mass-squared differences, namely, $\Delta m^2_{\rm sol}\simeq 7.5\times 10^{-5}$ eV$^2$ and $\Delta m^2_{\rm atm}\simeq 2.5\times 10^{-3}$ eV$^2$~\cite{Esteban:2024eli}, indicating that at least two of the three neutrino mass eigenvalues must be non-zero. 
Despite this breakthrough, the absolute neutrino mass scale, i.e., whether the lightest neutrino mass is non-zero, 
remains unknown. 
A direct model-independent laboratory constraint
on the absolute neutrino mass comes from the kinematics
of beta decay or electron capture. Currently, KATRIN sets the best direct limit of $m_\nu<0.45$ eV at 90\% confidence level (CL)~\cite{Katrin:2024cdt}, with a design sensitivity of 0.2 eV~\cite{KATRIN:2022ayy},  
while the proposed Project 8 experiment aims for 0.04 eV~\cite{ Project8:2022wqh}.

Precision cosmology provides an alternative indirect way to measure the \emph{total} neutrino mass.
In the standard cosmological model, neutrinos decouple from the thermal bath during Big Bang Nucleosynthesis (BBN) at a temperature around $1$~MeV. Afterward, neutrinos continue to free-stream until their temperature falls below their mass, making them non-relativistic.
Relativistic neutrinos suppress the growth of structure because they do not cluster and hinder the clustering of other matter through radiation pressure~\cite{Lesgourgues:2006nd}. 
This characteristic suppression of the matter power spectrum makes cosmological probes such as the Cosmic Microwave Background (CMB) and the Baryon Acoustic Oscillations (BAO) sensitive to neutrino mass.  
The strongest constraint to date on the total neutrino mass, $\sum m_\nu < 0.0642~{\rm eV}$ at 95\% CL, comes from Planck + DESI-DR2 BAO analysis~\cite{Planck:2018vyg,DESI:2025ejh}, surpassing current laboratory constraints by an order of magnitude.

Cosmological measurements of neutrino mass are growingly in tension with neutrino oscillation experiments, which require $\sum m_\nu > 0.059~(0.10)~{\rm eV}$ for the normal (inverted) mass ordering~\cite{Esteban:2024eli}. Thus, current cosmological constraints strongly disfavor the inverted ordering, and the parameter space remaining for normal ordering is progressively diminishing.  Several proposals have been made to alleviate this tension, including neutrino decays~\cite{Aalberts:2018obr,Chacko:2019nej,Chacko:2020hmh,Escudero:2020ped,Barenboim:2020vrr,FrancoAbellan:2021hdb,Chen:2022idm}, 
time-varying neutrino masses~\cite{Fardon:2003eh,Lorenz:2018fzb,Lorenz:2021alz,Sen:2023uga,Sen:2024pgb}, 
additional interactions of neutrinos~\cite{He:2020zns, Berbig:2020wve, Esteban:2021ozz,Green:2021gdc, Esteban:2022rjk, Foroughi-Abari:2025upe}, 
and the production of dark radiation from neutrinos~\cite{Beacom:2004yd,Farzan:2015pca,Escudero:2022gez, Li:2023puz}. Additionally, the cosmological data shows a slight preference for `negative' neutrino mass, which is attributed to excess lensing amplitude in CMB~\cite{Craig:2024tky,Green:2024xbb,Naredo-Tuero:2024sgf,Cozzumbo:2025ewt} and the smaller value of $\Omega_m$ (fractional matter density) measured by DESI~\cite{DESI:2024mwx,Loverde:2024nfi,DESI:2025ejh} and only a handful of beyond-$\Lambda$CDM cosmologies can address this tension~\cite{Craig:2024tky,Lynch:2025ine,Graham:2025fdt,Elbers:2024sha,Namikawa:2025doa}.  

Cosmological observables are also sensitive to the free-streaming properties of neutrinos and any interactions that impede this free-streaming behavior. 
In particular, some cosmological data 
show an intriguing preference of beyond-the-Standard Model (BSM) interactions
mediated by a heavy mediator~\cite{Cyr-Racine:2013jua,Oldengott:2014qra,Lancaster:2017ksf,Oldengott:2017fhy,Kreisch:2019yzn,Barenboim:2019tux,Ghosh:2019tab,Das:2020xke,RoyChoudhury:2020dmd,Brinckmann:2020bcn,RoyChoudhury:2022rva,Das:2023npl,Venzor:2023aka,Camarena:2023cku,He:2023oke,Pal:2024yom,Poudou:2025qcx,ACT:2025tim,He:2025jwp}.  Such self-interactions have also been shown to alleviate cosmological discrepancies, such as the $H_0$ and $S_8$ tensions, and, can also partially address the `negative' neutrino mass tension as shown later. Additionally, they provide characteristic imprints in the matter power spectrum which can be probed via Large Scale Structure (LSS) measurements~\cite{Kumar:2022vee,Libanore:2025ack,Camarena:2023cku,He:2023oke,Pal:2024yom,Poudou:2025qcx,He:2025jwp,Noriega:2025ulc}.

For flavor-universal self-interactions, CMB and LSS, including Lyman-$\alpha$, hint at the existence of moderately self-interacting (MI) neutrinos~\cite{Camarena:2023cku,He:2025jwp,Pal:2024yom,Poudou:2025qcx}. For flavor-specific self-interactions, where only one or two neutrino flavors participate, data also accommodate strong self-interaction (SI) among neutrinos with a statistical significance similar to that of $\Lambda$CDM model with the Planck CMB data alone~\cite{Das:2020xke,Brinckmann:2020bcn,RoyChoudhury:2022rva}. Using the CMB data from ACT DR4 further increases the significance of this SI mode~\cite{Kreisch:2022zxp,Das:2023npl}. 

However, BSM neutrino self-interactions are subject to stringent laboratory constraints,  
including meson decays and double beta decay~\cite{Lessa:2007up,Agostini:2015nwa,Pasquini:2015fjv,Blum:2018ljv,Berryman:2018ogk,Brune:2018sab, deGouvea:2019qaz, Brdar:2020nbj, PIENU:2021clt,NA62:2021bji,Kharusi:2021jez, Berryman:2022hds, Dev:2024twk, Zhang:2024meg}, as well as constraints from supernovae~\cite{Manohar:1987ec,Farzan:2002wx, Heurtier:2016otg,Das:2017iuj,Shalgar:2019rqe, Akita:2022etk, Chang:2022aas, Fiorillo:2022cdq, Akita:2023iwq}, high-energy astrophysical neutrino sources~\cite{Ioka:2014kca,Ng:2014pca,Bustamante:2020mep,Esteban:2021tub} and BBN~\cite{Huang:2017egl, Huang:2021dba}.  
Consequently, flavor-universal neutrino self-interactions are robustly excluded~\cite{Blinov:2019gcj, Lyu:2020lps}. While flavor-specific self-interactions (particularly those involving only $\tau$ neutrinos) are less constrained~\cite{RoyChoudhury:2022rva, Das:2023npl},  the viable parameter space remains quite limited, and there is no realistic model for this.

To address these seemingly irreconcilable requirements from cosmology and laboratory constraints, we present a novel mechanism where strongly interacting dark radiation (DR) masquerades as self-interacting neutrinos.  
In this scenario, active neutrinos resonantly convert into strongly self-interacting DR between the BBN and CMB epochs. Note that such neutrino conversion has been used previously to relax the cosmological neutrino mass bound~\cite{Beacom:2004yd,Farzan:2015pca,Escudero:2022gez} or to generate sub-MeV dark matter~\cite{
Berlin:2017ftj, Berlin:2018ztp, Aloni:2023tff, Benso:2024qrg}. However, the self-interacting DR playing the role of either flavor-specific or flavor-universal self-interacting neutrinos in this context has not been studied before.

After the neutrino-DR conversion is complete, the residual $N_{\rm eff}$ from neutrinos becomes (see~{\it Supplemental} Sec.~I) 
\be\label{eq:Neffnu}
N_{\rm eff}^\nu \simeq N_{\rm eff}^{\rm tot}\frac{g_\nu}{g_\nu+g_{\chi}}\;,
\ee
where 
$g_\nu=\frac78\times 2\times 3$ represents the neutrino degrees of freedom (d.o.f.), $g_{\chi}=\frac78\times 2\times n_\chi$ denotes the fermionic DR d.o.f., where $n_\chi$ is the number of DR flavors.
We achieve this via a type-I seesaw-motivated model augmented by a scalar and massless DR, with the former coupled to the heavy sterile neutrinos.
As shown in Eq.~\eqref{eq:Neffnu}, 
thermalization between neutrinos and DR reduces neutrino energy density, thus relaxing the cosmological neutrino mass constraint. Furthermore, the same interaction that produces the DR also makes the DR strongly self-interacting, which, to cosmological observables, appears identical to self-interacting neutrinos. Lastly, the active neutrinos have negligible non-standard interactions at late times and are thus free from the laboratory constraints. In the rest of the {\it Letter}, we will demonstrate how this model can address several cosmological tensions and can be realized free from terrestrial and BBN constraints.

\medskip
\noindent
{\textbf{\textit{The Mechanism.--}}}
Here we provide a proof-of-principle gauge-invariant model 
based on the type-I seesaw mechanism for neutrino mass generation~\cite{Minkowski:1977sc, Mohapatra:1979ia,Yanagida:1979as, Gell-Mann:1979vob,  Schechter:1980gr}, where three generations of left-handed Majorana sterile neutrinos $N^c$~\cite{deGouvea:2016qpx} are introduced to mediate between the SM and the dark sector (DS) via neutrino-Higgs Yukawa interactions: 
\be\label{eq:T1vari}
\mathcal{L} \supset \left( y^{\alpha i}L_{\alpha} N^c_i H-\frac12 M_N^{ii} N_i^c N_i^c + {\rm H.c.}\right)
+\mathcal{L}_{\rm dark}~. 
\ee
Here, $\alpha= e,\mu,\tau$ denotes lepton flavor and $i=1,2,3$ labels the sterile neutrino generations. 
The lepton number is explicitly broken by the Majorana mass term of $N^c$. 
After electroweak symmetry breaking, the Higgs vacuum expectation value $v_H$ introduces a Dirac mass between the left-handed neutrinos $\nu_\alpha$ and $N^c$ 
with a mass matrix $m_D^{\alpha i}\equiv y^{\alpha i}\,v_H/\sqrt2 $. 
In the seesaw limit ${\bm m_D}\ll {\bm M_N}$, the mixing between the interacting states $\{\nu_\alpha, N_i^c\}$ and the mass eigenstates $\{\nu_l,\nu_h\}$ is approximately given by~\cite{Casas:2001sr} 
\be\label{eq:mixing}
\begin{split}
&\nu_\alpha \approx U_{\alpha j} \left( {\nu_l}+i \sqrt{{\bm M_N}^{-1}}\sqrt{{\bm m_\nu}}\,\,  {\nu_h} \right)_j,\\
&N_j^c \approx  \left(\nu_h -i  \sqrt{{\bm M_N}^{-1}}\sqrt{{\bm m_\nu}}\,  \nu_l\right)_j,
\end{split}
\ee
where $U_{\alpha j}$ is the PMNS matrix and ${\bm m_\nu}$ (${\bm M_N}$) is the diagonal light (heavy) neutrino mass matrix. For simplicity, we set the complex orthogonal Casas-Ibarra matrix $R= I$ in obtaining \eqref{eq:mixing}, but our results are not sensitive to this choice for sufficiently small active-sterile mixing. 

The DS Lagrangian $\mathcal{L}_{\rm dark}$ consists of a massive scalar $\phi$ and massless fermion(s) $\chi$:
\be\label{eq:dark}
\mathcal{L}_{\rm dark}\supset - \frac12 \lambda_{\phi N}\, \phi\, N^c_i N^c_i  -\frac12 \lambda_{\phi\chi}\, \phi\,\chi^2 +{\rm H.c.} ,
\ee
where the couplings $ \lambda_{\phi N}, \lambda_{\phi\chi}$ are chosen to be real without loss of generality. 
In principle, additional terms like $\lambda \phi N_i\chi$ and $\lambda_{\phi H} \phi^2 H^\dagger H$ (where $H$ is the SM Higgs doublet) are allowed in the model, but these couplings are highly constrained:  $|\lambda|\lesssim 8\times 10^{-15}\left(\frac{m_N}{10~{\rm MeV}}\right)^{3/2}$ by requiring that the $N_i\to \chi \phi$ decay rate must be smaller than the $N_i\to {\rm SM}$ decay rates, in order not to exacerbate the BBN and CMB bounds on $N_i$~\cite{Dev:2025pru} and $|\lambda_{\phi H}|\lesssim 5\times 10^{-3}$ by requiring that the $H\to \phi\phi$ invisible branching ratio must not exceed the current LHC limit~\cite{ATLAS:2023tkt}. Therefore, we do not include them in Eq.~\eqref{eq:dark}. 
The small admixture of $\nu_l$ in $N^c$ generates interactions between active neutrinos $\nu_l$
and the scalar $\phi$ with a coupling given by 
\be \label{eq:la_phinu1}
%\lambda_{\nu_l \nu_l}
\lambda_{\phi\nu}\approx \lambda_{\phi N}\,\frac{m_{\nu}}{M_{N}}~.
\ee
Assuming the hierarchy $M_N\gtrsim \mathcal{O}(\mathrm{MeV})\gg m_\phi\gg m_\nu$,  
$\lambda_{\phi\nu}$ is typically $\sim 10^{-9}$ for our benchmark parameters, ensuring compatibility with laboratory bounds (see~{\it Supplemental} Sec.~II) while allowing for efficient cosmological effects. 
The equilibration of $\nu-\chi$ can occur via  
the $s$-channel $\nu_l\nu_l\to\phi\to\chi\chi$, and the $t$-channel $\nu_l\nu_l\to \phi\phi$ followed by the immediate $\phi\to \chi\chi$ decay. 
Additionally, an $\mathcal{O}(1)$ $\lambda_{\phi\chi}$ 
allows for large self-interaction among the $\chi$ particles mediated by $\phi$.  
When the temperature is much below the $\phi$ mass, the DR $\chi$'s self-interaction term assumes the form of an effective 4-Fermi operator\footnote{Such a choice reads $G_{\rm eff}(\frac12 \bar\Psi_\chi  \Psi_\chi)(\frac12 \bar\Psi_\chi \Psi_\chi)$ in the equivalent 4-component spinor representation where $\Psi_\chi\equiv(\chi~ \chi^\dagger)^{\rm T}$.}  
$\frac14G_{\rm eff}(\chi\chi+\chi^\dagger\chi^\dagger)^2$,   where 
\be\label{eq:Geff}
G_{\rm eff}\equiv\frac{\lambda_{\phi\chi}^2}{m_\phi^2}\sim \frac{0.01}{\,\rm MeV^2} \left(\frac{\rm 0.01MeV}{m_\phi}\right)^2 \left(\frac{\lambda_{\phi\chi}}{10^{-3}}\right)^2~.
\ee

\begin{figure}[t]
    %\centering
    \includegraphics[width=0.45\textwidth]{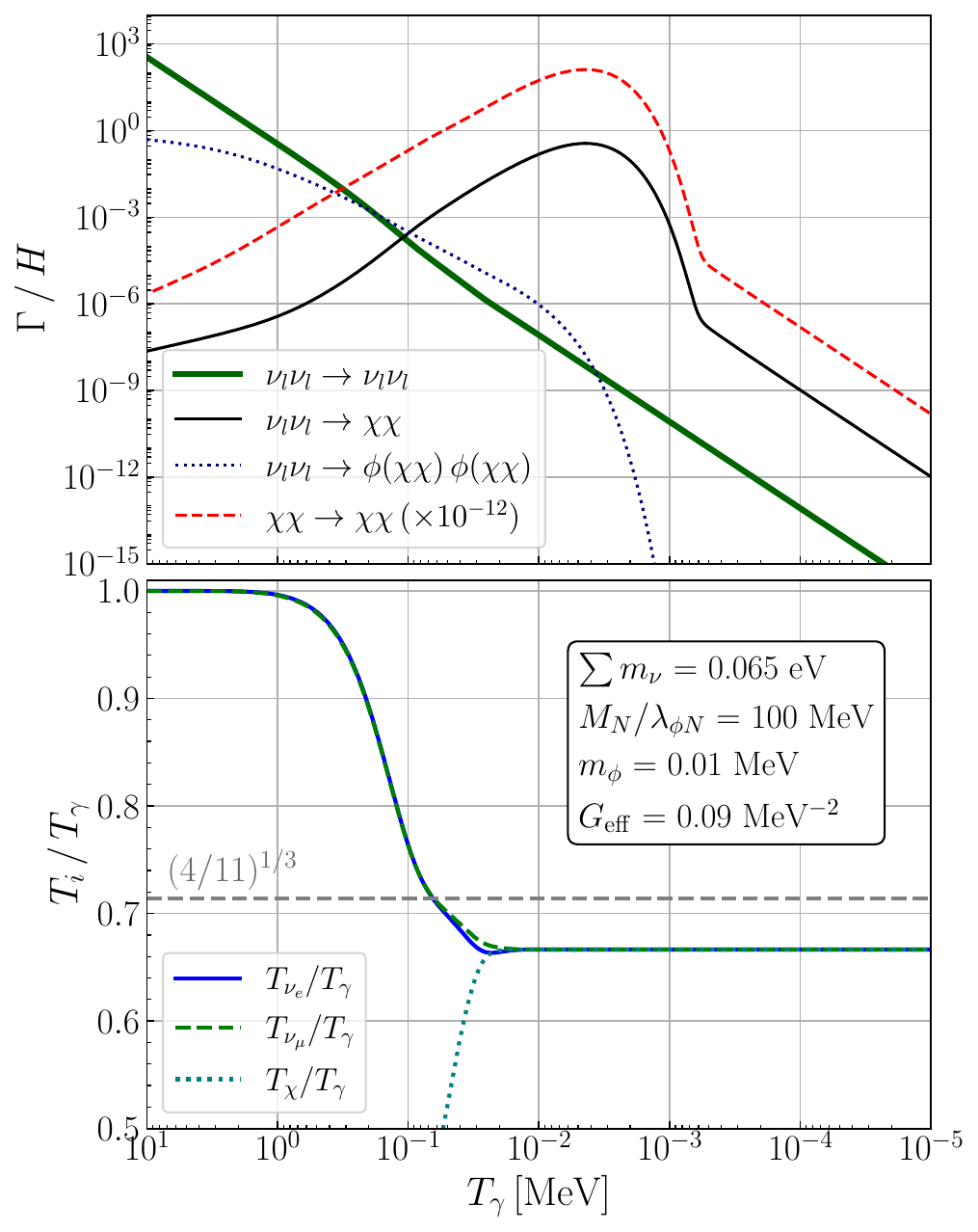}
    \captionsetup{justification=Justified}
    \caption{Evolution of interaction rates (top) and temperatures (bottom) for a benchmark point with degenerate neutrino masses.  %with $m_\phi=0.01$ MeV.
    The resonant $\nu-\chi$ conversion occurs when $T_\nu\sim m_\phi$, leading to efficient neutrino cooling (i.e. final $N_{\rm eff}^\nu=2.28$, $N_{\rm eff}^{\rm tot}=3.04$). 
    }
    \label{fig:benchmark}
\end{figure}

To illustrate how active neutrino energy is transferred to dark radiation, cf.~Eq.~\eqref{eq:Neffnu}, we plot evolutions of thermally averaged interaction rates (see~{\it Supplemental} Sec.~III) and temperatures for a benchmark point with $m_\phi=0.01\,$MeV, $M_N/\lambda_{\phi N}=100$ MeV, $\sum m_\nu=0.065$ eV and $G_{\rm eff}=0.09$ MeV$^{-2}$ in Fig.~\ref{fig:benchmark} assuming degenerate neutrino masses\footnote{The mass hierarchy of neutrinos has negligible effect in the overall cooling of neutrinos.  Benchmarks for normal ordering are shown in~{\it Supplemental} Sec.~III).}.
Clearly, the $t$-channel process dominates the production of $\chi$ at early times when the bath temperature $T_\gamma$ is higher than the mediator mass $m_\phi$. Afterward, the $s$-channel process becomes resonantly enhanced when $T_\nu\simeq m_\phi$, resulting in an efficient conversion of SM neutrinos into the DR $\chi$, thus cooling the neutrinos.
This is reflected in the rapid drop of $T_{\nu}$ when $T_\nu\simeq m_\phi$. For the chosen benchmark, we obtain $N_{\rm eff}^{\rm tot}=3.04$ after equilibration of $\nu$ and $\chi$, with contribution from neutrinos given by $N_{\rm eff}^\nu=2.28$. 

After $\chi$ kinetically decouples from $\nu$, it maintains the same temperature as $\nu$ but exhibits a much stronger self-interaction, since for the chosen benchmark 
%\be\label{eq:compSI}
$\lambda^2_{\phi\nu}/\lambda^2_{\phi \chi}\sim m^2_{\nu}
/\left(\left(M_N/\lambda_{\phi N}\right)^2 G_{\rm eff} m_\phi^2\right)\sim 10^{-15}$. 
%\ee 
This benchmark successfully reproduces a flavor-specific self-interacting neutrino cosmology with $1/4$ of active neutrinos having a self-interaction strength $G_{\rm eff}=0.09$ MeV$^{-2}$. 

\begin{figure*}[t]
    %\centering
    \includegraphics[width=0.47\linewidth]{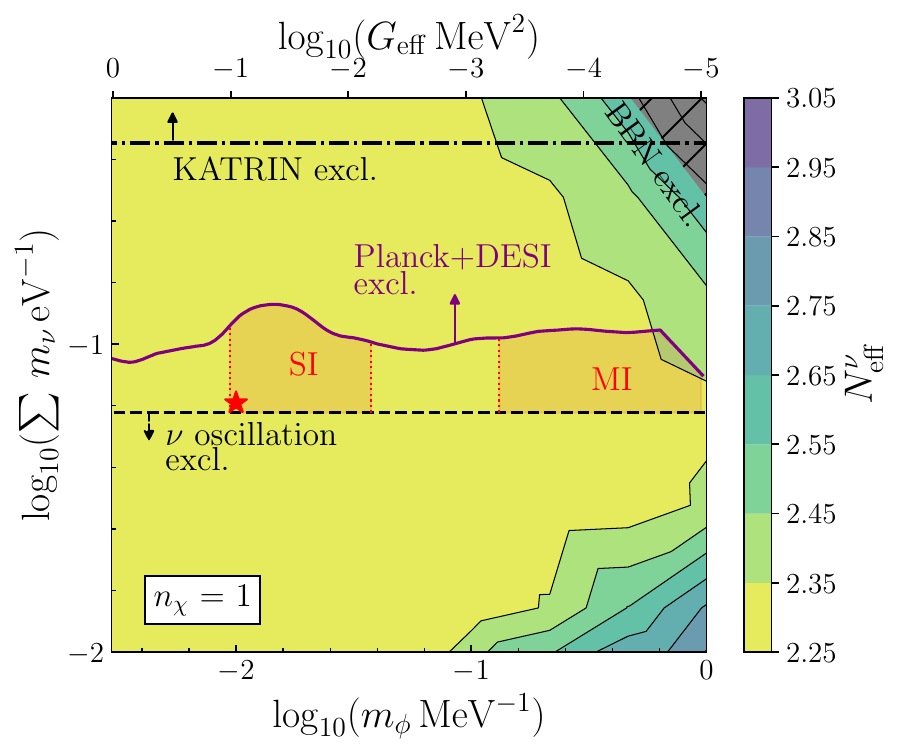}
    \includegraphics[width=0.47\linewidth]{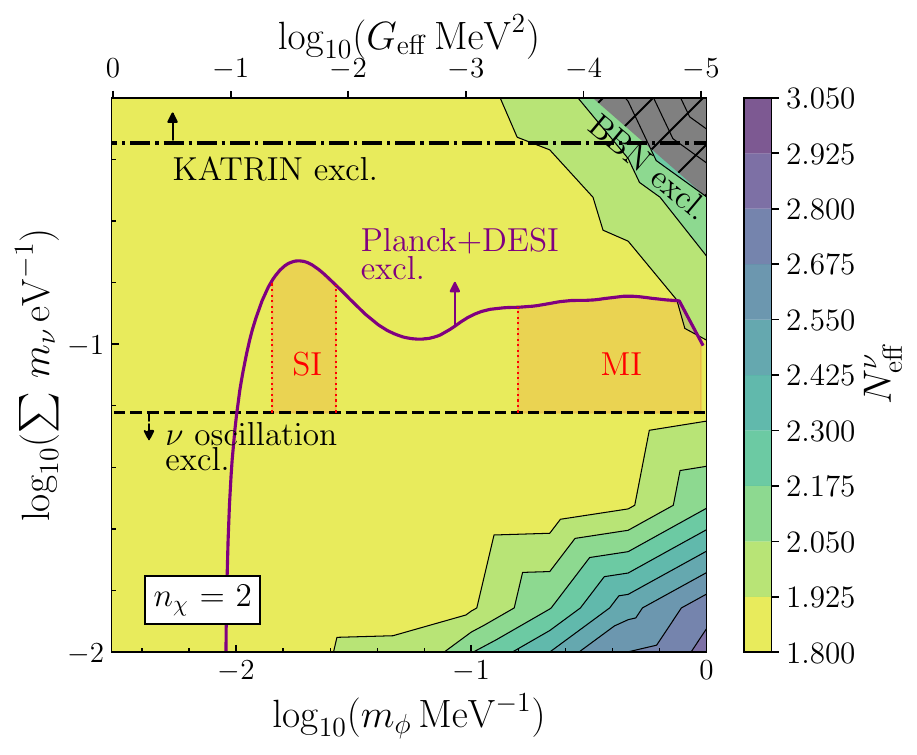}
    \captionsetup{justification=Justified}
    \caption{$\left(\,\sum m_\nu,\,m_\phi\,\right)$ plane for one $(\textbf{Left Panel})$ and two $(\textbf{Right Panel})$ flavors of DR with $\lambda_{\phi\chi}=0.003$, $M_N/\lambda_{\phi N}= 100$ MeV assuming degenerate $m_\nu$. 
    The colored contours show $N_{\rm eff}^\nu$ at $T_\gamma\sim 10^{-4}$ MeV. 
    The upper right corner (gray hatched) is excluded by BBN where $\Delta N_{\rm eff}^{\rm tot} > 0.4$. Terrestrial neutrino mass bounds from $\nu$ oscillation (black dashed) and KATRIN (black dot-dashed) are overlaid. Magenta contours show the marginalized $2\sigma$ upper limit on $\sum m_\nu$ as a function of $G_{\rm eff}$ from Fig.~\ref{fig:mcmc}. $1\sigma$ cosmological preferred SI and MI modes from Table~\ref{tab:numbers} are shown in red bands in the region not excluded by $\nu$ oscillation and cosmology. The red star is the benchmark point used in Fig.~\ref{fig:benchmark}.
    }
    \label{fig:1e-1scan}
\end{figure*}

To demonstrate that neutrino cooling can be realized in this model generically, we solve Boltzmann equations varying the total neutrino mass $\sum m_\nu$ and the mediator mass $m_\phi$ (see~{\it Supplemental} Sec.~III) with $\lambda_{\phi\chi}=0.003$, $M_N/\lambda_{\phi N}=100$ MeV. 
The resultant $N_{\rm eff}^\nu$ at $T_\gamma\sim 10^{-4}$ MeV, after $\chi-\nu$ decoupling, is shown in Fig.~\ref{fig:1e-1scan}. 
It is clear that for a large parameter space, the equilibration happens in time where neutrinos are maximally cooled, i.e. $N_{\rm eff}^\nu = 3(1+n_\chi/3)^{-1}$. When $n_\chi=1~(2)$, neutrinos together with DR behave as if $1/4~(2/5)$ of active neutrinos have significant self-interaction.  
One can convert more fractions of active neutrinos into self-interacting DR by increasing $n_\chi$ (see~{\it Supplemental} Sec.~III). When $n_\chi \gtrsim 40$, the neutrino contribution to $N_{\rm eff}$ is drastically reduced to $N_{\rm eff}^\nu \lesssim 0.2 $, with self-interacting DR constituting almost all $N^{\rm tot}_{\rm eff}$. This will mimic the flavor-universal neutrino self-interaction. 

\medskip
\noindent
{\textbf{\textit{Cosmological Phenomenology.--}}}
\begin{figure}[t]
    \centering
    \includegraphics[width=\linewidth]{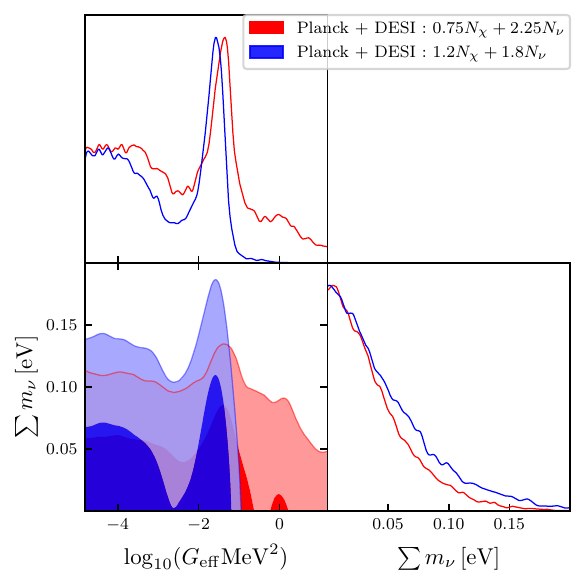}
\captionsetup{justification=Justified}
    \caption{1D marginalized posterior and 2D marginalized contours for $\log_{10} G_{\rm eff}$ and $\sum m_\nu$ with Planck + DESI dataset for $0.75 N_\chi$ and $1.2 N_\chi$ scenarios. The darker and lighter bands in the 2D plot represent $1\sigma$ and $2\sigma$ allowed regions, respectively. 
    }
    \label{fig:mcmc}
\end{figure}
\label{sec:cosmo} 
Having established that the model indeed allows for sufficient cooling of neutrinos, we now examine how this mechanism manifests in cosmological observations.

We perform a Bayesian analysis of the cosmological impact of our model with Planck-2018 dataset (temperature, polarization, and lensing)~\cite{Planck:2019nip,Planck:2018lbu} and DESI year-I data~\cite{DESI:2024mwx}.  We focus on two scenarios, assuming one or two flavors of $\chi$ with maximal neutrino cooling (cf.~Fig.~\ref{fig:1e-1scan}). These two benchmarks are cosmologically equivalent to flavor-specific neutrino self-interactions, and we comment on flavor-universal interactions later. From Eq.~\eqref{eq:Neffnu}, 
when the thermalization of $\nu-\chi$ is efficient, their contributions to $N_{\rm eff}$ after equilibration are given by\footnote{$N_\chi$ and $N_\nu$ are abbreviations of the effective number of d.o.f. $N_{\rm eff}^\chi$ and $N_{\rm eff}^\nu$, respectively. In the cosmological analysis, we fixed the total $N_{\rm eff}^{\rm tot} = 3.044$ following state-of-the-art calculations~\cite{Akita:2020szl, Froustey:2020mcq, Bennett:2020zkv}. For notational brevity, we treat $N_{\rm eff}^{\rm tot} = 3$ in Eq.~\eqref{eq:N2}. 
}
\begin{equation}
(N_\chi, N_\nu) = \begin{cases}
    (0.75, 2.25)~{\rm for}~ n_\chi = 1\\
    (1.20, 1.80)~{\rm for}~ n_\chi = 2
\end{cases} .
\label{eq:N2}
\end{equation}
In the cosmological analysis, $N_\nu$ neutrinos are free-streaming and $N_\chi$ dark radiations are self-interacting with the self-interaction rate  $\Gamma_\chi \propto G_{\rm eff}^2T_\chi^5$. See~{\it Supplemental} Sec. IV for details of the cosmological analysis.  In Fig.~\ref{fig:mcmc} and Table~\ref{tab:numbers}, we summarize the results.

In Fig.~\ref{fig:mcmc} we show the triangle plot for $\log_{10}(G_{\rm eff})$ and $\sum m_\nu$ from the Planck+DESI dataset for a Nested sampling analysis of the entire parameter space. The 1D posterior of $\log_{10}(G_{\rm eff}{\rm MeV}^2)$ shows two distinct modes: the SI (strongly interacting) mode ($\log_{10}(G_{\rm eff}{\rm MeV}^2) \approx -2$) and the MI (moderately interacting) mode ($\log_{10}(G_{\rm eff}{\rm MeV}^2) \approx -4$). 
The SI mode is largely preferred over the MI mode, which can be seen from the 1D posterior. The DESI data drives the SI mode preference since it prefers a smaller $\Omega_m$ than $\Lambda$CDM, thus also partially alleviating the `negative' neutrino mass tension. Additionally, the SI mode produces a slightly larger $H_0$, partially addressing the Hubble tension (\cite{supp}  Section IV). Thus, it is evident that the combined dataset favors cooler neutrinos and strongly self-interacting DR.

To quantify the preference of the SI mode, we performed a separate set of analyses restricting $\log_{10}(G_{\rm eff}{\rm MeV}^2)$ to the MI and SI regions by the following choice of priors on $\log_{10}(G_{\rm eff}{\rm MeV}^2)$ : $[-5, -2.5]$ and $[-2.5, 0.0]$, respectively. Table~\ref{tab:numbers} shows the relevant parameter limits (marginalized) and $\chi^2$ comparison for Planck and Planck + DESI datasets. The $\chi^2$ comparison shows that the SI mode is preferred over the MI mode for Planck + DESI dataset which exhibits the preference for strong neutrino interaction in the combined dataset. 

\renewcommand{\arraystretch}{1.1}
\begin{table*}[]
    \centering
    \begin{tabular}{|c|c|c|c|c|c|c|c|}
    \hline
        & &\multicolumn{3}{c|}{Planck} & \multicolumn{3}{c|}{Planck+DESI} \\
         \hline
      $N_{\rm eff}$ & Mode &  $\sum m_\nu~[{\rm eV}]$ & $\log_{10}(G_{\rm eff}\,{\rm MeV}^2)$ & $\Delta \chi^2$& $\sum m_\nu~[{\rm eV}]$ &  $\log_{10}(G_{\rm eff}\,{\rm MeV}^2)$ & $\Delta \chi^2$ \\
        \hline
        \multirow{2}{*}{$0.75N_\chi+2.25N_\nu$} & MI & $<0.302$ & $< -3.34$ &  $-3.36$ &$<0.104$ & $< -3.28$  & $-0.74$ \\
        & SI &$<0.290$ & $< -1.23$  & $-0.26$ &$<0.114$ & $-1.27^{+0.28}_{-0.92}$  & $-1.34$\\
        \hline
        \multirow{2}{*}{$1.20N_\chi+1.80N_\nu$} & MI & $<0.392$ & $< -3.48$  &$-0.78$ &$<0.129$ & $ < -3.44$  &  $-0.34$\\
        & SI &$<0.385$ & $-1.66^{+0.34}_{-0.24}$ &  $-0.38$ &$<0.149$ &$-1.65^{+0.30}_{-0.24}$  & $-1.66$\\
        \hline
    \end{tabular}
    \captionsetup{justification=Justified}
    \caption{$1 \sigma$ constraints on the interaction parameter and  $2\sigma$ upper bound on the sum of  neutrino masses for the separate SI and MI mode analyses. We also show the $\chi^2$ improvement of each mode over $\Lambda$CDM (with varying neutrino mass) for Planck and Planck+DESI datasets.
    }
    \label{tab:numbers}
\end{table*}

\medskip
\noindent
{\textit {Relaxation of neutrino mass bound--}}
The 2D marginalized contours from Fig.~\ref{fig:mcmc} show a large relaxation of the upper limit of $\sum m_\nu$. In this mechanism, various effects lead to the relaxation of the neutrino mass bound.
Conversion of massive neutrinos to massless $\chi$ reduces the effects of neutrino mass on cosmological observables due to a decrease in neutrino number density~\cite{Beacom:2004yd,Farzan:2015pca,Escudero:2022gez}. Using Eq.~\eqref{eq:Neffnu}, the amount of the relaxation can be computed as~\cite{Escudero:2022gez}
\begin{equation}
    \label{eq:massrelax}
    \sum m_\nu^{(N_\chi,N_\nu)} < \sum m_\nu^{(\Lambda{\rm CDM})}\left( 1 + \dfrac{N_\chi}{3}\right)\;, 
\end{equation}
where $\sum m_\nu^{(N_\chi,N_\nu)}$ and $\sum m_\nu^{(\Lambda{\rm CDM})}$ are the bounds in this cosmology and in $\Lambda{\rm CDM}$ cosmology $(N_\nu = 3, N_\chi = 0)$ with massive neutrinos. This is the primary mechanism for the relaxation of the neutrino mass bound in the MI region. In the SI mode region, there is additional relaxation of the neutrino mass. Large values of $G_{\rm eff}$ enhance the DR density perturbation which amplifies the CMB spectra at small scale similar to the SI mode in neutrino self-interaction~\cite{Kreisch:2019yzn}. This enhancement is partially compensated by increasing the neutrino mass. This effect leads to further relaxation of the SI mode mass bound as shown in Fig.~\ref{fig:mcmc}. 
Specifically, our model allows $\sum m_\nu$ up to 0.149 eV (0.385 eV) for the SI mode with Planck$+$DESI (Planck only) datasets. Thus the model relaxes the neutrino mass constraints for both SI and MI regions as well as partially addresses the `negative' neutrino mass tension.%\footnote{\textcolor{red}{A detailed quantitative study of this model for `negative' neutrino mass tension will require either extending the (effective) neutrino mass prior to negative values or some profile-likelihood analysis, which we leave for future work.}}}
% \textcolor{red}{As we have shown in the supplemental material, the SI mode favors a smaller $\Omega_m$ which can partially address the 'negative neutrino mass' tension. }

\medskip
\noindent
{\textit{Combined parameter space--}}
In the $\sum m_\nu$-$m_\phi$ plane (Fig.~\ref{fig:1e-1scan}), we overlay the regions favored by cosmological data and those constrained by terrestrial and cosmological probes over the $N_{\rm eff}^\nu$ contours.  
For a fixed value of $\lambda_{\phi \chi}$, the $m_\phi$ can be mapped to a value of $G_{\rm eff}$ following Eq.~\eqref{eq:Geff}. We overlay the $2\sigma$ contour (magenta) from Fig.~\ref{fig:mcmc} which denotes the marginalized $2\sigma$ upper limit on $\sum m_\nu$ as a function of $G_{\rm eff}$. With reference to terrestrial experiments, we also show the minimum $\sum m_\nu > 0.059~\text{eV} $ (black dashed) from the oscillation data~\cite{Esteban:2024eli} and the KATRIN upper bound $ \sum m_\nu < 0.45~\text{eV} $ (dot-dashed)~\cite{Katrin:2024cdt}. 
We show the $1\sigma$ ranges of cosmologically preferred SI and MI regions in red bands from Table~\ref{tab:numbers}.
Note that, for the choice of fixed parameters, both SI and MI regions lie in the parameter space of efficient conversion (yellow region)
and showcase \emph{viable} parameter space where neutrino mass is relaxed. Additionally, the SI region is cosmologically preferred over $\Lambda$CDM  which corresponds to the limit $\log_{10}(G_{\rm eff}{\rm MeV}^2) \lesssim -5$.  

\medskip
\noindent
{\textit{Flavor-universal self-interactions--}}
So far, we have discussed how to reproduce the flavor-specific self-interaction scenario with DR. In the limit of large DR flavors, for instance, $n_\chi \gtrsim 40$, the active neutrino number density is drastically reduced: $N_\nu \lesssim 0.2 $. Meanwhile, self-interacting DR number density increases: $N_\chi \gtrsim  2.8$,  contributing to almost all of the $N_{\rm eff}^{\rm tot}$. This scenario mimics flavor-universal neutrino self-interactions, with the role of neutrinos played by DR. Because of very low neutrino number density, the neutrino mass constraint from cosmological analysis is significantly relaxed in this case (see~Eq.~\eqref{eq:massrelax}), which is within the reach of future direct search experiments like Project 8~\cite{Project8:2022wqh} or even KATRIN~\cite{KATRIN:2022ayy}. 
Although the SI mode in flavor-universal interactions is not favored by CMB and LSS data, the MI mode is allowed by data at a large significance~\cite{RoyChoudhury:2020dmd,He:2025jwp,Poudou:2025qcx}. In our framework, both SI and MI modes can be realized free from those constraints (see~{\it Supplemental} Sec.~III).
%%%%%%%%%%%%%%%%%%%%%%

\medskip
\noindent
{\textbf{\textit{Conclusion.--}}}
This work highlights a broader paradigm in neutrino cosmology, where the degeneracy between neutrinos and neutrino-like dark radiation opens up new avenues for addressing several cosmological tensions while respecting BBN and terrestrial constraints. 
We demonstrated this with a simple type-I seesaw framework that provides a neutrino-mediator coupling that is weak enough to evade current laboratory bounds while impacting cosmology effectively via a strongly self-interacting dark sector. 
In particular, the specific choice of our model parameters predicts an MeV-scale sterile neutrino, which is cosmologically allowed for sufficiently low reheating temperatures  $T_{\rm rh}\lesssim 1$ GeV, while remaining    
accessible to laboratory probes. 
Future precision measurements of the high-$\ell$ CMB modes~\cite{Das:2023npl}, LSS, and 21-cm probes~\cite{Libanore:2025ack}, as well as laboratory searches for absolute neutrino mass~\cite{KATRIN:2022ayy, Project8:2022wqh} and sterile neutrinos~\cite{Abdullahi:2022jlv}, will be crucial for distinguishing conventional neutrinos from such self-interacting dark sector scenarios.
%%%%%%%%%%%%%%%%%%%%%
\acknowledgments
\medskip
\noindent
{\textbf{\textit{Acknowledgments.--}}}
We would like to thank Cristina Benso, Nikita Blinov, Kimberly Boddy, Bhaskar Dutta, David Imig, Can Kilic, Thomas Schwetz-Mangold, Jessie Shelton, and Yuhsin Tsai for useful discussions. This work used the high-performance computing service at the University of Notre Dame, managed by the Center for Research Computing (CRC) (\url{https://crc.nd.edu}). The authors acknowledge the Texas Advanced Computing Center (TACC) at The University of Texas at Austin for providing computational resources that have contributed to the research results reported within this {\it Letter} (URL: \url{http://www.tacc.utexas.edu}). A.D. was supported by the Government of India DAE project No. RSI 4001. The work of B.D. was partly supported by the U.S. Department of Energy under Grant No.~DE-SC0017987, and by a Humboldt Fellowship from the Alexander von Humboldt Foundation. C.G. acknowledges support from the NSFC (Grant No. 25201009). S.G. acknowledges support from the NSF under Grant No. PHY-2413016.

\bibliography{main}

@article{Agostini:2015nwa,
    author = "Agostini, M. and others",
    title = "{Results on $\beta \beta $ decay with emission of two neutrinos or Majorons in$^{76}$ Ge from GERDA Phase I}",
    eprint = "1501.02345",
    archivePrefix = "arXiv",
    primaryClass = "nucl-ex",
    doi = "10.1140/epjc/s10052-015-3627-y",
    journal = "Eur. Phys. J. C",
    volume = "75",
    number = "9",
    pages = "416",
    year = "2015"
}

@article{Ghosh:2019tab,
    author = "Ghosh, Subhajit and Khatri, Rishi and Roy, Tuhin S.",
    title = "{Can dark neutrino interactions phase out the Hubble tension?}",
    eprint = "1908.09843",
    archivePrefix = "arXiv",
    primaryClass = "hep-ph",
    reportNumber = "TIFR/TH/19-31",
    doi = "10.1103/PhysRevD.102.123544",
    journal = "Phys. Rev. D",
    volume = "102",
    number = "12",
    pages = "123544",
    year = "2020"
}

@article{Naredo-Tuero:2024sgf,
    author = "Naredo-Tuero, Daniel and Escudero, Miguel and Fern{\'a}ndez-Mart{\'\i}nez, Enrique and Marcano, Xabier and Poulin, Vivian",
    title = "{Critical look at the cosmological neutrino mass bound}",
    eprint = "2407.13831",
    archivePrefix = "arXiv",
    primaryClass = "astro-ph.CO",
    reportNumber = "CERN-TH-2024-115, IFT-UAM/CSIC-24-106",
    doi = "10.1103/PhysRevD.110.123537",
    journal = "Phys. Rev. D",
    volume = "110",
    number = "12",
    pages = "123537",
    year = "2024"
}

@article{Blum:2018ljv,
    author = "Blum, Kfir and Nir, Yosef and Shavit, Michal",
    title = "{Neutrinoless double-beta decay with massive scalar emission}",
    eprint = "1802.08019",
    archivePrefix = "arXiv",
    primaryClass = "hep-ph",
    doi = "10.1016/j.physletb.2018.08.022",
    journal = "Phys. Lett. B",
    volume = "785",
    pages = "354--361",
    year = "2018"
}

@article{Lessa:2007up,
    author = "Lessa, A. P. and Peres, O. L. G.",
    title = "{Revising limits on neutrino-Majoron couplings}",
    eprint = "hep-ph/0701068",
    archivePrefix = "arXiv",
    doi = "10.1103/PhysRevD.75.094001",
    journal = "Phys. Rev. D",
    volume = "75",
    pages = "094001",
    year = "2007"
}

@article{Kreisch:2019yzn,
    author = "Kreisch, Christina D. and Cyr-Racine, Francis-Yan and Dor\'e, Olivier",
    title = "{Neutrino puzzle: Anomalies, interactions, and cosmological tensions}",
    eprint = "1902.00534",
    archivePrefix = "arXiv",
    primaryClass = "astro-ph.CO",
    doi = "10.1103/PhysRevD.101.123505",
    journal = "Phys. Rev. D",
    volume = "101",
    number = "12",
    pages = "123505",
    year = "2020"
}

@article{He:2023oke,
    author = "He, Adam and An, Rui and Ivanov, Mikhail M. and Gluscevic, Vera",
    title = "{Self-interacting neutrinos in light of large-scale structure data}",
    eprint = "2309.03956",
    archivePrefix = "arXiv",
    primaryClass = "astro-ph.CO",
    reportNumber = "MIT-CTP/5608",
    doi = "10.1103/PhysRevD.109.103527",
    journal = "Phys. Rev. D",
    volume = "109",
    number = "10",
    pages = "103527",
    year = "2024"
}

@article{Kreisch:2022zxp,
    author = "Kreisch, Christina D. and others",
    title = "{Atacama Cosmology Telescope: The persistence of neutrino self-interaction in cosmological measurements}",
    eprint = "2207.03164",
    archivePrefix = "arXiv",
    primaryClass = "astro-ph.CO",
    doi = "10.1103/PhysRevD.109.043501",
    journal = "Phys. Rev. D",
    volume = "109",
    number = "4",
    pages = "043501",
    year = "2024"
}

@article{Lyu:2020lps,
    author = "Lyu, Kun-Feng and Stamou, Emmanuel and Wang, Lian-Tao",
    title = "{Self-interacting neutrinos: Solution to Hubble tension versus experimental constraints}",
    eprint = "2004.10868",
    archivePrefix = "arXiv",
    primaryClass = "hep-ph",
    doi = "10.1103/PhysRevD.103.015004",
    journal = "Phys. Rev. D",
    volume = "103",
    number = "1",
    pages = "015004",
    year = "2021"
}

@article{KATRIN:2022ayy,
    author = "Aker, M. and others",
    collaboration = "KATRIN",
    title = "{KATRIN: status and prospects for the neutrino mass and beyond}",
    eprint = "2203.08059",
    archivePrefix = "arXiv",
    primaryClass = "nucl-ex",
    doi = "10.1088/1361-6471/ac834e",
    journal = "J. Phys. G",
    volume = "49",
    number = "10",
    pages = "100501",
    year = "2022"
}

@article{Thomas:2019ran,
    author = "Thomas, Luke C. and Dezen, Ted and Grohs, Evan B. and Kishimoto, Chad T.",
    title = "{Electron-Positron Annihilation Freeze-Out in the Early Universe}",
    eprint = "1910.14050",
    archivePrefix = "arXiv",
    primaryClass = "hep-ph",
    doi = "10.1103/PhysRevD.101.063507",
    journal = "Phys. Rev. D",
    volume = "101",
    number = "6",
    pages = "063507",
    year = "2020"
}

@article{Blinov:2019gcj,
    author = "Blinov, Nikita and Kelly, Kevin James and Krnjaic, Gordan Z and McDermott, Samuel D",
    title = "{Constraining the Self-Interacting Neutrino Interpretation of the Hubble Tension}",
    eprint = "1905.02727",
    archivePrefix = "arXiv",
    primaryClass = "astro-ph.CO",
    reportNumber = "FERMILAB-PUB-19-175-A-T",
    doi = "10.1103/PhysRevLett.123.191102",
    journal = "Phys. Rev. Lett.",
    volume = "123",
    number = "19",
    pages = "191102",
    year = "2019"
}

@article{Lesgourgues:2006nd,
    author = "Lesgourgues, Julien and Pastor, Sergio",
    title = "{Massive neutrinos and cosmology}",
    eprint = "astro-ph/0603494",
    archivePrefix = "arXiv",
    reportNumber = "LAPTH-1131-05, IFIC-05-59",
    doi = "10.1016/j.physrep.2006.04.001",
    journal = "Phys. Rept.",
    volume = "429",
    pages = "307--379",
    year = "2006"
}

@article{Minkowski:1977sc,
    author = "Minkowski, Peter",
    title = "{$\mu \to e\gamma$ at a Rate of One Out of $10^{9}$ Muon Decays?}",
    reportNumber = "Print-77-0182 (BERN)",
    doi = "10.1016/0370-2693(77)90435-X",
    journal = "Phys. Lett. B",
    volume = "67",
    pages = "421--428",
    year = "1977"
}

@article{Mohapatra:1979ia,
    author = "Mohapatra, Rabindra N. and Senjanovic, Goran",
    title = "{Neutrino Mass and Spontaneous Parity Nonconservation}",
    reportNumber = "MDDP-TR-80-060, MDDP-PP-80-105, CCNY-HEP-79-10",
    doi = "10.1103/PhysRevLett.44.912",
    journal = "Phys. Rev. Lett.",
    volume = "44",
    pages = "912",
    year = "1980"
}

@article{Gell-Mann:1979vob,
    author = "Gell-Mann, Murray and Ramond, Pierre and Slansky, Richard",
    title = "{Complex Spinors and Unified Theories}",
    eprint = "1306.4669",
    archivePrefix = "arXiv",
    primaryClass = "hep-th",
    reportNumber = "PRINT-80-0576",
    journal = "Conf. Proc. C",
    volume = "790927",
    pages = "315--321",
    year = "1979"
}

@article{Yanagida:1979as,
    author = "Yanagida, Tsutomu",
    editor = "Sawada, Osamu and Sugamoto, Akio",
    title = "{Horizontal gauge symmetry and masses of neutrinos}",
    reportNumber = "KEK-79-18-95",
    journal = "Conf. Proc. C",
    volume = "7902131",
    pages = "95--99",
    year = "1979"
}

@article{Schechter:1980gr,
    author = "Schechter, J. and Valle, J. W. F.",
    title = "{Neutrino Masses in SU(2) x U(1) Theories}",
    reportNumber = "SU-4217-167, COO-3533-167",
    doi = "10.1103/PhysRevD.22.2227",
    journal = "Phys. Rev. D",
    volume = "22",
    pages = "2227",
    year = "1980"
}

@article{Deppisch:2020sqh,
    author = "Deppisch, Frank F. and Graf, Lukas and Rodejohann, Werner and Xu, Xun-Jie",
    title = "{Neutrino Self-Interactions and Double Beta Decay}",
    eprint = "2004.11919",
    archivePrefix = "arXiv",
    primaryClass = "hep-ph",
    doi = "10.1103/PhysRevD.102.051701",
    journal = "Phys. Rev. D",
    volume = "102",
    number = "5",
    pages = "051701",
    year = "2020"
}

@article{Fiorillo:2023ytr,
    author = "Fiorillo, Damiano F. G. and Raffelt, Georg G. and Vitagliano, Edoardo",
    title = "{Large Neutrino Secret Interactions Have a Small Impact on Supernovae}",
    eprint = "2307.15115",
    archivePrefix = "arXiv",
    primaryClass = "hep-ph",
    doi = "10.1103/PhysRevLett.132.021002",
    journal = "Phys. Rev. Lett.",
    volume = "132",
    number = "2",
    pages = "021002",
    year = "2024"
}

@article{Casas:2001sr,
    author = "Casas, J. A. and Ibarra, A.",
    title = "{Oscillating neutrinos and $\mu \to e, \gamma$}",
    eprint = "hep-ph/0103065",
    archivePrefix = "arXiv",
    reportNumber = "IEM-FT-211-01, OUTP-01-11P, IFT-UAM-CSIC-01-08",
    doi = "10.1016/S0550-3213(01)00475-8",
    journal = "Nucl. Phys. B",
    volume = "618",
    pages = "171--204",
    year = "2001"
}

@article{He:2020zns,
    author = "He, Hong-Jian and Ma, Yin-Zhe and Zheng, Jiaming",
    title = "{Resolving Hubble Tension by Self-Interacting Neutrinos with Dirac Seesaw}",
    eprint = "2003.12057",
    archivePrefix = "arXiv",
    primaryClass = "hep-ph",
    doi = "10.1088/1475-7516/2020/11/003",
    journal = "JCAP",
    volume = "11",
    pages = "003",
    year = "2020"
}

@article{Berlin:2018ztp,
    author = "Berlin, Asher and Blinov, Nikita",
    title = "{Thermal neutrino portal to sub-MeV dark matter}",
    eprint = "1807.04282",
    archivePrefix = "arXiv",
    primaryClass = "hep-ph",
    reportNumber = "SLAC-PUB-17278",
    doi = "10.1103/PhysRevD.99.095030",
    journal = "Phys. Rev. D",
    volume = "99",
    number = "9",
    pages = "095030",
    year = "2019"
}

@inproceedings{Project8:2022wqh,
    author = "Esfahani, A. Ashtari and others",
    collaboration = "Project 8",
    title = "{The Project 8 Neutrino Mass Experiment}",
    booktitle = "{Snowmass 2021}",
    eprint = "2203.07349",
    archivePrefix = "arXiv",
    primaryClass = "nucl-ex",
    month = "3",
    year = "2022"
}

@article{Virtanen:2019joe,
    author = "Virtanen, Pauli and others",
    title = "{SciPy 1.0--Fundamental Algorithms for Scientific Computing in Python}",
    eprint = "1907.10121",
    archivePrefix = "arXiv",
    primaryClass = "cs.MS",
    doi = "10.1038/s41592-019-0686-2",
    journal = "Nature Meth.",
    volume = "17",
    pages = "261",
    year = "2020"
}

@article{EscuderoAbenza:2020cmq,
    author = "Escudero Abenza, Miguel",
    title = "{Precision early universe thermodynamics made simple: $N_{\rm eff}$ and neutrino decoupling in the Standard Model and beyond}",
    eprint = "2001.04466",
    archivePrefix = "arXiv",
    primaryClass = "hep-ph",
    reportNumber = "KCL-2019-85",
    doi = "10.1088/1475-7516/2020/05/048",
    journal = "JCAP",
    volume = "05",
    pages = "048",
    year = "2020"
}

@article{Escudero:2022gez,
    author = "Escudero, Miguel and Schwetz, Thomas and Terol-Calvo, Jorge",
    title = "{A seesaw model for large neutrino masses in concordance with cosmology}",
    eprint = "2211.01729",
    archivePrefix = "arXiv",
    primaryClass = "hep-ph",
    reportNumber = "CERN-TH-2022-180",
    doi = "10.1007/JHEP02(2023)142",
    journal = "JHEP",
    volume = "02",
    pages = "142",
    year = "2023",
    note = "[Addendum: JHEP 06, 119 (2024)]"
}

@article{deGouvea:2016qpx,
    author = "de Gouv\^ea, Andr\'e",
    title = "{Neutrino Mass Models}",
    doi = "10.1146/annurev-nucl-102115-044600",
    journal = "Ann. Rev. Nucl. Part. Sci.",
    volume = "66",
    pages = "197--217",
    year = "2016"
}

@article{Brdar:2020nbj,
    author = "Brdar, Vedran and Lindner, Manfred and Vogl, Stefan and Xu, Xun-Jie",
    title = "{Revisiting neutrino self-interaction constraints from $Z$ and $\tau$ decays}",
    eprint = "2003.05339",
    archivePrefix = "arXiv",
    primaryClass = "hep-ph",
    doi = "10.1103/PhysRevD.101.115001",
    journal = "Phys. Rev. D",
    volume = "101",
    number = "11",
    pages = "115001",
    year = "2020"
}

@article{Dev:2024twk,
    author = "Dev, P. S. Bhupal and Kim, Doojin and Sathyan, Deepak and Sinha, Kuver and Zhang, Yongchao",
    title = "{New Laboratory Constraints on Neutrinophilic Mediators}",
    eprint = "2407.12738",
    archivePrefix = "arXiv",
    primaryClass = "hep-ph",
    reportNumber = "CETUP-2024-005",
    doi = "10.1016/j.physletb.2025.139765",
    journal = "Phys. Lett. B",
    volume = "868",
    pages = "139765",
    year = "2025"
}

@article{Berlin:2017ftj,
    author = "Berlin, Asher and Blinov, Nikita",
    title = "{Thermal Dark Matter Below an MeV}",
    eprint = "1706.07046",
    archivePrefix = "arXiv",
    primaryClass = "hep-ph",
    reportNumber = "SLAC-PUB-16989",
    doi = "10.1103/PhysRevLett.120.021801",
    journal = "Phys. Rev. Lett.",
    volume = "120",
    number = "2",
    pages = "021801",
    year = "2018"
}

@article{Venzor:2023aka,
    author = "Venzor, Jorge and Garcia-Arroyo, Gabriela and De-Santiago, Josue and P\'erez-Lorenzana, Abdel",
    title = "{Resonant neutrino self-interactions and the H0 tension}",
    eprint = "2303.12792",
    archivePrefix = "arXiv",
    primaryClass = "astro-ph.CO",
    doi = "10.1103/PhysRevD.108.043536",
    journal = "Phys. Rev. D",
    volume = "108",
    number = "4",
    pages = "043536",
    year = "2023"
}

@article{KATRIN:2024cdt,
    author = "Aker, Max and others",
    collaboration = "KATRIN",
    title = "{Direct neutrino-mass measurement based on 259 days of KATRIN data}",
    eprint = "2406.13516",
    archivePrefix = "arXiv",
    primaryClass = "nucl-ex",
    doi = "10.1126/science.adq9592",
    journal = "Science",
    volume = "388",
    number = "6743",
    pages = "adq9592",
    year = "2025"
}

@article{Craig:2024tky,
    author = "Craig, Nathaniel and Green, Daniel and Meyers, Joel and Rajendran, Surjeet",
    title = "{No \ensuremath{\nu}s is Good News}",
    eprint = "2405.00836",
    archivePrefix = "arXiv",
    primaryClass = "astro-ph.CO",
    reportNumber = "FERMILAB-PUB-24-0492-SQMS-V",
    doi = "10.1007/JHEP09(2024)097",
    journal = "JHEP",
    volume = "09",
    pages = "097",
    year = "2024"
}

@article{Gelmini:2004ah,
    author = "Gelmini, Graciela and Palomares-Ruiz, Sergio and Pascoli, Silvia",
    title = "{Low reheating temperature and the visible sterile neutrino}",
    eprint = "astro-ph/0403323",
    archivePrefix = "arXiv",
    reportNumber = "UCLA-04-TEP-5",
    doi = "10.1103/PhysRevLett.93.081302",
    journal = "Phys. Rev. Lett.",
    volume = "93",
    pages = "081302",
    year = "2004"
}

@article{Gondolo:1990dk,
    author = "Gondolo, Paolo and Gelmini, Graciela",
    title = "{Cosmic abundances of stable particles: Improved analysis}",
    reportNumber = "UCLA-90-TEP-68",
    doi = "10.1016/0550-3213(91)90438-4",
    journal = "Nucl. Phys. B",
    volume = "360",
    pages = "145--179",
    year = "1991"
}

@article{Dolgov:1997mb,
    author = "Dolgov, A. D. and Hansen, S. H. and Semikoz, D. V.",
    title = "{Nonequilibrium corrections to the spectra of massless neutrinos in the early universe}",
    eprint = "hep-ph/9703315",
    archivePrefix = "arXiv",
    reportNumber = "TAC-1997-010",
    doi = "10.1016/S0550-3213(97)00479-3",
    journal = "Nucl. Phys. B",
    volume = "503",
    pages = "426--444",
    year = "1997"
}

@article{Esteban:2024eli,
    author = "Esteban, Ivan and Gonzalez-Garcia, M. C. and Maltoni, Michele and Martinez-Soler, Ivan and Pinheiro, Jo\~ao Paulo and Schwetz, Thomas",
    title = "{NuFit-6.0: updated global analysis of three-flavor neutrino oscillations}",
    eprint = "2410.05380",
    archivePrefix = "arXiv",
    primaryClass = "hep-ph",
    reportNumber = "IFT-UAM/CSIC-24-140, YITP-SB-2024-24, IPPP/24/64, IPPP/24/64, IFT-UAM/CSIC-24-140, YITP-SB-2024-24",
    doi = "10.1007/JHEP12(2024)216",
    journal = "JHEP",
    volume = "12",
    pages = "216",
    year = "2024"
}

@article{DESI:2024mwx,
    author = "Adame, A. G. and others",
    collaboration = "DESI",
    title = "{DESI 2024 VI: cosmological constraints from the measurements of baryon acoustic oscillations}",
    eprint = "2404.03002",
    archivePrefix = "arXiv",
    primaryClass = "astro-ph.CO",
    reportNumber = "FERMILAB-PUB-24-0154-PPD",
    doi = "10.1088/1475-7516/2025/02/021",
    journal = "JCAP",
    volume = "02",
    pages = "021",
    year = "2025"
}

@article{DESI:2025ejh,
    author = "Elbers, W. and others",
    collaboration = "DESI",
    title = "{Constraints on Neutrino Physics from DESI DR2 BAO and DR1 Full Shape}",
    eprint = "2503.14744",
    archivePrefix = "arXiv",
    primaryClass = "astro-ph.CO",
    reportNumber = "FERMILAB-PUB-25-0168-PPD",
    doi = "10.1103/w9pk-xsk7",
    journal = "Phys. Rev. D",
    volume = "112",
    number = "8",
    pages = "083513",
    year = "2025"
}

@article{Planck:2018vyg,
    author = "Aghanim, N. and others",
    collaboration = "Planck",
    title = "{Planck 2018 results. VI. Cosmological parameters}",
    eprint = "1807.06209",
    archivePrefix = "arXiv",
    primaryClass = "astro-ph.CO",
    doi = "10.1051/0004-6361/201833910",
    journal = "Astron. Astrophys.",
    volume = "641",
    pages = "A6",
    year = "2020",
    note = "[Erratum: Astron.Astrophys. 652, C4 (2021)]"
}

@article{Brune:2018sab,
    author = {Brune, Tim and P\"as, Heinrich},
    title = "{Massive Majorons and constraints on the Majoron-neutrino coupling}",
    eprint = "1808.08158",
    archivePrefix = "arXiv",
    primaryClass = "hep-ph",
    reportNumber = "DO-TH 18/23",
    doi = "10.1103/PhysRevD.99.096005",
    journal = "Phys. Rev. D",
    volume = "99",
    number = "9",
    pages = "096005",
    year = "2019"
}

@article{Shalgar:2019rqe,
    author = "Shalgar, Shashank and Tamborra, Irene and Bustamante, Mauricio",
    title = "{Core-collapse supernovae stymie secret neutrino interactions}",
    eprint = "1912.09115",
    archivePrefix = "arXiv",
    primaryClass = "astro-ph.HE",
    doi = "10.1103/PhysRevD.103.123008",
    journal = "Phys. Rev. D",
    volume = "103",
    number = "12",
    pages = "123008",
    year = "2021"
}

@article{Kolb:1987qy,
    author = "Kolb, Edward W. and Turner, Michael S.",
    title = "{Supernova SN 1987a and the Secret Interactions of Neutrinos}",
    reportNumber = "FERMILAB-PUB-87-110-A",
    doi = "10.1103/PhysRevD.36.2895",
    journal = "Phys. Rev. D",
    volume = "36",
    pages = "2895",
    year = "1987"
}

@article{DESI:2025zgx,
    author = "Abdul Karim, M. and others",
    collaboration = "DESI",
    title = "{DESI DR2 Results II: Measurements of Baryon Acoustic Oscillations and Cosmological Constraints}",
    eprint = "2503.14738",
    archivePrefix = "arXiv",
    primaryClass = "astro-ph.CO",
    reportNumber = "FERMILAB-PUB-25-0169-PPD",
    doi = "10.1103/tr6y-kpc6",
    journal = "Phys. Rev. D",
    volume = "112",
    number = "8",
    pages = "083515",
    year = "2025"
}

@article{Zhang:2024meg,
    author = "Zhang, Yue",
    title = "{Neutrino Self-interaction and Weak Mixing Angle Measurements}",
    eprint = "2411.05070",
    archivePrefix = "arXiv",
    primaryClass = "hep-ph",
    doi = "10.1103/7y8g-56fz",
    journal = "Phys. Rev. D",
    volume = "112",
    number = "3",
    pages = "035027",
    year = "2025"
}

@article{Berryman:2018ogk,
    author = "Berryman, Jeffrey M. and De Gouv\^ea, Andr\'e and Kelly, Kevin J. and Zhang, Yue",
    title = "{Lepton-Number-Charged Scalars and Neutrino Beamstrahlung}",
    eprint = "1802.00009",
    archivePrefix = "arXiv",
    primaryClass = "hep-ph",
    reportNumber = "NUHEP-TH-18-03, FERMILAB-PUB-18-020-T",
    doi = "10.1103/PhysRevD.97.075030",
    journal = "Phys. Rev. D",
    volume = "97",
    number = "7",
    pages = "075030",
    year = "2018"
}

@article{deGouvea:2019qaz,
    author = "de Gouv\^ea, Andr\'e and Dev, P. S. Bhupal and Dutta, Bhaskar and Ghosh, Tathagata and Han, Tao and Zhang, Yongchao",
    title = "{Leptonic Scalars at the LHC}",
    eprint = "1910.01132",
    archivePrefix = "arXiv",
    primaryClass = "hep-ph",
    reportNumber = "PITT-PACC 1909, MI-TH-1936",
    doi = "10.1007/JHEP07(2020)142",
    journal = "JHEP",
    volume = "07",
    pages = "142",
    year = "2020"
}

@article{Berbig:2020wve,
    author = "Berbig, Maximilian and Jana, Sudip and Trautner, Andreas",
    title = "{The Hubble tension and a renormalizable model of gauged neutrino self-interactions}",
    eprint = "2004.13039",
    archivePrefix = "arXiv",
    primaryClass = "hep-ph",
    doi = "10.1103/PhysRevD.102.115008",
    journal = "Phys. Rev. D",
    volume = "102",
    number = "11",
    pages = "115008",
    year = "2020"
}

@article{Das:2020xke,
    author = "Das, Anirban and Ghosh, Subhajit",
    title = "{Flavor-specific interaction favors strong neutrino self-coupling in the early universe}",
    eprint = "2011.12315",
    archivePrefix = "arXiv",
    primaryClass = "astro-ph.CO",
    reportNumber = "SLAC-PUB-17547",
    doi = "10.1088/1475-7516/2021/07/038",
    journal = "JCAP",
    volume = "07",
    pages = "038",
    year = "2021"
}

@article{Das:2023npl,
    author = "Das, Anirban and Ghosh, Subhajit",
    title = "{The magnificent ACT of flavor-specific neutrino self-interaction}",
    eprint = "2303.08843",
    archivePrefix = "arXiv",
    primaryClass = "astro-ph.CO",
    reportNumber = "SLAC-PUB-17708",
    doi = "10.1088/1475-7516/2023/09/042",
    journal = "JCAP",
    volume = "09",
    pages = "042",
    year = "2023"
}

@article{Berryman:2022hds,
    author = "Berryman, Jeffrey M. and others",
    title = "{Neutrino self-interactions: A white paper}",
    eprint = "2203.01955",
    archivePrefix = "arXiv",
    primaryClass = "hep-ph",
    reportNumber = "CERN-TH-2022-024, DESY-22-035, FERMILAB-PUB-22-099-T",
    doi = "10.1016/j.dark.2023.101267",
    journal = "Phys. Dark Univ.",
    volume = "42",
    pages = "101267",
    year = "2023"
}

@manual{python,
    title={{Python: A dynamic, open source programming language}},
    author={{Python Core Team}},
    organization={{Python Software Foundation}},
    year={2019},
    url={https://www.python.org/},
}

@article{Harris:2020xlr,
    author = "Harris, Charles R. and others",
    title = "{Array programming with NumPy}",
    eprint = "2006.10256",
    archivePrefix = "arXiv",
    primaryClass = "cs.MS",
    doi = "10.1038/s41586-020-2649-2",
    journal = "Nature",
    volume = "585",
    number = "7825",
    pages = "357--362",
    year = "2020"
}

@misc{Mathematica,
  author = {Wolfram Research{,} Inc.},
  title = {Mathematica, {V}ersion 14.0},
  note = {Champaign, IL, 2024}
}

@article{Shtabovenko:2023idz,
    author = "Shtabovenko, Vladyslav and Mertig, Rolf and Orellana, Frederik",
    title = "{FeynCalc 10: Do multiloop integrals dream of computer codes?}",
    eprint = "2312.14089",
    archivePrefix = "arXiv",
    primaryClass = "hep-ph",
    reportNumber = "P3H-23-089, TTP23-056, SI-HEP-2023-27",
    doi = "10.1016/j.cpc.2024.109357",
    journal = "Comput. Phys. Commun.",
    volume = "306",
    pages = "109357",
    year = "2025"
}

@article{Alloul:2013bka,
    author = "Alloul, Adam and Christensen, Neil D. and Degrande, C\'eline and Duhr, Claude and Fuks, Benjamin",
    title = "{FeynRules  2.0 - A complete toolbox for tree-level phenomenology}",
    eprint = "1310.1921",
    archivePrefix = "arXiv",
    primaryClass = "hep-ph",
    reportNumber = "CERN-PH-TH-2013-239, MCNET-13-14, IPPP-13-71, DCPT-13-142, PITT-PACC-1308",
    doi = "10.1016/j.cpc.2014.04.012",
    journal = "Comput. Phys. Commun.",
    volume = "185",
    pages = "2250--2300",
    year = "2014"
}

@article{feynarts,
    author = {T.~Hahn},
    title = {{FeynArts} 3.11 User's Guide},
    journal = {Comput. Phys. Commun.},
    volume = {140},
    pages = {418--430},
    year = {2001},
    doi = {10.1016/S0010-4655(00)00252-0},
    eprint = {hep-ph/0012260},
}

@article{Brinckmann:2018cvx,
    author = "Brinckmann, Thejs and Lesgourgues, Julien",
    title = "{MontePython 3: boosted MCMC sampler and other features}",
    eprint = "1804.07261",
    archivePrefix = "arXiv",
    primaryClass = "astro-ph.CO",
    reportNumber = "TTK-18-15",
    doi = "10.1016/j.dark.2018.100260",
    journal = "Phys. Dark Univ.",
    volume = "24",
    pages = "100260",
    year = "2019"
}

@article{Ma:1995ey,
    author = "Ma, Chung-Pei and Bertschinger, Edmund",
    title = "{Cosmological perturbation theory in the synchronous and conformal Newtonian gauges}",
    eprint = "astro-ph/9506072",
    archivePrefix = "arXiv",
    doi = "10.1086/176550",
    journal = "Astrophys. J.",
    volume = "455",
    pages = "7--25",
    year = "1995"
}

@article{Brinckmann:2022ajr,
    author = "Brinckmann, Thejs and Chang, Jae Hyeok and Du, Peizhi and LoVerde, Marilena",
    title = "{Confronting interacting dark radiation scenarios with cosmological data}",
    eprint = "2212.13264",
    archivePrefix = "arXiv",
    primaryClass = "astro-ph.CO",
    doi = "10.1103/PhysRevD.107.123517",
    journal = "Phys. Rev. D",
    volume = "107",
    number = "12",
    pages = "123517",
    year = "2023"
}

@article{Blas:2011rf,
    author = "Blas, Diego and Lesgourgues, Julien and Tram, Thomas",
    title = "{The Cosmic Linear Anisotropy Solving System (CLASS) II: Approximation schemes}",
    eprint = "1104.2933",
    archivePrefix = "arXiv",
    primaryClass = "astro-ph.CO",
    reportNumber = "CERN-PH-TH-2011-082, LAPTH-010-11",
    doi = "10.1088/1475-7516/2011/07/034",
    journal = "JCAP",
    volume = "07",
    pages = "034",
    year = "2011"
}

@article{Lesgourgues:2011re,
    author = "Lesgourgues, Julien",
    title = "{The Cosmic Linear Anisotropy Solving System (CLASS) I: Overview}",
    eprint = "1104.2932",
    archivePrefix = "arXiv",
    primaryClass = "astro-ph.IM",
    month = "4",
    year = "2011"
}

@article{Feroz:2007kg,
    author = "Feroz, Farhan and Hobson, M. P.",
    title = "{Multimodal nested sampling: an efficient and robust alternative to MCMC methods for astronomical data analysis}",
    eprint = "0704.3704",
    archivePrefix = "arXiv",
    primaryClass = "astro-ph",
    doi = "10.1111/j.1365-2966.2007.12353.x",
    journal = "Mon. Not. Roy. Astron. Soc.",
    volume = "384",
    pages = "449",
    year = "2008"
}

@article{Feroz:2008xx,
    author = "Feroz, F. and Hobson, M. P. and Bridges, M.",
    title = "{MultiNest: an efficient and robust Bayesian inference tool for cosmology and particle physics}",
    eprint = "0809.3437",
    archivePrefix = "arXiv",
    primaryClass = "astro-ph",
    doi = "10.1111/j.1365-2966.2009.14548.x",
    journal = "Mon. Not. Roy. Astron. Soc.",
    volume = "398",
    pages = "1601--1614",
    year = "2009"
}

@article{Feroz:2013hea,
    author = "Feroz, F. and Hobson, M. P. and Cameron, E. and Pettitt, A. N.",
    title = "{Importance Nested Sampling and the MultiNest Algorithm}",
    eprint = "1306.2144",
    archivePrefix = "arXiv",
    primaryClass = "astro-ph.IM",
    doi = "10.21105/astro.1306.2144",
    journal = "Open J. Astrophys.",
    volume = "2",
    number = "1",
    pages = "10",
    year = "2019"
}

@article{Buchner:2014nha,
    author = "Buchner, J. and Georgakakis, A. and Nandra, K. and Hsu, L. and Rangel, C. and Brightman, M. and Merloni, A. and Salvato, M. and Donley, J. and Kocevski, D.",
    title = "{X-ray spectral modelling of the AGN obscuring region in the CDFS: Bayesian model selection and catalogue}",
    eprint = "1402.0004",
    archivePrefix = "arXiv",
    primaryClass = "astro-ph.HE",
    doi = "10.1051/0004-6361/201322971",
    journal = "Astron. Astrophys.",
    volume = "564",
    pages = "A125",
    year = "2014"
}

@article{Audren:2012wb,
    author = "Audren, Benjamin and Lesgourgues, Julien and Benabed, Karim and Prunet, Simon",
    title = "{Conservative Constraints on Early Cosmology: an illustration of the Monte Python cosmological parameter inference code}",
    eprint = "1210.7183",
    archivePrefix = "arXiv",
    primaryClass = "astro-ph.CO",
    reportNumber = "CERN-PH-TH-2012-290, LAPTH-048-12",
    doi = "10.1088/1475-7516/2013/02/001",
    journal = "JCAP",
    volume = "02",
    pages = "001",
    year = "2013"
}

@article{Planck:2019nip,
    author = "Aghanim, N. and others",
    collaboration = "Planck",
    title = "{Planck 2018 results. V. CMB power spectra and likelihoods}",
    eprint = "1907.12875",
    archivePrefix = "arXiv",
    primaryClass = "astro-ph.CO",
    doi = "10.1051/0004-6361/201936386",
    journal = "Astron. Astrophys.",
    volume = "641",
    pages = "A5",
    year = "2020"
}

@article{Planck:2018lbu,
    author = "Aghanim, N. and others",
    collaboration = "Planck",
    title = "{Planck 2018 results. VIII. Gravitational lensing}",
    eprint = "1807.06210",
    archivePrefix = "arXiv",
    primaryClass = "astro-ph.CO",
    doi = "10.1051/0004-6361/201833886",
    journal = "Astron. Astrophys.",
    volume = "641",
    pages = "A8",
    year = "2020"
}

@article{Herold:2024nvk,
    author = "Herold, Laura and Kamionkowski, Marc",
    title = "{Revisiting the impact of neutrino mass hierarchies on neutrino mass constraints in light of recent DESI data}",
    eprint = "2412.03546",
    archivePrefix = "arXiv",
    primaryClass = "astro-ph.CO",
    doi = "10.1103/PhysRevD.111.083518",
    journal = "Phys. Rev. D",
    volume = "111",
    number = "8",
    pages = "083518",
    year = "2025"
}

@article{Bennett:2020zkv,
    author = "Bennett, Jack J. and Buldgen, Gilles and De Salas, Pablo F. and Drewes, Marco and Gariazzo, Stefano and Pastor, Sergio and Wong, Yvonne Y. Y.",
    title = "{Towards a precision calculation of $N_{\rm eff}$ in the Standard Model II: Neutrino decoupling in the presence of flavour oscillations and finite-temperature QED}",
    eprint = "2012.02726",
    archivePrefix = "arXiv",
    primaryClass = "hep-ph",
    reportNumber = "CPPC-2020-10",
    doi = "10.1088/1475-7516/2021/04/073",
    journal = "JCAP",
    volume = "04",
    pages = "073",
    year = "2021"
}

@article{Cyr-Racine:2013jua,
    author = "Cyr-Racine, Francis-Yan and Sigurdson, Kris",
    title = "{Limits on Neutrino-Neutrino Scattering in the Early Universe}",
    eprint = "1306.1536",
    archivePrefix = "arXiv",
    primaryClass = "astro-ph.CO",
    doi = "10.1103/PhysRevD.90.123533",
    journal = "Phys. Rev. D",
    volume = "90",
    number = "12",
    pages = "123533",
    year = "2014"
}

@article{Lancaster:2017ksf,
    author = "Lancaster, Lachlan and Cyr-Racine, Francis-Yan and Knox, Lloyd and Pan, Zhen",
    title = "{A tale of two modes: Neutrino free-streaming in the early universe}",
    eprint = "1704.06657",
    archivePrefix = "arXiv",
    primaryClass = "astro-ph.CO",
    doi = "10.1088/1475-7516/2017/07/033",
    journal = "JCAP",
    volume = "07",
    pages = "033",
    year = "2017"
}

@article{Chang:2022aas,
    author = "Chang, Po-Wen and Esteban, Ivan and Beacom, John F. and Thompson, Todd A. and Hirata, Christopher M.",
    title = "{Toward Powerful Probes of Neutrino Self-Interactions in Supernovae}",
    eprint = "2206.12426",
    archivePrefix = "arXiv",
    primaryClass = "hep-ph",
    doi = "10.1103/PhysRevLett.131.071002",
    journal = "Phys. Rev. Lett.",
    volume = "131",
    number = "7",
    pages = "071002",
    year = "2023"
}

@article{Farzan:2015pca,
    author = "Farzan, Yasaman and Hannestad, Steen",
    title = "{Neutrinos secretly converting to lighter particles to please both KATRIN and the cosmos}",
    eprint = "1510.02201",
    archivePrefix = "arXiv",
    primaryClass = "hep-ph",
    doi = "10.1088/1475-7516/2016/02/058",
    journal = "JCAP",
    volume = "02",
    pages = "058",
    year = "2016"
}

@article{Farzan:2002wx,
    author = "Farzan, Yasaman",
    title = "{Bounds on the coupling of the Majoron to light neutrinos from supernova cooling}",
    eprint = "hep-ph/0211375",
    archivePrefix = "arXiv",
    reportNumber = "SLAC-PUB-9543, SISSA-69-2002-EP",
    doi = "10.1103/PhysRevD.67.073015",
    journal = "Phys. Rev. D",
    volume = "67",
    pages = "073015",
    year = "2003"
}

@article{Heurtier:2016otg,
    author = "Heurtier, Lucien and Zhang, Yongchao",
    title = "{Supernova Constraints on Massive (Pseudo)Scalar Coupling to Neutrinos}",
    eprint = "1609.05882",
    archivePrefix = "arXiv",
    primaryClass = "hep-ph",
    reportNumber = "ULB-TH-16-16",
    doi = "10.1088/1475-7516/2017/02/042",
    journal = "JCAP",
    volume = "02",
    pages = "042",
    year = "2017"
}

@article{Das:2017iuj,
    author = "Das, Anirban and Dighe, Amol and Sen, Manibrata",
    title = "{New effects of non-standard self-interactions of neutrinos in a supernova}",
    eprint = "1705.00468",
    archivePrefix = "arXiv",
    primaryClass = "hep-ph",
    reportNumber = "TIFR-TH-17-18",
    doi = "10.1088/1475-7516/2017/05/051",
    journal = "JCAP",
    volume = "05",
    pages = "051",
    year = "2017"
}

@article{Manohar:1987ec,
    author = "Manohar, Aneesh",
    title = "{A Limit on the Neutrino-neutrino Scattering Cross-section From the Supernova}",
    reportNumber = "MIT-CTP-1461",
    doi = "10.1016/0370-2693(87)91171-3",
    journal = "Phys. Lett. B",
    volume = "192",
    pages = "217",
    year = "1987"
}

@article{Esteban:2021tub,
    author = "Esteban, Ivan and Pandey, Sujata and Brdar, Vedran and Beacom, John F.",
    title = "{Probing secret interactions of astrophysical neutrinos in the high-statistics era}",
    eprint = "2107.13568",
    archivePrefix = "arXiv",
    primaryClass = "hep-ph",
    reportNumber = "FERMILAB-PUB-21-328-T, nuhep-th/21-06",
    doi = "10.1103/PhysRevD.104.123014",
    journal = "Phys. Rev. D",
    volume = "104",
    number = "12",
    pages = "123014",
    year = "2021"
}

@article{Ioka:2014kca,
    author = "Ioka, Kunihto and Murase, Kohta",
    title = "{IceCube PeV\textendash{}EeV neutrinos and secret interactions of neutrinos}",
    eprint = "1404.2279",
    archivePrefix = "arXiv",
    primaryClass = "astro-ph.HE",
    reportNumber = "KEK-TH-1723, KEK-COSMO-141",
    doi = "10.1093/ptep/ptu090",
    journal = "PTEP",
    volume = "2014",
    number = "6",
    pages = "061E01",
    year = "2014"
}

@article{Huang:2017egl,
    author = "Huang, Guo-yuan and Ohlsson, Tommy and Zhou, Shun",
    title = "{Observational Constraints on Secret Neutrino Interactions from Big Bang Nucleosynthesis}",
    eprint = "1712.04792",
    archivePrefix = "arXiv",
    primaryClass = "hep-ph",
    doi = "10.1103/PhysRevD.97.075009",
    journal = "Phys. Rev. D",
    volume = "97",
    number = "7",
    pages = "075009",
    year = "2018"
}

@article{Pasquini:2015fjv,
    author = "Pasquini, P. S. and Peres, O. L. G.",
    title = "{Bounds on Neutrino-Scalar Yukawa Coupling}",
    eprint = "1511.01811",
    archivePrefix = "arXiv",
    primaryClass = "hep-ph",
    doi = "10.1103/PhysRevD.93.053007",
    journal = "Phys. Rev. D",
    volume = "93",
    number = "5",
    pages = "053007",
    year = "2016",
    note = "[Erratum: Phys.Rev.D 93, 079902 (2016)]"
}

@article{Ng:2014pca,
    author = "Ng, Kenny C. Y. and Beacom, John F.",
    title = "{Cosmic neutrino cascades from secret neutrino interactions}",
    eprint = "1404.2288",
    archivePrefix = "arXiv",
    primaryClass = "astro-ph.HE",
    doi = "10.1103/PhysRevD.90.065035",
    journal = "Phys. Rev. D",
    volume = "90",
    number = "6",
    pages = "065035",
    year = "2014",
    note = "[Erratum: Phys.Rev.D 90, 089904 (2014)]"
}

@article{Bustamante:2020mep,
    author = "Bustamante, Mauricio and Rosenstr\o{}m, Charlotte and Shalgar, Shashank and Tamborra, Irene",
    title = "{Bounds on secret neutrino interactions from high-energy astrophysical neutrinos}",
    eprint = "2001.04994",
    archivePrefix = "arXiv",
    primaryClass = "astro-ph.HE",
    doi = "10.1103/PhysRevD.101.123024",
    journal = "Phys. Rev. D",
    volume = "101",
    number = "12",
    pages = "123024",
    year = "2020"
}

@article{RoyChoudhury:2020dmd,
    author = "Roy Choudhury, Shouvik and Hannestad, Steen and Tram, Thomas",
    title = "{Updated constraints on massive neutrino self-interactions from cosmology in light of the $H_0$ tension}",
    eprint = "2012.07519",
    archivePrefix = "arXiv",
    primaryClass = "astro-ph.CO",
    doi = "10.1088/1475-7516/2021/03/084",
    journal = "JCAP",
    volume = "03",
    pages = "084",
    year = "2021"
}

@article{Fiorillo:2022cdq,
    author = "Fiorillo, Damiano F. G. and Raffelt, Georg G. and Vitagliano, Edoardo",
    title = "{Strong Supernova 1987A Constraints on Bosons Decaying to Neutrinos}",
    eprint = "2209.11773",
    archivePrefix = "arXiv",
    primaryClass = "hep-ph",
    doi = "10.1103/PhysRevLett.131.021001",
    journal = "Phys. Rev. Lett.",
    volume = "131",
    number = "2",
    pages = "021001",
    year = "2023"
}

@article{PIENU:2017wbj,
    author = "Aguilar-Arevalo, A. and others",
    collaboration = "PIENU",
    title = "{Improved search for heavy neutrinos in the decay $\pi\rightarrow e\nu$}",
    eprint = "1712.03275",
    archivePrefix = "arXiv",
    primaryClass = "hep-ex",
    doi = "10.1103/PhysRevD.97.072012",
    journal = "Phys. Rev. D",
    volume = "97",
    number = "7",
    pages = "072012",
    year = "2018"
}

@article{NA62:2020mcv,
    author = "Cortina Gil, Eduardo and others",
    collaboration = "NA62",
    title = "{Search for heavy neutral lepton production in K+ decays to positrons}",
    eprint = "2005.09575",
    archivePrefix = "arXiv",
    primaryClass = "hep-ex",
    reportNumber = "CERN-EP-2020-089",
    doi = "10.1016/j.physletb.2020.135599",
    journal = "Phys. Lett. B",
    volume = "807",
    pages = "135599",
    year = "2020"
}

@article{GERDA:2020xhi,
    author = "Agostini, M. and others",
    collaboration = "GERDA",
    title = "{Final Results of GERDA on the Search for Neutrinoless Double-$\beta$ Decay}",
    eprint = "2009.06079",
    archivePrefix = "arXiv",
    primaryClass = "nucl-ex",
    doi = "10.1103/PhysRevLett.125.252502",
    journal = "Phys. Rev. Lett.",
    volume = "125",
    number = "25",
    pages = "252502",
    year = "2020"
}

@article{KamLAND-Zen:2024eml,
    author = "Abe, S. and others",
    collaboration = "KamLAND-Zen",
    title = "{Search for Majorana Neutrinos with the Complete KamLAND-Zen Dataset}",
    eprint = "2406.11438",
    archivePrefix = "arXiv",
    primaryClass = "hep-ex",
    doi = "10.1103/jkf6-48j8",
    journal = "Phys. Rev. Lett.",
    volume = "135",
    number = "26",
    pages = "262501",
    year = "2025"
}

@article{Bolton:2019pcu,
    author = "Bolton, Patrick D. and Deppisch, Frank F. and Dev, P. S. Bhupal",
    title = "{Neutrinoless double beta decay versus other probes of heavy sterile neutrinos}",
    eprint = "1912.03058",
    archivePrefix = "arXiv",
    primaryClass = "hep-ph",
    doi = "10.1007/JHEP03(2020)170",
    journal = "JHEP",
    volume = "03",
    pages = "170",
    year = "2020"
}

@article{Abdullahi:2022jlv,
    author = "Abdullahi, Asli M. and others",
    title = "{The present and future status of heavy neutral leptons}",
    eprint = "2203.08039",
    archivePrefix = "arXiv",
    primaryClass = "hep-ph",
    reportNumber = "FERMILAB-CONF-22-184-T-V",
    doi = "10.1088/1361-6471/ac98f9",
    journal = "J. Phys. G",
    volume = "50",
    number = "2",
    pages = "020501",
    year = "2023"
}

@article{Kharusi:2021jez,
    author = "Kharusi, S. Al and others",
    title = "{Search for Majoron-emitting modes of $^{136}$Xe double beta decay with the complete EXO-200 dataset}",
    eprint = "2109.01327",
    archivePrefix = "arXiv",
    primaryClass = "hep-ex",
    doi = "10.1103/PhysRevD.104.112002",
    journal = "Phys. Rev. D",
    volume = "104",
    number = "11",
    pages = "112002",
    year = "2021"
}

@article{PIENU:2021clt,
    author = "Aguilar-Arevalo, A. and others",
    collaboration = "PIENU",
    title = "{Search for three body pion decays ${\pi}^+{\to}l^+{\nu}X$}",
    eprint = "2101.07381",
    archivePrefix = "arXiv",
    primaryClass = "hep-ex",
    doi = "10.1103/PhysRevD.103.052006",
    journal = "Phys. Rev. D",
    volume = "103",
    number = "5",
    pages = "052006",
    year = "2021"
}

@article{NA62:2021bji,
    author = "Cortina Gil, Eduardo and others",
    collaboration = "NA62",
    title = "{Search for $K^+$ decays to a muon and invisible particles}",
    eprint = "2101.12304",
    archivePrefix = "arXiv",
    primaryClass = "hep-ex",
    reportNumber = "CERN-EP-2021-018",
    doi = "10.1016/j.physletb.2021.136259",
    journal = "Phys. Lett. B",
    volume = "816",
    pages = "136259",
    year = "2021"
}

@article{RoyChoudhury:2022rva,
    author = "Roy Choudhury, Shouvik and Hannestad, Steen and Tram, Thomas",
    title = "{Massive neutrino self-interactions and inflation}",
    eprint = "2207.07142",
    archivePrefix = "arXiv",
    primaryClass = "astro-ph.CO",
    doi = "10.1088/1475-7516/2022/10/018",
    journal = "JCAP",
    volume = "10",
    pages = "018",
    year = "2022"
}

@article{Poudou:2025qcx,
    author = "Poudou, Ad\`ele and Simon, Th\'eo and Montandon, Thomas and Teixeira, Elsa M. and Poulin, Vivian",
    title = "{Self-interacting neutrinos in light of recent CMB and LSS data}",
    eprint = "2503.10485",
    archivePrefix = "arXiv",
    primaryClass = "astro-ph.CO",
    doi = "10.1103/mljb-42fm",
    journal = "Phys. Rev. D",
    volume = "112",
    number = "10",
    pages = "103535",
    year = "2025"
}

@article{He:2025jwp,
    author = "He, Adam and Ivanov, Mikhail M. and Bird, Simeon and An, Rui and Gluscevic, Vera",
    title = "{A Fresh Look at Neutrino Self-Interactions With the Lyman-$\alpha$ Forest: Constraints from EFT and PRIYA}",
    eprint = "2503.15592",
    archivePrefix = "arXiv",
    primaryClass = "astro-ph.CO",
    reportNumber = "MIT-CTP/5854",
    doi = "10.1103/wzpy-p7w8",
    journal = "Phys. Rev. D",
    volume = "112",
    number = "6",
    pages = "063540",
    year = "2025"
}

@article{Aloni:2023tff,
    author = "Aloni, Daniel and Joseph, Melissa and Schmaltz, Martin and Weiner, Neal",
    title = "{Dark Radiation from Neutrino Mixing after Big Bang Nucleosynthesis}",
    eprint = "2301.10792",
    archivePrefix = "arXiv",
    primaryClass = "astro-ph.CO",
    doi = "10.1103/PhysRevLett.131.221001",
    journal = "Phys. Rev. Lett.",
    volume = "131",
    number = "22",
    pages = "221001",
    year = "2023"
}

@article{Boyarsky:2020dzc,
    author = "Boyarsky, Alexey and Ovchynnikov, Maksym and Ruchayskiy, Oleg and Syvolap, Vsevolod",
    title = "{Improved big bang nucleosynthesis constraints on heavy neutral leptons}",
    eprint = "2008.00749",
    archivePrefix = "arXiv",
    primaryClass = "hep-ph",
    doi = "10.1103/PhysRevD.104.023517",
    journal = "Phys. Rev. D",
    volume = "104",
    number = "2",
    pages = "023517",
    year = "2021"
}

@article{Sabti:2020yrt,
    author = "Sabti, Nashwan and Magalich, Andrii and Filimonova, Anastasiia",
    title = "{An Extended Analysis of Heavy Neutral Leptons during Big Bang Nucleosynthesis}",
    eprint = "2006.07387",
    archivePrefix = "arXiv",
    primaryClass = "hep-ph",
    reportNumber = "KCL-2020-09",
    doi = "10.1088/1475-7516/2020/11/056",
    journal = "JCAP",
    volume = "11",
    pages = "056",
    year = "2020"
}

@article{Chen:2024cla,
    author = "Chen, Yu-Ming and Zhang, Yue",
    title = "{BBN Constraint on Heavy Neutrino Production and Decay}",
    eprint = "2410.07343",
    archivePrefix = "arXiv",
    primaryClass = "hep-ph",
    doi = "10.1103/741s-211w",
    journal = "Phys. Rev. D",
    volume = "111",
    number = "12",
    pages = "123024",
    year = "2025"
}

@article{Aalberts:2018obr,
    author = "Aalberts, Jelle L. and others",
    title = "{Precision constraints on radiative neutrino decay with CMB spectral distortion}",
    eprint = "1803.00588",
    archivePrefix = "arXiv",
    primaryClass = "astro-ph.CO",
    doi = "10.1103/PhysRevD.98.023001",
    journal = "Phys. Rev. D",
    volume = "98",
    pages = "023001",
    year = "2018"
}

@article{Barenboim:2020vrr,
    author = "Barenboim, Gabriela and Chen, Joe Zhiyu and Hannestad, Steen and Oldengott, Isabel M. and Tram, Thomas and Wong, Yvonne Y. Y.",
    title = "{Invisible neutrino decay in precision cosmology}",
    eprint = "2011.01502",
    archivePrefix = "arXiv",
    primaryClass = "astro-ph.CO",
    doi = "10.1088/1475-7516/2021/03/087",
    journal = "JCAP",
    volume = "03",
    pages = "087",
    year = "2021"
}

@article{Chacko:2019nej,
    author = "Chacko, Zackaria and Dev, Abhish and Du, Peizhi and Poulin, Vivian and Tsai, Yuhsin",
    title = "{Cosmological Limits on the Neutrino Mass and Lifetime}",
    eprint = "1909.05275",
    archivePrefix = "arXiv",
    primaryClass = "hep-ph",
    doi = "10.1007/JHEP04(2020)020",
    journal = "JHEP",
    volume = "04",
    pages = "020",
    year = "2020"
}

@article{Chacko:2020hmh,
    author = "Chacko, Zackaria and Dev, Abhish and Du, Peizhi and Poulin, Vivian and Tsai, Yuhsin",
    title = "{Determining the Neutrino Lifetime from Cosmology}",
    eprint = "2002.08401",
    archivePrefix = "arXiv",
    primaryClass = "astro-ph.CO",
    doi = "10.1103/PhysRevD.103.043519",
    journal = "Phys. Rev. D",
    volume = "103",
    number = "4",
    pages = "043519",
    year = "2021"
}

@article{Escudero:2020ped,
    author = "Escudero, Miguel and Lopez-Pavon, Jacobo and Rius, Nuria and Sandner, Stefan",
    title = "{Relaxing Cosmological Neutrino Mass Bounds with Unstable Neutrinos}",
    eprint = "2007.04994",
    archivePrefix = "arXiv",
    primaryClass = "hep-ph",
    reportNumber = "KCL-2020-27, FTUV-20-0625.4735, IFIC/20-33",
    doi = "10.1007/JHEP12(2020)119",
    journal = "JHEP",
    volume = "12",
    pages = "119",
    year = "2020"
}

@article{FrancoAbellan:2021hdb,
    author = "Franco Abell\'an, Guillermo and Chacko, Zackaria and Dev, Abhish and Du, Peizhi and Poulin, Vivian and Tsai, Yuhsin",
    title = "{Improved cosmological constraints on the neutrino mass and lifetime}",
    eprint = "2112.13862",
    archivePrefix = "arXiv",
    primaryClass = "hep-ph",
    reportNumber = "FERMILAB-PUB-21-779-T",
    doi = "10.1007/JHEP08(2022)076",
    journal = "JHEP",
    volume = "08",
    pages = "076",
    year = "2022"
}

@article{Metropolis:1953am,
    author = "Metropolis, N. and Rosenbluth, A. W. and Rosenbluth, M. N. and Teller, A. H. and Teller, E.",
    title = "{Equation of state calculations by fast computing machines}",
    doi = "10.1063/1.1699114",
    journal = "J. Chem. Phys.",
    volume = "21",
    pages = "1087--1092",
    year = "1953"
}

@article{Hastings:1970aa,
    author = "Hastings, W. K.",
    title = "{Monte Carlo Sampling Methods Using Markov Chains and Their Applications}",
    doi = "10.1093/biomet/57.1.97",
    journal = "Biometrika",
    volume = "57",
    pages = "97--109",
    year = "1970"
}

@article{Benso:2024qrg,
    author = "Benso, Cristina and Schwetz, Thomas and Vatsyayan, Drona",
    title = "{Large neutrino mass in cosmology and keV sterile neutrino dark matter from a dark sector}",
    eprint = "2410.23926",
    archivePrefix = "arXiv",
    primaryClass = "hep-ph",
    doi = "10.1088/1475-7516/2025/04/054",
    journal = "JCAP",
    volume = "04",
    pages = "054",
    year = "2025"
}

@article{Fardon:2003eh,
    author = "Fardon, Rob and Nelson, Ann E. and Weiner, Neal",
    title = "{Dark energy from mass varying neutrinos}",
    eprint = "astro-ph/0309800",
    archivePrefix = "arXiv",
    reportNumber = "UW-PT-03-22",
    doi = "10.1088/1475-7516/2004/10/005",
    journal = "JCAP",
    volume = "10",
    pages = "005",
    year = "2004"
}

@article{Lorenz:2021alz,
    author = {Lorenz, Christiane S. and Funcke, Lena and L\"offler, Matthias and Calabrese, Erminia},
    title = "{Reconstruction of the neutrino mass as a function of redshift}",
    eprint = "2102.13618",
    archivePrefix = "arXiv",
    primaryClass = "astro-ph.CO",
    doi = "10.1103/PhysRevD.104.123518",
    journal = "Phys. Rev. D",
    volume = "104",
    number = "12",
    pages = "123518",
    year = "2021"
}

@article{Lorenz:2018fzb,
    author = "Lorenz, Christiane S. and Funcke, Lena and Calabrese, Erminia and Hannestad, Steen",
    title = "{Time-varying neutrino mass from a supercooled phase transition: current cosmological constraints and impact on the $\Omega_m$-$\sigma_8$ plane}",
    eprint = "1811.01991",
    archivePrefix = "arXiv",
    primaryClass = "astro-ph.CO",
    doi = "10.1103/PhysRevD.99.023501",
    journal = "Phys. Rev. D",
    volume = "99",
    number = "2",
    pages = "023501",
    year = "2019"
}

@article{Camarena:2023cku,
    author = "Camarena, David and Cyr-Racine, Francis-Yan and Houghteling, John",
    title = "{Confronting self-interacting neutrinos with the full shape of the galaxy power spectrum}",
    eprint = "2309.03941",
    archivePrefix = "arXiv",
    primaryClass = "astro-ph.CO",
    doi = "10.1103/PhysRevD.108.103535",
    journal = "Phys. Rev. D",
    volume = "108",
    number = "10",
    pages = "103535",
    year = "2023"
}

@article{Oldengott:2014qra,
    author = "Oldengott, Isabel M. and Rampf, Cornelius and Wong, Yvonne Y. Y.",
    title = "{Boltzmann hierarchy for interacting neutrinos I: formalism}",
    eprint = "1409.1577",
    archivePrefix = "arXiv",
    primaryClass = "astro-ph.CO",
    doi = "10.1088/1475-7516/2015/04/016",
    journal = "JCAP",
    volume = "04",
    pages = "016",
    year = "2015"
}

@article{Oldengott:2017fhy,
    author = "Oldengott, Isabel M. and Tram, Thomas and Rampf, Cornelius and Wong, Yvonne Y. Y.",
    title = "{Interacting neutrinos in cosmology: exact description and constraints}",
    eprint = "1706.02123",
    archivePrefix = "arXiv",
    primaryClass = "astro-ph.CO",
    doi = "10.1088/1475-7516/2017/11/027",
    journal = "JCAP",
    volume = "11",
    pages = "027",
    year = "2017"
}

@article{ACT:2025tim,
    author = "Calabrese, Erminia and others",
    collaboration = "ACT",
    title = "{The Atacama Cosmology Telescope: DR6 Constraints on Extended Cosmological Models}",
    eprint = "2503.14454",
    archivePrefix = "arXiv",
    primaryClass = "astro-ph.CO",
    reportNumber = "FERMILAB-PUB-25-0157-PPD",
    doi = "10.1088/1475-7516/2025/11/063",
    journal = "JCAP",
    volume = "11",
    pages = "063",
    year = "2025"
}

@article{Beacom:2004yd,
    author = "Beacom, John F. and Bell, Nicole F. and Dodelson, Scott",
    title = "{Neutrinoless universe}",
    eprint = "astro-ph/0404585",
    archivePrefix = "arXiv",
    reportNumber = "FERMILAB-PUB-04-050-A",
    doi = "10.1103/PhysRevLett.93.121302",
    journal = "Phys. Rev. Lett.",
    volume = "93",
    pages = "121302",
    year = "2004"
}

@article{Green:2021gdc,
    author = "Green, Daniel and Kaplan, David E. and Rajendran, Surjeet",
    title = "{Neutrino interactions in the late universe}",
    eprint = "2108.06928",
    archivePrefix = "arXiv",
    primaryClass = "hep-ph",
    doi = "10.1007/JHEP11(2021)162",
    journal = "JHEP",
    volume = "11",
    pages = "162",
    year = "2021"
}

@article{Libanore:2025ack,
    author = "Libanore, Sarah and Ghosh, Subhajit and Kovetz, Ely D. and Boddy, Kimberly K. and Raccanelli, Alvise",
    title = "{Joint 21-cm and CMB Forecasts for Constraining Self-Interacting Massive Neutrinos}",
    eprint = "2504.15348",
    archivePrefix = "arXiv",
    primaryClass = "astro-ph.CO",
    doi = "10.1103/tdms-6n76",
    journal = "Phys. Rev. D",
    volume = "112",
    number = "6",
    pages = "063502",
    year = "2025"
}

@article{Chen:2022idm,
    author = "Chen, Joe Zhiyu and Oldengott, Isabel M. and Pierobon, Giovanni and Wong, Yvonne Y. Y.",
    title = "{Weaker yet again: mass spectrum-consistent cosmological constraints on the neutrino lifetime}",
    eprint = "2203.09075",
    archivePrefix = "arXiv",
    primaryClass = "hep-ph",
    doi = "10.1140/epjc/s10052-022-10518-3",
    journal = "Eur. Phys. J. C",
    volume = "82",
    number = "7",
    pages = "640",
    year = "2022"
}

@article{Esteban:2021ozz,
    author = "Esteban, Ivan and Salvado, Jordi",
    title = "{Long Range Interactions in Cosmology: Implications for Neutrinos}",
    eprint = "2101.05804",
    archivePrefix = "arXiv",
    primaryClass = "hep-ph",
    doi = "10.1088/1475-7516/2021/05/036",
    journal = "JCAP",
    volume = "05",
    pages = "036",
    year = "2021"
}

@article{Esteban:2022rjk,
    author = "Esteban, Ivan and Mena, Olga and Salvado, Jordi",
    title = "{Nonstandard neutrino cosmology dilutes the lensing anomaly}",
    eprint = "2202.04656",
    archivePrefix = "arXiv",
    primaryClass = "astro-ph.CO",
    doi = "10.1103/PhysRevD.106.083516",
    journal = "Phys. Rev. D",
    volume = "106",
    number = "8",
    pages = "083516",
    year = "2022"
}

@article{Sen:2023uga,
    author = "Sen, Manibrata and Smirnov, Alexei Y.",
    title = "{Refractive neutrino masses, ultralight dark matter and cosmology}",
    eprint = "2306.15718",
    archivePrefix = "arXiv",
    primaryClass = "hep-ph",
    doi = "10.1088/1475-7516/2024/01/040",
    journal = "JCAP",
    volume = "01",
    pages = "040",
    year = "2024"
}

@article{Sen:2024pgb,
    author = "Sen, Manibrata and Smirnov, Alexei Y.",
    title = "{Neutrinos with refractive masses and the DESI baryon acoustic oscillation results}",
    eprint = "2407.02462",
    archivePrefix = "arXiv",
    primaryClass = "hep-ph",
    doi = "10.1103/d9hh-b3r9",
    journal = "Phys. Rev. D",
    volume = "111",
    number = "10",
    pages = "103048",
    year = "2025"
}

@article{Brinckmann:2020bcn,
    author = "Brinckmann, Thejs and Chang, Jae Hyeok and LoVerde, Marilena",
    title = "{Self-interacting neutrinos, the Hubble parameter tension, and the cosmic microwave background}",
    eprint = "2012.11830",
    archivePrefix = "arXiv",
    primaryClass = "astro-ph.CO",
    reportNumber = "YITP-SB-2020-40",
    doi = "10.1103/PhysRevD.104.063523",
    journal = "Phys. Rev. D",
    volume = "104",
    number = "6",
    pages = "063523",
    year = "2021"
}

@article{Foroughi-Abari:2025upe,
    author = "Foroughi-Abari, Saeid and Kelly, Kevin J. and Rai, Mudit and Zhang, Yue",
    title = "{Enabling Strong Neutrino Self-Interaction with an Unparticle Mediator}",
    eprint = "2501.02049",
    archivePrefix = "arXiv",
    primaryClass = "hep-ph",
    reportNumber = "MI-HET-847",
    doi = "10.1103/PhysRevLett.134.181001",
    journal = "Phys. Rev. Lett.",
    volume = "134",
    number = "18",
    pages = "181001",
    year = "2025"
}

@article{Huang:2021dba,
    author = "Huang, Guo-yuan and Rodejohann, Werner",
    title = "{Solving the Hubble tension without spoiling Big Bang Nucleosynthesis}",
    eprint = "2102.04280",
    archivePrefix = "arXiv",
    primaryClass = "hep-ph",
    doi = "10.1103/PhysRevD.103.123007",
    journal = "Phys. Rev. D",
    volume = "103",
    pages = "123007",
    year = "2021"
}

@article{Li:2023puz,
    author = "Li, Shao-Ping and Xu, Xun-Jie",
    title = "{N$_{eff}$ constraints on light mediators coupled to neutrinos: the dilution-resistant effect}",
    eprint = "2307.13967",
    archivePrefix = "arXiv",
    primaryClass = "hep-ph",
    doi = "10.1007/JHEP10(2023)012",
    journal = "JHEP",
    volume = "10",
    pages = "012",
    year = "2023"
}

@article{ATLAS:2023tkt,
    author = "Aad, Georges and others",
    collaboration = "ATLAS",
    title = "{Combination of searches for invisible decays of the Higgs boson using 139 fb{\ensuremath{-}}1 of proton-proton collision data at s=13 TeV collected with the ATLAS experiment}",
    eprint = "2301.10731",
    archivePrefix = "arXiv",
    primaryClass = "hep-ex",
    reportNumber = "CERN-EP-2022-289",
    doi = "10.1016/j.physletb.2023.137963",
    journal = "Phys. Lett. B",
    volume = "842",
    pages = "137963",
    year = "2023"
}

@article{Hannestad:2004px,
    author = "Hannestad, Steen",
    title = "{What is the lowest possible reheating temperature?}",
    eprint = "astro-ph/0403291",
    archivePrefix = "arXiv",
    doi = "10.1103/PhysRevD.70.043506",
    journal = "Phys. Rev. D",
    volume = "70",
    pages = "043506",
    year = "2004"
}

@article{Dev:2025pru,
    author = "Dev, P. S. Bhupal and Wu, Quan-feng and Xu, Xun-Jie",
    title = "{No Hiding in the Dark: Cosmological Bounds on Heavy Neutral Leptons with Dark Decay Channels}",
journal = "",
    eprint = "2507.12270",
    archivePrefix = "arXiv",
    primaryClass = "hep-ph",
    reportNumber = "CETUP-2025-002",
    month = "7",
    year = "2025"
}

@article{Pal:2024yom,
    author = "Pal, Sourav and Samanta, Rickmoy and Pal, Supratik",
    title = "{Exploring neutrino interactions in light of present and upcoming galaxy surveys}",
    eprint = "2409.03712",
    archivePrefix = "arXiv",
    primaryClass = "astro-ph.CO",
    doi = "10.1088/1475-7516/2025/03/047",
    journal = "JCAP",
    volume = "03",
    pages = "047",
    year = "2025"
}

@article{Akita:2020szl,
    author = "Akita, Kensuke and Yamaguchi, Masahide",
    title = "{A precision calculation of relic neutrino decoupling}",
    eprint = "2005.07047",
    archivePrefix = "arXiv",
    primaryClass = "hep-ph",
    doi = "10.1088/1475-7516/2020/08/012",
    journal = "JCAP",
    volume = "08",
    pages = "012",
    year = "2020"
}

@article{Froustey:2020mcq,
    author = "Froustey, Julien and Pitrou, Cyril and Volpe, Maria Cristina",
    title = "{Neutrino decoupling including flavour oscillations and primordial nucleosynthesis}",
    eprint = "2008.01074",
    archivePrefix = "arXiv",
    primaryClass = "hep-ph",
    doi = "10.1088/1475-7516/2020/12/015",
    journal = "JCAP",
    volume = "12",
    pages = "015",
    year = "2020"
}

@article{Akita:2022etk,
    author = "Akita, Kensuke and Im, Sang Hui and Masud, Mehedi",
    title = "{Probing non-standard neutrino interactions with a light boson from next galactic and diffuse supernova neutrinos}",
    eprint = "2206.06852",
    archivePrefix = "arXiv",
    primaryClass = "hep-ph",
    reportNumber = "CTPU-PTC-22-13",
    doi = "10.1007/JHEP12(2022)050",
    journal = "JHEP",
    volume = "12",
    pages = "050",
    year = "2022"
}

@article{Akita:2023iwq,
    author = "Akita, Kensuke and Im, Sang Hui and Masud, Mehedi and Yun, Seokhoon",
    title = "{Limits on heavy neutral leptons, Z' bosons and majorons from high-energy supernova neutrinos}",
    eprint = "2312.13627",
    archivePrefix = "arXiv",
    primaryClass = "hep-ph",
    reportNumber = "CTPU-PTC-23-55",
    doi = "10.1007/JHEP07(2024)057",
    journal = "JHEP",
    volume = "07",
    pages = "057",
    year = "2024"
}

@article{Barenboim:2019tux,
    author = "Barenboim, Gabriela and Denton, Peter B. and Oldengott, Isabel M.",
    title = "{Constraints on inflation with an extended neutrino sector}",
    eprint = "1903.02036",
    archivePrefix = "arXiv",
    primaryClass = "astro-ph.CO",
    doi = "10.1103/PhysRevD.99.083515",
    journal = "Phys. Rev. D",
    volume = "99",
    number = "8",
    pages = "083515",
    year = "2019"
}

@article{Noriega:2025ulc,
    author = "Noriega, Hern{\'a}n E. and De-Santiago, Josue and Garcia-Arroyo, Gabriela and Venzor, Jorge and P{\'e}rez-Lorenzana, Abdel",
    title = "{Resonant neutrino self-interactions: Insights from the full shape galaxy power spectrum}",
    eprint = "2506.07994",
    archivePrefix = "arXiv",
    primaryClass = "astro-ph.CO",
    doi = "10.1103/b9x4-hnqn",
    journal = "Phys. Rev. D",
    volume = "112",
    number = "6",
    pages = "063509",
    year = "2025"
}

@article{Green:2024xbb,
    author = "Green, Daniel and Meyers, Joel",
    title = "{Cosmological preference for a negative neutrino mass}",
    eprint = "2407.07878",
    archivePrefix = "arXiv",
    primaryClass = "astro-ph.CO",
    doi = "10.1103/PhysRevD.111.083507",
    journal = "Phys. Rev. D",
    volume = "111",
    number = "8",
    pages = "083507",
    year = "2025"
}

@article{Loverde:2024nfi,
    author = "Loverde, Marilena and Weiner, Zachary J.",
    title = "{Massive neutrinos and cosmic composition}",
    eprint = "2410.00090",
    archivePrefix = "arXiv",
    primaryClass = "astro-ph.CO",
    doi = "10.1088/1475-7516/2024/12/048",
    journal = "JCAP",
    volume = "12",
    pages = "048",
    year = "2024"
}

@article{Lynch:2025ine,
    author = "Lynch, Gabriel P. and Knox, Lloyd",
    title = "{What{\textquoteright}s the matter with {\ensuremath{\Sigma}}m{\ensuremath{\nu}}?}",
    eprint = "2503.14470",
    archivePrefix = "arXiv",
    primaryClass = "astro-ph.CO",
    doi = "10.1103/613p-pph2",
    journal = "Phys. Rev. D",
    volume = "112",
    number = "8",
    pages = "083543",
    year = "2025"
}

@article{Graham:2025fdt,
    author = "Graham, Peter W. and Green, Daniel and Meyers, Joel",
    title = "{Dark Forces Gathering}",
    eprint = "2508.20999",
    archivePrefix = "arXiv",
    primaryClass = "astro-ph.CO",
    doi = "10.1103/1bqb-qlrj",
    journal = "Phys. Rev. D",
    volume = "113",
    number = "4",
    pages = "043514",
    year = "2026"
}

@article{Cozzumbo:2025ewt,
    author = "Cozzumbo, Andrea and Atzori Corona, Mattia and Murgia, Riccardo and Archidiacono, Maria and Cadeddu, Matteo",
    title = "{A short blanket for cosmology: the CMB lensing anomaly behind the preference for a negative neutrino mass}",
    eprint = "2511.01967",
    archivePrefix = "arXiv",
    primaryClass = "astro-ph.CO",
    month = "11",
    year = "2025"
}

@article{Elbers:2024sha,
    author = "Elbers, Willem and Frenk, Carlos S. and Jenkins, Adrian and Li, Baojiu and Pascoli, Silvia",
    title = "{Negative neutrino masses as a mirage of dark energy}",
    eprint = "2407.10965",
    archivePrefix = "arXiv",
    primaryClass = "astro-ph.CO",
    doi = "10.1103/PhysRevD.111.063534",
    journal = "Phys. Rev. D",
    volume = "111",
    number = "6",
    pages = "063534",
    year = "2025"
}

@article{Namikawa:2025doa,
    author = "Namikawa, Toshiya",
    title = "{Resolving the Negative Effective Neutrino Mass Parameter with Cosmic Birefringence}",
    eprint = "2506.22999",
    archivePrefix = "arXiv",
    primaryClass = "astro-ph.CO",
    doi = "10.1103/qgnn-6hsf",
    journal = "Phys. Rev. Lett.",
    volume = "135",
    number = "16",
    pages = "161004",
    year = "2025"
}

@article{Kumar:2022vee,
    author = "Kumar, Suresh and Nunes, Rafael C. and Yadav, Priya",
    title = "{Updating non-standard neutrinos properties with Planck-CMB data and full-shape analysis of BOSS and eBOSS galaxies}",
    eprint = "2205.04292",
    archivePrefix = "arXiv",
    primaryClass = "astro-ph.CO",
    doi = "10.1088/1475-7516/2022/09/060",
    journal = "JCAP",
    volume = "09",
    pages = "060",
    year = "2022"
}

@article{Wright:2025xka,
    author = "Wright, Angus H. and others",
    title = "{KiDS-Legacy: Cosmological constraints from cosmic shear with the complete Kilo-Degree Survey}",
    eprint = "2503.19441",
    archivePrefix = "arXiv",
    primaryClass = "astro-ph.CO",
    doi = "10.1051/0004-6361/202554908",
    journal = "Astron. Astrophys.",
    volume = "703",
    pages = "A158",
    year = "2025"
}

@article{Pantos:2026koc,
    author = "Pantos, Ioannis and Perivolaropoulos, Leandros",
    title = "{Status of the $S_8$ Tension: A 2026 Review of Probe Discrepancies}",
    eprint = "2602.12238",
    archivePrefix = "arXiv",
    primaryClass = "astro-ph.CO",
    month = "2",
    year = "2026"
}

@article{DES:2026fyc,
    author = "Abbott, T. M. C. and others",
    collaboration = "DES",
    title = "{Dark Energy Survey Year 6 Results: Cosmological Constraints from Galaxy Clustering and Weak Lensing}",
    eprint = "2601.14559",
    archivePrefix = "arXiv",
    primaryClass = "astro-ph.CO",
    reportNumber = "DES-2025-0929, FERMILAB-PUB-26-0026-PPD",
    month = "1",
    year = "2026"
}

@misc{supp,
howpublished="See \href{https://journals.aps.org/prl/accepted/10.1103/jprg-jll6} 
{\it Supplemental Material}, which includes additional Refs.~\cite{EscuderoAbenza:2020cmq,
Thomas:2019ran,
Deppisch:2020sqh, 
Kolb:1987qy,
Fiorillo:2023ytr,
PIENU:2017wbj, 
NA62:2020mcv,
GERDA:2020xhi,
KamLAND-Zen:2024eml,
Bolton:2019pcu,
Sabti:2020yrt, 
Boyarsky:2020dzc, 
Chen:2024cla,
Hannestad:2004px,
Gelmini:2004ah, 
Dolgov:1997mb,
Harris:2020xlr,
Virtanen:2019joe,
Shtabovenko:2023idz,
feynarts,
Alloul:2013bka,
Gondolo:1990dk,
Ma:1995ey,
Brinckmann:2022ajr,
Lesgourgues:2011re,
Blas:2011rf,
Feroz:2007kg,
Feroz:2008xx,
Feroz:2013hea,
Buchner:2014nha,
Audren:2012wb,
Brinckmann:2018cvx,
Herold:2024nvk,
Metropolis:1953am, 
Hastings:1970aa,
DESI:2025zgx,
Wright:2025xka,
Pantos:2026koc,
DES:2026fyc}." }
\bibliographystyle{JHEP}

\appendix 
\onecolumngrid
\medskip
\hrule
%\medskip
\begin{center}
{\it \large Supplemental Material}
% \\[5pt]
% {\bf \large Impostor Among $\nu$s: Dark Radiation Masquerading as Self-Interacting Neutrinos
% }\\[5pt]
% Anirban Das, P.~S.~Bhupal Dev, Christina Gao, Subhajit Ghosh, Taegyun Kim
\end{center}

\setcounter{equation}{0}
\setcounter{figure}{0}
\setcounter{table}{0}
\makeatletter
\renewcommand{\theequation}{S\arabic{equation}}
\renewcommand{\thefigure}{S\arabic{figure}}
\renewcommand{\thetable}{S\arabic{table}}

\section{I. Neutrino Cooling from Interaction with Dark Sector}
\label{app:nucooling}
We focus on the period of cosmology after neutrino decoupling ($T\lesssim$ MeV) and before photon decoupling ($T \gtrsim$ 0.2 eV). 
In the radiation-dominated era, the total energy density is given by
\be\label{eq:Neff}
\rho_{\rm rad}=\rho_\gamma \left[1+\frac78 \left(\frac{T^{\rm SM}_\nu}{T}\right)^4 N_{\rm eff}\right]~,
\ee
where $T$ is the photon temperature, and $\rho_\gamma=\frac{\pi^2}{15}T^4$. 
Since photon numbers are not conserved in chemical processes, the chemical potential of photon is zero. Here $T_\nu^{\rm SM}$ is the same as $T$ before electron-positron annihilation, but equal to $\left(\frac{4}{11} \right)^{1/3}T$ afterward.  We invert Eq.~\eqref{eq:Neff} to obtain an expression for $N_{\rm eff}$:
\be
N_{\rm eff}= \frac{\rho_\nu+\rho_{\rm DS} }{\rho_\gamma} \frac87\left(\frac{T}{T^{\rm SM}_\nu}\right)^4~.
\ee
For the discussion below,  we largely follow Refs.~\cite{Escudero:2022gez,EscuderoAbenza:2020cmq}. 

\subsubsection{Neutrino-DS Equilibrium}

Let the dark sector (DS) contains 1 scalar boson $\phi$ and $N_\chi$ Majorana fermions. 
Let's assume that after neutrinos decouple from photons, DS comes into equilibrium via decay and inverse decay processes $\nu\nu\leftrightarrow \phi\leftrightarrow \chi\chi$. Such processes do not conserve entropy, but conserve energy, i.e.,
\be\label{eq:energy_conserve}
(\rho_\nu(\mu_\nu,T_\nu)+\rho_\phi(\mu_\phi,T_{\rm DS})+\rho_\chi(\mu_\chi,T_{\rm DS}))a^4={\rm constant} .
\ee
Furthermore, the particle number also obeys a conservation law,
\be\label{eq:number_conserve}
(n_\nu(\mu_\nu,T_\nu)+2n_\phi(\mu_\phi,T_{\rm DS})+n_\chi(\mu_\chi,T_{\rm DS}))a^3={\rm constant} \,.
\ee 
Rescaling Eq.~\eqref{eq:number_conserve} to cancel $a$ in Eq.~\eqref{eq:energy_conserve}, let us compare the time of neutrino decoupling $T_\nu^0$ and later the time of neutrino-DS equilibrium $T_\nu^{\rm eq}(\gg m_\phi\gg m_\chi,m_\nu)$:
\be\label{eq:equilibrium}
\begin{split}
\frac{\rho_\nu(0,T_\nu^0)}{n_\nu(0,T_\nu^0)^{4/3}}&=\frac{\rho_\nu(\mu^{\rm eq}_\nu,T^{\rm eq}_\nu)+\rho_\phi(\mu^{\rm eq}_\phi,T^{\rm eq}_\nu)+\rho_\chi(\mu^{\rm eq}_\chi,T^{\rm eq}_\nu)}{\left(n_\nu(\mu^{\rm eq}_\nu,T_\nu^{\rm eq})+2n_\phi(\mu^{\rm eq}_\phi,T_\nu^{\rm eq})+n_\chi(\mu_\chi^{\rm eq},T_\nu^{\rm eq})\right)^{4/3}}\, ,\\
\frac{7 \pi ^{14/3}}{90\ 6^{2/3} \zeta (3)^{4/3}}&=-\frac{3 \pi ^{2/3} \left(2 (N_\chi+3) \text{Li}_4(-z_{\rm eq})-\text{Li}_4\left(z_{\rm eq}^2\right)\right)}{2 \sqrt[3]{2} \left(\text{Li}_3\left(z_{\rm eq}^2\right)-(N_\chi+3) \text{Li}_3(-z_{\rm eq})\right){}^{4/3}}\, ,
\end{split}
\ee
where $z_{\rm eq}\equiv \exp(\mu_\nu^{\rm eq}/T_\nu^{\rm eq})$. To get the second line of Eq.~\eqref{eq:equilibrium} we have also imposed chemical equilibrium
\be
\mu_\nu^{\rm eq}=\mu_\chi^{\rm eq}=\frac12 \mu_\phi^{\rm eq}~.
\ee
Numerically solving Eq.~\eqref{eq:equilibrium} gives us a value of $z^{\rm eq}$ which only depends on $N_\chi$.

Separately, the photon and electron bath (later the photon bath after electron-positron annihilation) has to conserve entropy, i.e.
\be\label{eq:entropy_conserve}
g_\gamma(T)T^3a^3={\rm constant}\, .
\ee
Rescaling Eq.~\eqref{eq:entropy_conserve} to cancel $a$ in Eq.~\eqref{eq:number_conserve}, we get: 
 \be\label{eq:equilibrium2}
 \begin{split}
\frac{n_\nu(0,T_\nu^0)}{g_\gamma(T^0)(T^0)^3}&=\frac{n_\nu(\mu^{\rm eq}_\nu,T_\nu^{\rm eq})+2n_\phi(2\mu^{\rm eq}_\nu,T_\nu^{\rm eq})+n_\chi(\mu^{\rm eq}_\nu,T_\nu^{\rm eq})}{g_\gamma(T^{\rm eq})(T^{\rm eq})^3}\, , \\
%\frac{9 {\xi^0_\nu}^3 \zeta (3)}{2 \pi ^2g_\gamma(T^0)}&=\frac{2 {\xi_\nu^{\rm eq}}^3 \left(\text{Li}_3\left({z^{\rm eq}}^2\right)-(N_\chi+3) \text{Li}_3(-z^{\rm eq})\right)}{\pi ^2g_\gamma(T^{\rm eq})}\\
\xi_{\nu}^{\rm eq} &=\xi^0_\nu \left(\frac{g_\gamma(T^{\rm eq})}{g_\gamma(T^0)} \right)^{1/3}
\left(\frac{\frac94 \zeta (3) }{\text{Li}_3\left(z_{\rm eq}^2\right)-(N_\chi+3) \text{Li}_3(-z_{\rm eq})}\right)^{1/3}.
\end{split}
 \ee
where $\xi^\nu \equiv T_\nu/T$ and $\text{Li}_n(z)$ are the Polylogarithms of order $n$. 

For the photon bath, depending on whether the neutrino-DS equilibrium happens before or after the electron-positron-annihilation decoupling  from the bath which happens at $T^{e^+e^-}\approx 16$ keV~\cite{Thomas:2019ran}, we have
\be
\frac{g_\gamma(T^{\rm eq})}{g_\gamma(T^0)}  \approx \left\{
 \begin{array}{cc}
 1 \, , & T^{\rm eq}>T^{e^+e^-}\\
\frac{2}{\frac78\times 4+2}=\frac{4}{11} \, ,  &T^{\rm eq}<T^{e^+e^-}
\end{array}
 \right. .
\ee
Therefore, the last line of Eq.~\eqref{eq:equilibrium2} can also be written as
\be
\xi_{\nu}^{\rm eq} =\xi^{\rm SM}_\nu 
\left(\frac{\frac94 \zeta (3) }{\text{Li}_3\left(z_{\rm eq}^2\right)-(N_\chi+3) \text{Li}_3(-z_{\rm eq})}\right)^{1/3}~.
\ee
At equilibrium, the energy density taken by the mediator $\phi$ is approximately given by
\be
\frac{\rho_\phi}{\rho_{\rm tot}}\approx 
\frac{45 (\xi_{\nu}^{\rm eq}) ^4 \text{Li}_4\left(z_{\rm eq}^2\right)}{-90 (\xi_{\nu}^{\rm eq})^4 (N_\chi+3) \text{Li}_4(-z_{\rm eq})+45 (\xi_{\nu}^{\rm eq})^4 \text{Li}_4\left(z_{\rm eq}^2\right)+\pi ^4}\xrightarrow{z_{\rm eq}\ll1}\frac{z_{\rm eq}}{2 N
_\chi+6}+O\left(z_{\rm eq}^{4/3}\right)~.
\ee
Numerical computation shows that $\phi$ occupies at most a few percent of the total energy during the process.
%In Ref.~\cite{Escudero:2022gez,EscuderoAbenza:2020cmq}, it is assumed that the neutrino-DS equilibrium is attained instantaneously, so that $T^0=T^{\rm eq}$ and hence Eq.~\eqref{eq:equilibrium2} becomes
%\be
%T_{\nu}^{\rm eq} =T^0_\nu \left(\frac{\frac94 \zeta (3) }{\text{Li}_3\left(z_{\rm eq}^2\right)-(N_\chi+3) \text{Li}_3(-z_{\rm eq})}\right)^{1/3}
%\ee

\subsubsection{Decoupling of the Mediator}

At a later time, the temperature drops well below the mediator mass and the scalar mediator decouples at around $T^f_\nu$. This happens within the thermal bath of the neutrino-DS, thus the entropy is conserved within that sector:
\be
(s_\nu(\mu_\nu,T_\nu)+s_\phi(2\mu_\nu,T_\nu)+s_\chi(\mu_\nu,T_\nu))a^3 ={\rm constant}\, ,
\ee
where 
\be
s(\mu,T)= \frac{\rho(\mu, T)+P(\mu, T)-\mu n(\mu, T)}{T}~.
\ee
Again, using Eq.~\eqref{eq:number_conserve} to cancel $a$, let us compare the time of equilibrium $T^{\rm eq}$ and a later time after $\phi$ decouples $T^f$:
\be\label{eq:decoupling1}
\begin{split}
\frac{s_\nu(\mu^{\rm eq}_\nu,T^{\rm eq}_\nu)+s_\phi(2\mu^{\rm eq}_\nu,T^{\rm eq}_\nu)+s_\chi(\mu^{\rm eq}_\nu,T^{\rm eq}_\nu)}{n_\nu(\mu^{\rm eq}_\nu,T_\nu^{\rm eq})+2n_\phi(2\mu^{\rm eq}_\nu,T_\nu^{\rm eq})+n_\chi(\mu^{\rm eq}_\nu,T_\nu^{\rm eq})}=\frac{s_\nu(\mu^f_\nu,T^f_\nu)+s_\chi(\mu^f_\nu,T^f_\nu)}{n_\nu(\mu^f_\nu,T_\nu^f)+n_\chi(\mu^f_\nu,T_\nu^f)}\, , \\
\frac{8 N_f \text{Li}_4(-z_{\rm e})-2N_f \text{Li}_3(-z_{\rm e}) \log (z_{\rm e})-4 \text{Li}_4\left(z_{\rm e}^2\right)+\text{Li}_3\left(z_{\rm e}^2\right) \log \left(z_{\rm e}^2\right)}{2 N_f \text{Li}_3(-z_{\rm e})-2 \text{Li}_3\left(z_{\rm e}^2\right)} 
=\frac{4 \text{Li}_4(-z_f)}{\text{Li}_3(-z_f)}-\log (z_f) \, , 
\end{split}
\ee
where $N_f=N_\chi+3$, $z_{\rm e}\equiv z_{\rm eq}$. Thus, given $z_{\rm eq}$, we can numerically solve Eq.~\eqref{eq:decoupling1} to obtain $z_f\equiv \exp(\mu^f_\nu/T^f_\nu)$. 

To find $T_\nu^f$,  rescaling Eq.~\eqref{eq:entropy_conserve} to cancel $a$ in Eq.~\eqref{eq:number_conserve}, we obtain
 \be\label{eq:decoupling}
 \begin{split}
\frac{n_\nu(\mu^f_\nu,T_\nu^f)+n_\chi(\mu^f_\nu,T_\nu^f)}{g_\gamma(T^f)(T^f)^3}&=\frac{n_\nu(\mu^{\rm eq}_\nu,T_\nu^{\rm eq})+2n_\phi(2\mu^{\rm eq}_\nu,T_\nu^{\rm eq})+n_\chi(\mu^{\rm eq}_\nu,T_\nu^{\rm eq})}{g_\gamma(T^{\rm eq}){(T^{\rm eq}})^3} \, ,
\\
%\frac{2 {\xi_\nu^{\rm eq}}^3 \left(\text{Li}_3\left(z_{\rm eq}^2\right)-(N_\chi+3) \text{Li}_3(-z_{\rm eq})\right)}{\pi ^2g_\gamma(T^{\rm eq})}&=-\frac{2 (N_\chi+3) {\xi_\nu^f}^3 \text{Li}_3(-z)}{\pi ^2g_\gamma(T^f)}\\
\xi_\nu^f=& \xi_\nu^{\rm eq} \left( \frac{g_\gamma(T^f)}{g_\gamma(T^{\rm eq})}\right)^{1/3} \left(\frac{\text{Li}_3\left(z_{\rm eq}^2\right)-(N_\chi+3) \text{Li}_3(-z_{\rm eq})}{-(N_\chi+3) \text{Li}_3(-z_f) }\right)^{1/3},\\
\xi_\nu^f=&\xi^{\rm SM}_\nu \left(\frac{\frac94 \zeta (3) }{-(N_\chi+3) \text{Li}_3(-z_f)}\right)^{1/3},
\end{split}
 \ee
where in the last line we used Eq.~\eqref{eq:equilibrium2}.

%At an even later time, when the cross-section of $\nu\nu \to \chi\chi$ becomes small compared to the expansion rate, $\chi$ kinetically decouples from the neutrino sector again. Just like the story of neutrino decoupling from the SM, the $\chi$ particles behave as dark radiation and maintain the same temperature as the neutrinos.

\subsubsection{$N_{\rm eff}$ from DS and neutrinos}
\begin{figure}
    %\centering
    \includegraphics[width=0.47\textwidth]{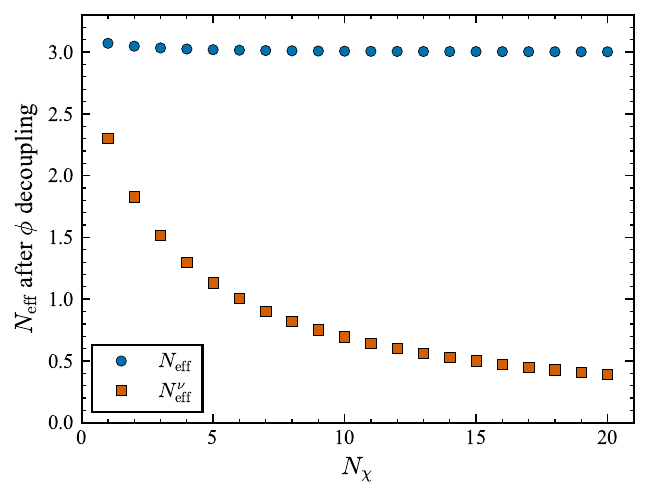}
    \captionsetup{justification=Justified}
    \caption{$N_{\rm eff}$ [Eq.~\eqref{eq:N_eff}] and $N_{\rm eff}^\nu$  as  functions of $N_\chi$ after $\phi$ decouples.
    }
    \label{fig:Neff}
\end{figure}

After the DS and neutrinos become equilibrated, 
\be
\begin{split}
N_{\rm eff}(T_\nu^{\rm eq})=\frac{\rho_\nu(\mu^{\rm eq}_\nu,T^{\rm eq}_\nu)+\rho_\phi(2\mu^{\rm eq}_\nu,T^{\rm eq}_\nu)+\rho_\chi(\mu^{\rm eq}_\nu,T^{\rm eq}_\nu) }{\frac{\pi^2}{15} (T^{\rm eq})^4} \frac87\frac1{(\xi_{\nu}^{\rm SM})^4}= 3 \, .
\end{split}
\ee
This is expected since we explicitly imposed energy conservation in obtaining the equilibrium solution.
After $\phi$ decouples, $N_{\rm eff}$ becomes
\be\label{eq:N_eff}
\begin{split}
N_{\rm eff}(T_\nu^f)=&\frac{\rho_\nu(\mu^f_\nu,T^f_\nu)+\rho_\chi(\mu^f_\nu,T^f_\nu) }{\frac{\pi^2}{15} (T^f)^4} \frac87\frac1{(\xi_{\nu}^{\rm SM})^4}\\
=&-\frac{810\ 2^{1/3} 3^{2/3} (N_\chi+3) \zeta (3)^{4/3} \text{Li}_4(-z_f)}{7 \pi ^4 (-(N_\chi+3) \text{Li}_3(-z_f)){}^{4/3}} \, .
\end{split}
\ee
Fig.~\ref{fig:Neff} shows the total $N_{\rm eff}$ Eq.~\eqref{eq:N_eff} and the contribution from neutrinos $N_{\rm eff}^\nu$ after the mediator $\phi$ decouples. It is clear that the total $N_{\rm eff}\approx 3$ irrespective of number of dark radiation species, whereas the contribution from neutrinos can be estimated as
\be
N_{\rm eff}^\nu\approx \frac{3}{3+N_\chi} N_{\rm eff}~.
\label{eq:Neffnu}
\ee
 
\section{II. Laboratory and BBN Constraints}
\label{sec:bounds}
 As summarized in Fig.~1 of Ref.~\cite{Berryman:2022hds}, 
various cosmological, astrophysical, and laboratory constraints have ruled out neutrino-mediator couplings larger than approximately $10^{-7}$ when the mediator mass is below 1 MeV. 
For the mediator mass in the keV--MeV range, 
the constraints on the neutrino-mediator coupling mainly come from laboratory measurements such as double-beta decay ($\phi\beta\beta$)~\cite{Agostini:2015nwa,
Brune:2018sab, Deppisch:2020sqh, Kharusi:2021jez} and the rare meson/$\tau/Z$ decays~\cite{Berryman:2018ogk, Brdar:2020nbj, Dev:2024twk}, with the strongest bound coming from $\phi\beta\beta$ requiring $\lambda_{\nu_e \nu_e}\lesssim 10^{-5}$~\cite{Brune:2018sab}. 
%prospective accelerator-based searches for $\nu$ scattering (DUNE~\cite{DUNE:2020lwj}, FASER$\nu$~\cite{FASER:2025myb}) and MET~\cite{deGouvea:2019qaz,Dev:2021axj}. 
Furthermore, there are astrophysical bounds from long-baseline observations, such as those of neutrinos from SN 1987A~\cite{Kolb:1987qy,Shalgar:2019rqe,Fiorillo:2023ytr} which require $\lambda_{\nu_\alpha \nu_\alpha}\lesssim 10^{-2}$ assuming flavor-universal interactions.  
In our model, the neutrino-mediator coupling [cf.~Eq.~(5)] %\eqref{eq:la_phinu1}) 
is approximately $\leq 10^{-9} $ for each neutrino mass eigenstate, and is consistent with the aforementioned constraints.
%exception of BBN, where we require $ \Delta N_\text{eff} \leq 0.4 $.

Additionally, our model introduces mixing between SM-like neutrinos and heavy sterile neutrinos, making bounds from laboratory searches for heavy neutral lepton (HNL) applicable.
For $M_N$ in the MeV range, precision measurements of meson decays~\cite{PIENU:2017wbj, NA62:2020mcv} and neutrinoless double beta decay searches~\cite{GERDA:2020xhi,KamLAND-Zen:2024eml} can probe the active-sterile mixing angle [cf.~Eq.~(3)], %(Eq.~\eqref{eq:mixing}), 
which is approximately given by $ m_\nu/M_N\lesssim 10^{-8}$ in our case.  
This extreme suppression arises naturally from the seesaw mechanism without any fine-tuning. Since we are not interested in the phenomenology of  HNLs in this work, we have verified that our benchmark $M_N = 10$~MeV is consistent with all existing terrestrial constraints~\cite{Bolton:2019pcu}. Nevertheless, future HNL searches in meson decays and beam dump experiments~\cite{Abdullahi:2022jlv} could provide an important test of our model.

\begin{figure}
    %\centering
    %\includegraphics[width=0.47\textwidth]{Rate_to_Hubble_HNL_production.pdf}
\includegraphics[width=0.47\textwidth]{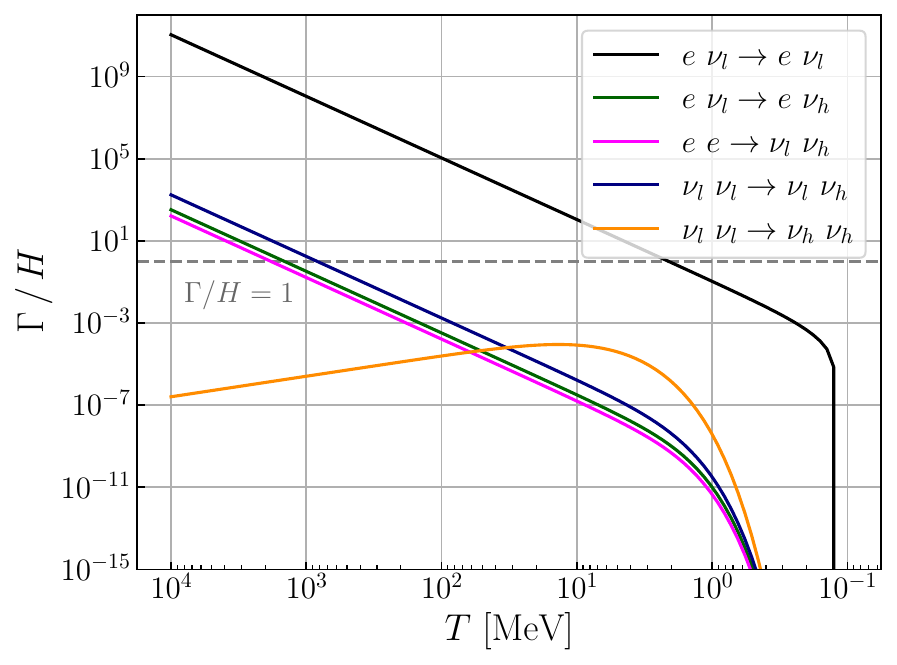}
\captionsetup{justification=Justified}
    \caption{Interaction rates over Hubble for dominant processes that keep the active neutrinos and the sterile neutrinos (or HNLs) in the thermal bath with the SM particles. The production of HNL via $\nu_l \nu_l\to \nu_l \nu_h$ (blue), $e \nu_l\to e \nu_h$ (green),   and $e e \to \nu_l \nu_h$ (magenta) only depend on the mixing angle between active neutrinos and sterile neutrinos; hence, their rates scale down by $\sim m_{\nu_l}/M_{N}$ with respect to the SM $e \nu_l\to e \nu_l$ (black). HNL pair production via $Z/\phi$ as mediators (orange) has $\Gamma/H \lesssim 10^{-3}$ at all times. Therefore, requiring $\Gamma/H < 1$ for HNL production at all times implies a low-reheating temperature of  $T_{\rm rh}\lesssim 1$ GeV.  We have taken $M_N = 10$~MeV for this benchmark case.
    }
    \label{fig:Rate_to_Hubble_nh}
\end{figure}

If the sterile neutrinos come into equilibrium prior to BBN and decay to DR, it may significantly affect $N_{\rm eff}$ and thus be severely constrained~\cite{Sabti:2020yrt, Boyarsky:2020dzc, Chen:2024cla}. Such constraints can however be avoided by choosing a low reheating temperature. 
In Fig.~\ref{fig:Rate_to_Hubble_nh}, we show the rate over Hubble for dominant processes that keep the active neutrinos and the sterile neutrinos in the thermal bath with the SM particles.
Sterile neutrino production happens through mixing with the active neutrino such as $e\,\nu_l \to e\,\nu_h$, or pair production via exchange of $Z$ or $\phi$ in $\nu_l\,\nu_l \to \nu_h\,\nu_h$. 
First, $e(\nu_l) \nu_l \to e(\nu_l)\nu_h$ yields the same temperature dependence as the SM $e\,\nu_l \to e\,\nu_l$, but scaled down by $m_{\nu_l}/M_N$ with a minimum energy requirement factor  $\sqrt{1-M_N^2/s}$. Therefore, $e\,\nu_l \to e\,\nu_h$ (green) and $\nu_l \nu_l \to \nu_l \nu_h$ (blue) have the same behavior as the SM (black). 
In addition, sterile neutrinos can be pair produced via exchange of $Z$ and $\phi$. $Z$ mediated channel is suppressed at low energy, yet, the contribution rises as the temperature increases following the trend of $e-\nu$ scattering. On the other hand, $\phi$ exchange process comprises of all $s-$, $t-$ and $u-$channel in cross-section [cf.~Eq.~\eqref{eq:2nlto2nh}], creating a long-tailed $\Gamma/H\sim T^{-1}$ behavior (orange). 
Estimates using $\Gamma/H$ show that as long as the reheating temperature is approximately below 1 GeV, which is perfectly consistent with the generic lower limit of $4$ MeV~\cite{Hannestad:2004px},  there is not enough time for the sterile neutrinos to equilibrate with the SM before BBN, and therefore, our scenario is safe from both BBN and CMB constraints~\cite{Gelmini:2004ah, Dev:2025pru}.

Unlike the model of strong neutrino self-interaction~\cite{Blinov:2019gcj}, the mediator $\phi$ in our model is not strongly constrained by BBN since the resonant conversion of neutrinos to DR takes place after BBN (before CMB) for the viable parameter space and the neutrino coupling to $\phi$ itself is tiny. If the conversion process takes place close to BBN ($m_\phi \sim {\rm MeV}$), $\Delta N_{\rm eff}$ can be significantly affected (cf. Fig.~\ref{fig:temp2} below). This is reflected in the upper right corner of Fig.~\ref{fig:1e-1scan}, 
which is excluded (gray hatched) since $\Delta N_{\rm eff}^{\rm tot} > 0.4$. 
There, thermalization of $\chi$ leads to cooling of all particles ($e^{\pm},\,\gamma$) that are in the thermal bath. Since $N_{\rm eff}^{\rm tot}$ is a measure of radiation energy density other than photons over the photon energy density, efficient cooling of all species can lead to an overall increase of $N_{\rm eff}^{\rm tot}$.

\section{III. Boltzmann Equations and Thermal History}\label{app:boltzmann_eq}

In this section, we derive the thermal history of the modified cosmology in our model by solving the relevant Boltzmann equations. 

\subsubsection{Setting up the Boltzmann Equations}\label{app:tevol}

Our main interest is to monitor how the temperature of neutrinos evolves as the DR gets produced. In order to do so, we start from the homogeneous and isotropic Boltzmann equation with a collision term:
\begin{equation}\label{eq:boltzm}
    \frac{\pd f}{\pd t}-Hp\frac{\pd f}{\pd p}=C[f]\, ,
\end{equation}
where $f=f(t,p)$ is the distribution function. Assuming a Maxwell-Boltzmann distribution (which is a good approximation in the radiation-dominated epoch), the collision term is
\be
    C[f_1]=-\frac{1}{2E_1}\int~ d\Pi_2\,d\Pi_3\,d\Pi_4 ~\delta^{(4)}(p_1+p_2-p_3-p_4)   \times \left( |\mathcal{M}_{1\,2\to3\,4}|^2 f_{1}f_{2} -|\mathcal{M}_{3\,4\to1\,2}|^2 f_{3}f_{4} \right)\, ,
\ee
where 
$
d\Pi_n = \frac{d^3p_{n}}{(2\pi)^3\,2E_n} 
$ is the phase-space factor, 
and
\be
|\mathcal{M}_{1\,2\to3\,4}|^2 = |\mathcal{M}_{3\,4\to1\,2}|^2 = \frac{1}{g_1} S \sum_\text{spin} |\mathcal{M}|^2\, .
\ee
Here $\sum_\text{spin} |\mathcal{M}|^2$ is summed over spins of all particles except the first one, $g_i$ is the internal degrees of freedom, and $S$ is the symmetrization factor which includes $\frac{1}{2!}$  for each pair of identical particles in
initial and final states and the factor 2 if there are 2 identical particles in the initial state~\cite{Dolgov:1997mb}.

We multiply both sides of Eq.~\eqref{eq:boltzm} by $\int  g_1 E_1 \frac{d^3p_1}{(2\pi)^3}$, and obtain an equation for the energy density evolution of particle 1:
\be
\begin{split}
%\frac{\partial}{\partial t}\int g_f \frac{d^3p}{(2\pi)^3} f -\int g_f \frac{d^3p}{(2\pi)^3} H p \frac{\partial f}{\partial p} 
 \frac{d \rho_1}{d t}+3H(\rho_1+P_1)
=&\int g_1 E_1\frac{d^3p_1}{(2\pi)^3}C[f_1]\\
=&-\int 2E_1 \left(\prod_{i=1}^4\,d\Pi_i \right) (2\pi)^4\,\delta^{(4)}\left(p_1+p_2-p_3-p_4\right) \times\frac{S}4\sum_\text{spin} |\mathcal{M}|^2 \left( f_{1}f_{2} - f_{3}f_{4} \right)\\
=&-\int \frac{d^3p_1}{(2\pi)^3} \frac{d^3p_2}{(2\pi)^3} 2E_1\sigma v \left( f_{1}f_{2} - f_{3}f_{4} \right)\equiv \frac{\delta\rho_{12\to34}}{\delta t} \, ,\label{eq:delrho_delt}
\end{split}
\ee
where $P_1$ is the fluid pressure, $\sigma$ is the cross-section, $v$ is the  M\o ller velocity, and the initial (final) state particles are assumed to be identical. Suppose that the initial (final) state  particles have a temperature $T~(T')$. The collision term can be further simplified as
\be\label{eq:delrho_delt2}
\frac{\delta\rho_{12\to34}}{\delta t} = -\frac{S}{(2\pi)^4} \int_{s_\text{min}}^\infty 
ds~ \sigma \,s^2 \left[TK_2\left(\frac{\sqrt s}{T}\right)-T'K_2\left(\frac{\sqrt s}{T'} \right)\right] \, ,
\ee
where $s$ is the center-of-mass energy and $K_2$ is the modified Bessel function of the second kind. 
Furthermore, the right-hand-side of the Eq.~(\ref{eq:delrho_delt}) needs to be a linear combination of all possible interactions that affect particle 1 except for the self-interaction. This accounts for all possible energy flows as a given particle type transfers into the other. For instance, in the $\nu_e$ sector, we may compartmentalize the total energy transfer term into SM-only term ($\frac{\delta \rho^{\text{SM}}_{\nu_e}}{\delta t}$) and new BSM terms ($\frac{\delta \rho^{\text{BSM}}_{\nu_e}}{\delta t}$). In temperature evolution equations~(\ref{eq:Tdiffeq1}--\ref{eq:Tdiffeq4}) below, energy transfer of BSM is explicitly presented.

We are interested in the cosmological era from BBN to CMB during which the photon, neutrino and the DS may come out or reach equilibrium, i.e. their temperatures $T_\gamma, T_{\nu_{e,\mu,\tau}}, T_\chi$ may evolve separately. Assuming $T_{\nu_\tau}=T_{\nu_\mu}$, there will be four coupled temperature evolution ($\dot T\equiv dT/dt$) equations, which can be obtained from the Boltzmann equations for density evolution as follows:  
\begin{align}\label{eq:Tdiffeq1}
\dot T_\gamma \left(\frac{\partial \rho_e}{\partial T_\gamma}+\frac{\partial \rho_\gamma}{\partial T_\gamma}\right)&+4H\rho_\gamma+3H(\rho_e+P_e)= -\frac{\delta \rho_{\nu_e}}{\delta t} -2\frac{\delta \rho_{\nu_\mu}}{\delta t} \, ,\\
\dot T_{\nu_e}\frac{\partial \rho_{\nu_e}}{\partial T_{\nu_e}}+4H\rho_{\nu_e}=&\frac{\delta \rho_{\nu_e\nu_e \to \chi\chi}}{\delta t} + \frac{\delta \rho_{\nu_e\nu_e \to 4\chi}}{\delta t} + \frac{\delta \rho_{\nu_e\nu_\mu \to 4\chi}}{\delta t} + \frac{\delta \rho_{\nu_e\nu_\mu \to \chi\chi}}{\delta t} + \frac{\delta \rho_{\nu_e}}{\delta t} \, ,\\
\dot T_{\nu_\mu}\frac{\partial \rho_{\nu_\mu}}{\partial T_{\nu_\mu}}+4H\rho_{\nu_\mu}=&\frac{\delta \rho_{\nu_\mu\nu_\mu \to\chi\chi}}{\delta t} + \frac{\delta \rho_{\nu_\mu\nu_\mu \to 4\chi}}{\delta t} + \frac{\delta \rho_{\nu_e\nu_\mu \to 4\chi}}{\delta t} + \frac{\delta \rho_{\nu_e\nu_\mu\to \chi\chi}}{\delta t} + \frac{\delta \rho_{\nu_\mu}}{\delta t} \, ,\\
\dot T_{\chi}\frac{\partial \rho_\chi}{\partial T_\chi}+4H\rho_\chi=& -\frac{\delta \rho_{\nu_e\nu_e \to 4\chi}}{\delta t} - 2\frac{\delta \rho_{\nu_\mu\nu_\mu \to 4\chi}}{\delta t} - 3\frac{\delta \rho_{\nu_e\nu_\mu \to 4\chi}}{\delta t} -\frac{\delta \rho_{\nu_e\nu_e \to \chi\chi}}{\delta t} -2 \frac{\delta \rho_{\nu_\mu\nu_\mu \to \chi\chi}}{\delta t} -3\frac{\delta \rho_{\nu_e\nu_\mu \to \chi\chi}}{\delta t}\, .\label{eq:Tdiffeq4}
\end{align}
Here the right hand sides are collision terms of different processes which are functions of temperatures only. Note that in our convention, $\frac{\delta\rho_{\nu_e}}{\delta t}$ is the energy transfer under the SM which receives contributions from $\nu_e\nu_e\leftrightarrow\nu_l\nu_l$, $\nu_e\nu_l\leftrightarrow\nu_e\nu_l$, $\nu_e e\leftrightarrow\nu_e e$, and $\nu_e\nu_e\leftrightarrow e e$ yielding the conventional neutrino decoupling in cosmology.

To obtain the left hand sides, we assume the form of ideal gases for each species and use the relation $\dot\rho_i=\frac{\partial \rho_i}{\partial T_i}\dot T_i $. Furthermore, $H$ can be traded for $\rho$ using the Friedmann equation $H^2=\frac{8\pi G \rho_{\rm tot}}3$, where $\rho_\text{tot}=\rho_\gamma+\rho_e+\rho_{\nu_e}+2\,\rho_{\nu_\mu}+\rho_\chi$. It can be easily verified that the total energy  is conserved i.e. $\dot\rho_{\rm tot}=-3H(\rho_{\rm tot}+P_{\rm tot})$. Note that we did not include $\phi$ in the DS bath, since for the range of $\phi$ mass and the temperature of interest it decays immediately to $\chi$. The boosted decay width ${\Gamma_\phi\,E_\phi}/{m_\phi}$ is much higher than the Hubble within our time range. Therefore, process of $\phi$ pair production is equivalent to the decay of $\phi\to\chi\chi$ assuming that the branching ratio of $\phi\to\nu\nu$ is suppressed by the $\nu-\phi$ coupling, $\lesssim 10^{-8}$, cf. Eq.~\eqref{eq:la_phinu}. The coupled differential equations Eq.~\eqref{eq:Tdiffeq1}-\eqref{eq:Tdiffeq4} are numerically solved using \texttt{python} with  \texttt{numpy}~\cite{Harris:2020xlr} and \texttt{odeint} library in \texttt{scipy}~\cite{Virtanen:2019joe} package.

\subsubsection{Cross-sections}\label{app:Xsec}

The energy density transfer in Eq.~(\ref{eq:delrho_delt2}) displays the explicit cross-section dependence. Below we give expressions of various cross-sections that are relevant in this work. We cross-checked the cross-section calculation through \texttt{Mathematica} with \texttt{FeynCalc}~\cite{Shtabovenko:2023idz}, \texttt{FeynArts}~\cite{feynarts} and \texttt{FeynRules}~\cite{Alloul:2013bka} pipeline.

The $s$-channel production of light fermions from the $i$-th neutrino through $\phi$ has a cross-section given by\footnote{We indicate $\nu_{l;i}$ as light neutrinos in mass eigenstates.}
\be \label{eq:sigma_nunuchichi}
\sigma_{\nu_{l;i}\, \nu_{l;i}\to\chi\chi} =
\frac{1}{16 \pi s}\lambda_{\phi\chi}^2
\left(\lambda_{\phi N}\,\frac{m_{\nu;i}}{M_{N;i}}\right)^2
 \frac{(s-4 m_\chi^2)\,(s-4 m_{\nu;i}^2)}{{(s-m_\phi^2)}^2+m_\phi^2\Gamma_\phi^2}~.
\ee
Assuming that $\lambda_{\phi\chi},\lambda_{\phi N}$ are of order 1, the cross-section is largely suppressed by the mass ratio of light and heavy neutrinos. The cross-section is enhanced when $s\approx m_\phi^2\gg m_\chi^2$, allowing us to use the narrow width approximation: 
\be
\sigma_{\nu_{l;i}\, \nu_{l;i}\to\chi\chi} \xrightarrow{s\to m_\phi^2} 
 \lambda_{\phi\chi}^2\,
\lambda_{\phi \nu_{l;i}}^2\, \frac 1{16 \pi \Gamma_\phi^2} 
%=\frac{\lambda_{\phi\chi}^2\,\lambda_{\phi N}^2\,m_{n_{l;i}}^2}{\pi\,{m_{N;i}^2} \,\Gamma_\phi^{\,2}} 
\approx \frac{\lambda_{\phi \nu_{l;i}}^2}{\lambda_{\phi\chi}^2} \frac{16\pi}{\,m_\phi^2} \, ,
\ee
where $\lambda_{\phi \nu_{l;i}}$ is the coupling between the SM-like neutrinos and the mediator, and is given by 
\be\label{eq:la_phinu}
\lambda_{\phi \nu_{l;i}}\equiv \lambda_{\phi N}\,\frac{m_{\nu;i}}{M_{N;i}}~\ll1 ~.
\ee
To get the final expression, we used the fact that $\phi$ dominantly decays to $\chi$ since $\lambda_{\phi\chi}\gg\lambda_{\phi \nu_{l;i}}$, and
\be
\Gamma_\phi\approx\frac{m_\phi \lambda_{\phi\chi}^2}{16\pi}~.
\ee
To get the cross-section in flavor basis $\nu_{\alpha_1}\nu_{\alpha_2}\to \chi\chi$, we simplify multiply $\sigma_{\nu_{l;i}\, \nu_{l;i}\to\chi\chi}$ by $|U_{\alpha_1 i}|^2\,|U_{\alpha_2 i}|^2$ from the PMNS matrix and sum over the mass eigenstates $i=1,2,3$. 
Note that this cross-section does not explicitly depend on neutrino masses, but on the ratio given in Eq~(\ref{eq:la_phinu}).

In addition, $\chi$ can be produced via  $\nu\nu\to \phi\phi$ by $t$ and $u$-channel exchange of heavy neutrinos $N^c$. The cross-section for ${\nu_{l;i}\, \nu_{l;i}\to\phi\phi}$ is equivalent to that of ${\nu_{l;i}\, \nu_{l;i}\to\phi(\chi\chi)\phi(\chi\chi)}$ since the branching ratio of $\phi\to\chi\chi$ is approximately 1. The cross-section is given by
\be\label{eq:nunuphiphi}
\begin{split}
\sigma_{\nu_{l;i}\, \nu_{l;i}\to\phi\phi} =&\frac{\lambda_{\phi \nu_{l;i}}^2 \lambda_{\phi N}^2}{128\pi s} 
\Bigg[ \frac{8f(s)}{1+\frac{s}{M_{N;i}^2}g_i(s)^2} -24 f(s) +\frac{16\left( 6g_i(s)^2+4g_i(s)+1+\frac{4M_{N;i}^2}{s} \right)}{(1+2g_i(s))}\tanh^{-1}\left(\frac{f(s)}{1+2g_i(s)}\right)  \Bigg]\\
& \xrightarrow{M_N^2\gg m^2_\phi,s } \frac{\lambda_{\phi \nu_{l;i}}^2 \lambda_{\phi N}^2}{8\pi M_{N;i}^2} \left(1-\frac{s}{M_{N;i}^2}\right) f(s) \, ,
%{\rm or} \xrightarrow{s\gg m^2_\phi,M_N^2 }  \frac{\lambda_{\phi \nu_{l;i}}^2 \lambda_{\phi N}^2}{8\pi s}\left( \tanh^{-1}\left(\frac{f(s)}{1+2g_i(s)}\right) -1 \right)
\end{split}
\ee
where $f(s)\equiv \sqrt{1-\frac{4 m_\phi^2}{s}} $ and $g_i(s)\equiv \frac{M_{N;i}^2-m_\phi^2}s$. Note that the expression takes a much simplified form in the limit of $M_N^2\gg m^2_\phi,s $. This process needs to be suppressed compared to the neutrino interaction via $W/Z$ bosons at early times in order to avoid populating $\chi$ sector prematurely (equivalent to avoiding $N_\text{eff}^\text{tot}$ bound). 
%As appearing in the upper panel of Fig.\,\ref{fig:benchmark}, $\chi$ sector receives two step heating $\nu\nu\to\phi(\chi\chi)\phi(\chi\chi)$ followed by $\nu\nu\to\chi\chi$. 
Our numerical scan comprises of parameter space where $\nu\nu\to \phi\phi$ is never important before BBN; see Fig. 1 top panel. This can always be achieved by choosing a large enough $M_N$.

The $\chi$ self-interaction cross-section is given by
\be
\sigma_{\chi\chi\to\chi\chi}=\frac{ \lambda_{\phi\chi}^4 }{16 \pi  s^2 \left(m_\phi^2-s\right)^2} \left[\frac{2 \left(5 m_\phi^8-9 m_\phi^6 s+4 m_\phi^2 s^3\right) \log \left(\frac{m_\phi^2}{m_\phi^2+s}\right)}{2 m_\phi^2+s}+\frac{s \left(5 m_\phi^6-9 m_\phi^4 s+6 s^3\right)}{m_\phi^2+s}\right]\, ,
\ee
assuming $m_\chi^2\ll s, m_\phi^2$. In the limit $m_\phi^2\gg s$,
\be
\sigma_{\chi\chi\to \chi\chi}\approx \frac{5 \lambda_{\phi\chi}^4 s}{96 \pi m_\phi^4}+\mathcal{O}\left(\frac{s^2}{m_\phi^6}\right) \, ,
\ee
which takes the form of $\frac s{12\pi} G_{\rm eff}^2 $, where $G_{\rm eff}$ characterizes the effective self-interaction among dark radiation. This has the same form as the cross-section of $\nu-\nu$ scattering in the SM assuming the interaction Lagrangian $G_F \bar \nu \nu \bar\nu \nu$.

Lastly, the heavy sterile neutrinos can be thermalized with the SM bath through mixing with the active neutrinos via processes like $e\,\nu_l \to e\,\nu_h$, or pair production, via exchange of $W/Z$ or $\phi$ in $\nu_l\,\nu_l \to \nu_h\,\nu_h$. The process
$e\,\nu_l \to e\,\nu_h$ has the same form of cross-section as the SM process $e\,\nu_l \to e\,\nu_l$, but scaled down by $m_{\nu_l}/M_N$ with a minimum energy requirement factor $\sqrt{1-M_N^2/s}$. 
On the other hand, for the $\nu_h$ pair production, the $Z$ mediated channel is suppressed at low energy, whereas the $\phi$-mediated channel is more important. Below we give the differential cross-section of $\phi$-exchanged pair production:
\begin{align}\label{eq:2nlto2nh}
    \frac{d\sigma_{\nu_{l;i}\nu_{l;i}\to\nu_h\nu_h}}{d\Omega} \approx  \frac{1}{16\pi^2s}\sqrt{\frac{s-4M_N^2}{s}} \,\lambda_{\phi N}^4\frac{m_{\nu_{l;i}}^2}{M_N^2} \Bigg[ & 3+M_N^4\left(\frac{1}{t^2}-\frac{1}{tu}+\frac{1}{u^2}\right)
    %\nonumber\\
    %&
    -M_N^2\left(\frac{u}{st}+\frac{t}{su}+\frac{2 s}{tu} +\frac2s -\frac1t -\frac1u\right)\nonumber\\
    & + \frac12 \bigg(\frac{s^2}{tu}+\frac{t^2}{su}+\frac{u^2}{st}-\frac ts -\frac st -\frac us -\frac su -\frac ut -\frac tu \bigg) \Bigg] \, .
\end{align}
%%%%%%%%%%%%%%%%%%
\subsubsection{Interaction Rates}\label{app:gamma_by_hubble}
To estimate whether a process comes into equilibrium with the thermal bath, the minimum requirement is to have them in contact with the bath particles. When this happens can be estimated using the ratio of the thermally averaged interaction rate to the Hubble rate, $\Gamma/H$. 

Consider a 2 $\to$ 2 process $p_1+p_2\to p_3+p_4$, where the two initial-state particles are of identical mass $m$.  Suppose that the initial-state particles are part of the early universe bath. To estimate when this process begins to become efficient, we compare the Hubble rate with the rate of this process~\cite{Gondolo:1990dk}, which is defined as
$
\Gamma\equiv n \langle \sigma v\rangle
$, 
where $n$ is the density of the initial state particles and $\langle \sigma v\rangle$ is the thermally averaged cross-section defined as:
\be
\langle \sigma v\rangle=
\frac{\int d^3p_1\int d^3p_2~\sigma v\, e^{-E_1/T-E_2/T}}{\int d^3p_1\int d^3p_2~e^{-E_1/T-E_2/T}} \, ,
\ee
where the denominator is simply given by 
$\left(4 \pi  m^2 T K_2\left(\frac{m}{T}\right)\right)^2$. 
To obtain an expression for the numerator, we express the relative velocity as
$
%=\frac{\sqrt{(p_1\cdot p_2)^2-m_1^2m_2^2}}{E_1E_2}
v=\frac1{2E_1E_2}\sqrt{s(s-4m^2)}
$. 
Defining $E_\pm =E_1\pm E_2$,  
the numerator can be simplified as
\be
\begin{split}
\int d^3p_1\int d^3p_2~\sigma v\, e^{-(E_1+E_2)/T}=&
 \pi^2\int_{4m^2}^{\infty} ds~ \sigma  \sqrt{s(s-4m^2)} \int_{\sqrt s}^{\infty} dE_+ e^{-E_+/T} \int dE_-\\
=&2\pi^2\int_{4m^2}^{\infty} ds~ \sigma  (s-4m^2) \int dE_+ e^{-E_+/T}\sqrt{E_+^2-s}\\
=&2\pi^2T \int_{4m^2}^{\infty} ds~ \sigma  (s-4m^2)\sqrt{s}K_1\left(\frac{\sqrt s}T\right).
\end{split}
\ee
where we used $|E_-|\leq \sqrt{1-\frac{4m^2}s} \sqrt{E_+^2-s}$. 
In summary, we obtain 
\be\label{eq:thermalrate}
\Gamma_{12\to 34}=
\frac{g_1}{16\pi^2m^2K_2\left(\frac{m}{T}\right)}  \int_{4m^2}^{\infty} ds~ \sigma_{12\to34}  (s-4m^2)\sqrt{s}K_1\left(\frac{\sqrt s}T\right)
\xrightarrow{\frac mT\to 0}\frac{g_1}{32\pi^2T^2}  \int_{4m^2}^{\infty} ds~ \sigma_{12\to34}  s^{3/2} K_1\left(\frac{\sqrt s}T\right) \, .
\ee

Here, we present the explicit  expressions for $\Gamma/H$ using cross-sections from the previous subsection. Below are the results assuming $m_\chi, m_\nu\ll \sqrt{s}$, 
\begin{align}
\Gamma_{\nu\nu\to\chi\chi}\approx & \frac{ \lambda_{\phi\chi}^2\left(\frac{\lambda_{\phi N} m_{\nu}}{M_{N}}\right)^2}{256\pi^3T^2}  \int_{(m_\phi-\Gamma_\phi)^2}^{(m_\phi+\Gamma_\phi)^2} ds~ \frac{ s^{1/2} K_1\left(\frac{\sqrt s}T\right)s^2}{{(s-m_\phi^2)}^2+m_\phi^2\Gamma_\phi^2}\approx  \left(\frac{\lambda_{\phi N} m_\nu}{M_{N}}\right)^2 K_1\left(\frac{m_\phi}T\right)\frac{m_\phi^3}{4\pi^2T^2} \, , \\
\Gamma_{\chi\chi\to\chi\chi}\approx&\frac{5 \lambda_{\phi\chi}^4}{16\pi^3T^2 96m_\phi^4}  \int_{4m_\chi^2}^{\infty} ds~s^{5/2} K_1\left(\frac{\sqrt s}T\right)\approx\lambda_{\phi\chi}^4 \frac{5 T^5 }{2\pi^3 m_\phi^4} \text{ when $m_\chi\ll \sqrt{s}\ll m_\phi$} \, , \\
\Gamma_{\nu\nu\to\phi\phi}\approx &\frac{\lambda_{\phi\chi}^2 m_{\nu}^2 \lambda_{\phi N}^4}{256\pi^3T^2M_N^6}  \int_{4m_\phi^2}^{\infty} ds~ s^{5/2}K_1\left(\frac{\sqrt s}T\right)\approx \lambda_{\phi\chi}^2\left(\frac{m_{\nu}}{M_{N}}\right)^2  \lambda_{\phi N}^4 \frac{3T^5}{\pi^3 M_N^4} \text{ when $m_\phi\ll \sqrt s \ll M_N $} \, , \\
\Gamma_{\phi\to\chi\chi}\approx & ~ \Gamma_\phi \frac{K_1(m_\phi/T)}{K_2(m_\phi/T)}\text{~~~ where $\Gamma_\phi$ is decay width of $\phi$ in rest frame} \, .
\end{align}

Assuming $T>m_\phi$, the ratio of the interaction rate of $\phi$ to Hubble rate reduces to 
\begin{align}
    &\frac{\Gamma_{\phi\to\chi\chi}}{H}\sim \lambda_{\phi\chi}^2 m_\phi \left(  \frac{m_\phi}{T} \right) \frac{M_{\rm pl}}{T^2}~.
\end{align}
This ratio is much bigger than 1 for all the model parameters and temperatures of interest in this work, which means that $\phi$ is never long-lived enough to be part of the bath. This is why we do not track the evolution of $\phi$ in the main text.

\begin{figure*}[!t]
    \centering \includegraphics[width=.8\linewidth]{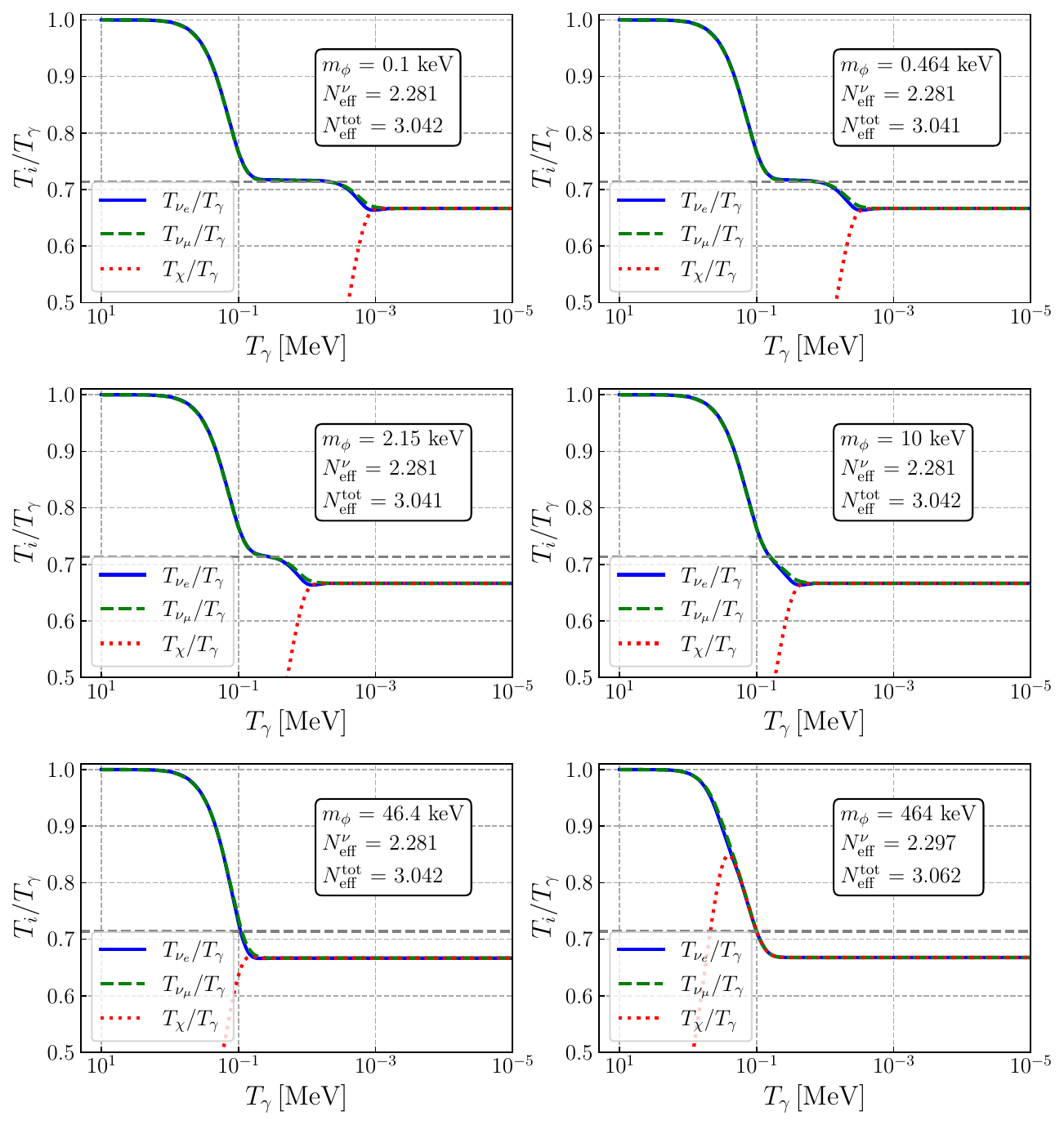}
    \captionsetup{justification=Justified}
    \caption{Temperature evolution of $\nu$ and $\chi$ when varying only  $m_\phi$. Here $\sum m_\nu=0.065$ eV and $n_\chi=1$. Other parameters are as in Fig.~\ref{fig:1e-1scan}. 
    Cooling is efficient for all values of $m_\phi$ but with different initialization times. 
    }
    \label{fig:temp1}
\end{figure*}

\subsubsection{Calculation of $N_{\rm eff}^\nu$}
In Fig.~\ref{fig:1e-1scan},
we show the resultant $N_{\rm eff}^\nu$ for $n_{\chi}=1$ (left panel) and $n_\chi=2$ (right panel) after $\nu-\chi$ decoupling from solving the  Boltzmann equations~(\ref{eq:Tdiffeq1}--\ref{eq:Tdiffeq4}). Here we give more details of this analysis.

In Fig.~\ref{fig:temp1} we show the temperature evolution for $\chi$ and $\nu$ at a fixed neutrino mass $\sum m_\nu=0.065$ eV and the number of DR flavor $n_\chi=1$, but with different $\phi$ masses. This corresponds to $G_{\rm eff}\in(10^{-5}~{\rm MeV}^{-2},10^{-1}~{\rm MeV}^{-2})$. It can be seen that $m_\phi$ only affects when neutrinos start to be cooled, since the cooling occurs resonantly through $\nu\nu\to \phi\to \chi\chi$ when $T_\nu\sim m_\phi$. It does not affect the final $N_{\rm eff}^\nu$ as long as the $\nu-\chi$ thermalization is efficient.

\begin{figure*}
    \centering
    \includegraphics[width=.8\linewidth]{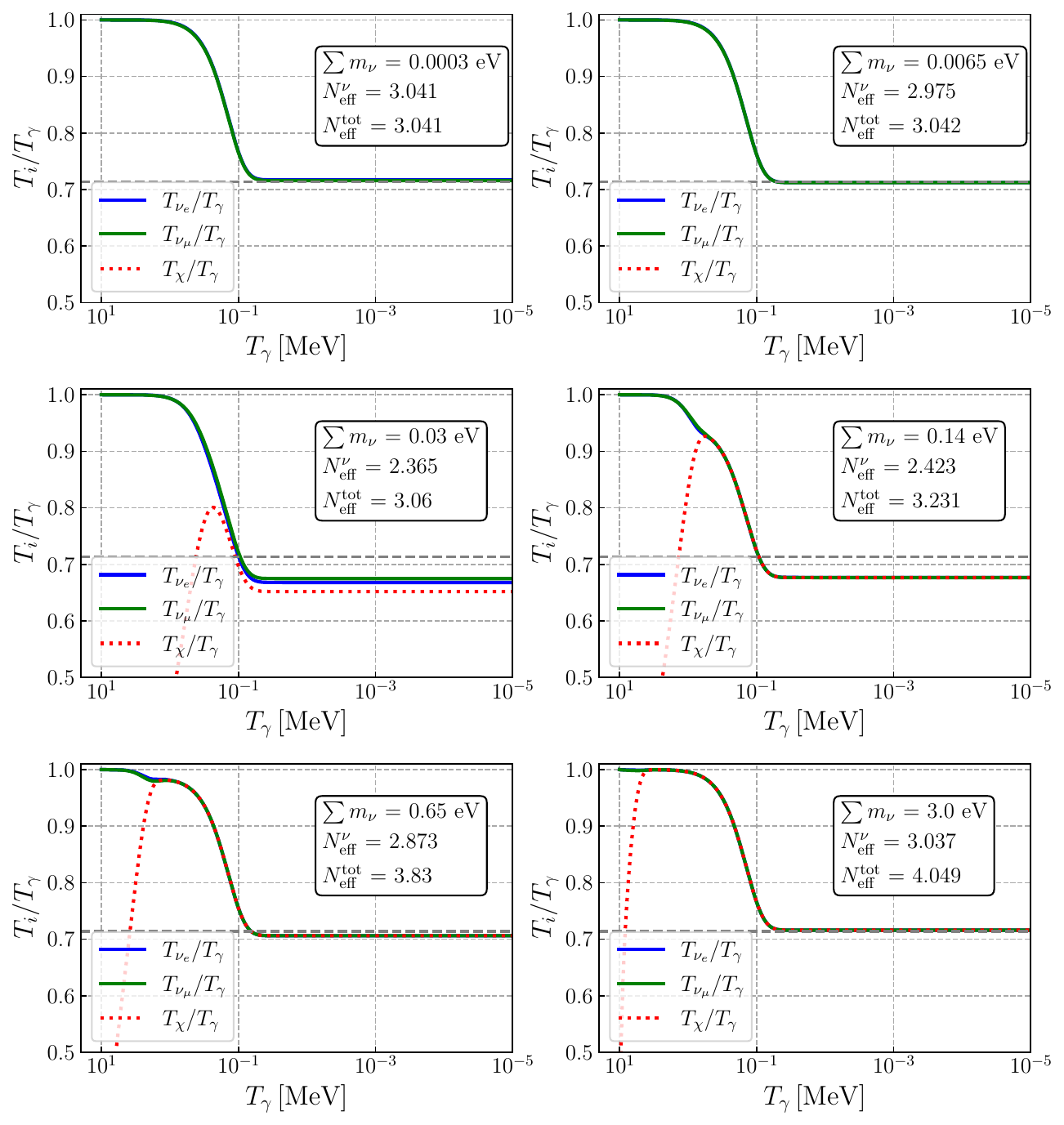}
    \captionsetup{justification=Justified}
    \caption{Temperature evolution of $\nu$ and $\chi$  when varying only $\sum m_{\nu}$. Here $m_\phi = 1~\text{MeV}$, $n_\chi=1$, and other parameters are as in Fig.~\ref{fig:1e-1scan}. 
    As $\sum m_\nu$ increases, the coupling of $\nu-\phi$ increases (cf. Eq.~\ref{eq:la_phinu1}
    ), hence increasing the $\nu\to\chi$ energy transfer efficiency. This results in an insufficient cooling for too small $\sum m_\nu$ in the first row. For large enough $\sum m_\nu$, the cooling immediately takes off at a temperature close to $\phi$ mass. This results in a significant increase in $N_{\rm eff}^{\rm tot}$ in the last row, which would be severely constrained by BBN.}
    \label{fig:temp2}
\end{figure*}

In Fig.~\ref{fig:temp2} we show the temperature evolution for $\chi$ and $\nu$ at a fixed $m_\phi=1$ MeV for difference choices of $\sum m_\nu$. This corresponds to $G_{\rm eff}\approx 10^{-5}~{\rm MeV}^{-2}$ in Fig.~\ref{fig:1e-1scan}. 
Since the neutrino--mediator coupling in Eq.~(\ref{eq:la_phinu}) is proportional to the light to heavy neutrino mass ratio, $\sum m_\nu$ directly controls the overall size of $\sigma(\nu\nu\to \chi\chi)$, and hence, the $\nu\to\chi$ energy transfer efficiency. For $\sum m_\nu< 0.03$ eV, cooling is insufficient as indicated in the lower right corner of Fig.~\ref{fig:1e-1scan}. 
This is reflected in the negligible difference between $N_{\rm eff}^\nu$ and $N_{\rm eff}^{\rm tot}$ in the first row of Fig.~\ref{fig:temp2}. 
For $\sum m_\nu > 0.65$ eV, $\sigma(\nu\nu\to \chi\chi)$ is large enough that the cooling takes off immediately when the temperature is close to the $\phi$ mass (MeV). 
Under this condition, 
the neutrino is still in thermal contact with $e^\pm$, implying that they are indirectly coupled with $\gamma$ as well. The production of $\chi$ lowers the neutrino temperature and this leads to energy flow of $e\to\nu$ to accommodate thermal equilibrium. Due to the strong coupling between $e-\gamma$ before/during BBN, photon gets cooled accordingly. As a result, the system after BBN includes small portion of extra radiation, $\chi$, which means that the total $N_\text{eff}$ is higher than $N_\text{eff}^\text{SM}$.

\begin{figure*}
    %\centering
    \includegraphics[width=.45\linewidth]{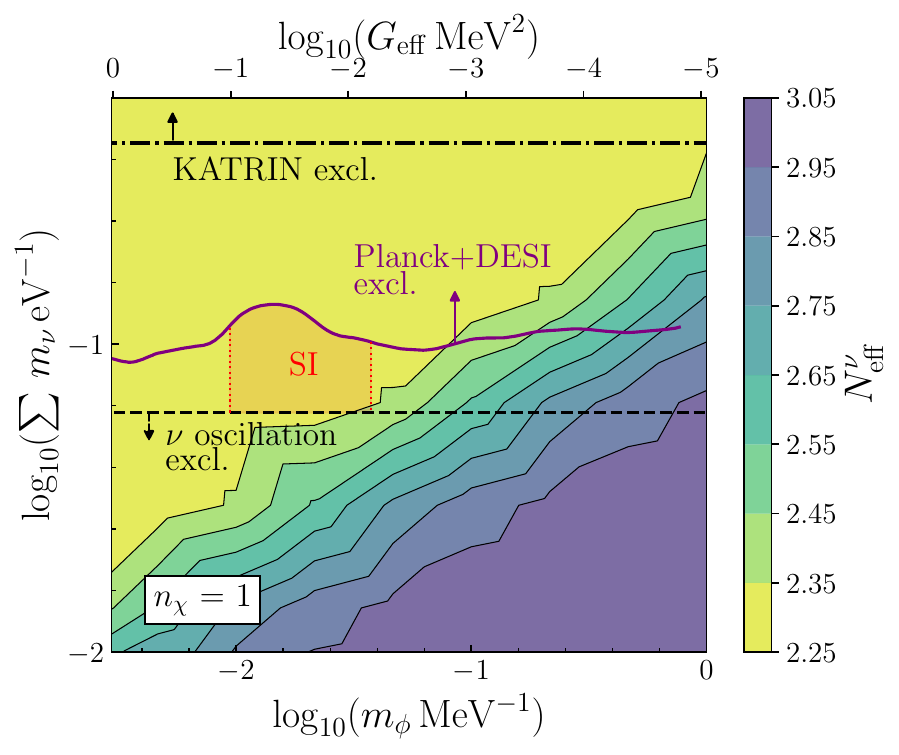}
    \includegraphics[width=.45\linewidth]{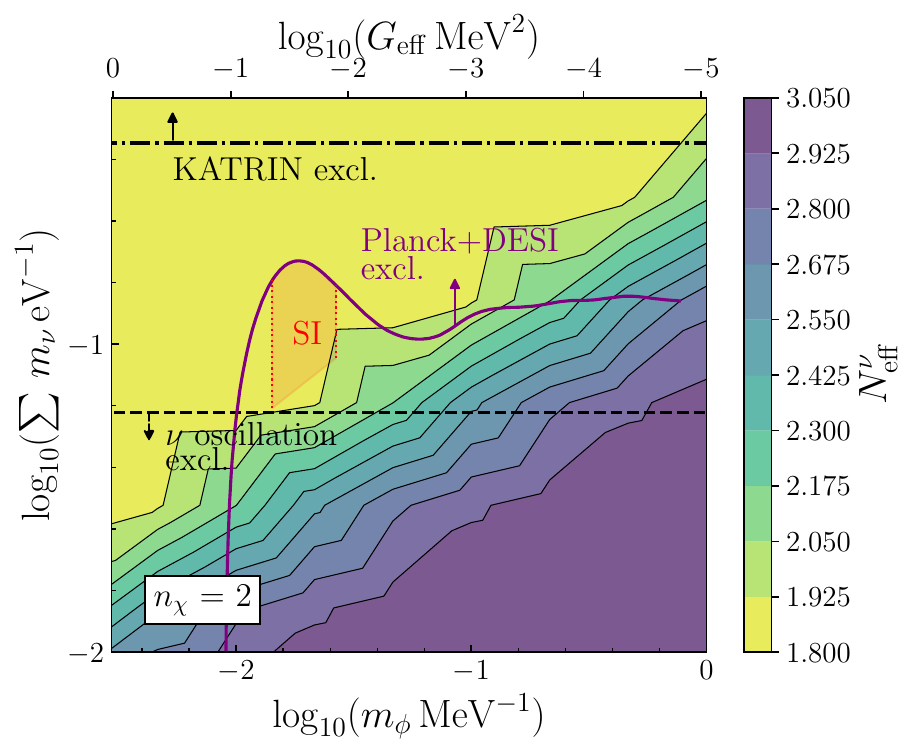}
    \captionsetup{justification=Justified}
    \caption{$\left(~\sum m_\nu,\,m_\phi~\right)$ parameter scan result with $n_\chi= 1~ (\textbf{Left})$ and $n_\chi= 2~ (\textbf{Right})$ with $m_N/\lambda_{\phi N}=1$ GeV, $\lambda_{\phi\chi}=0.003$, assuming degenerate neutrino masses. 
    }
    \label{fig:1e-2scan}
\end{figure*}

In Fig.~\ref{fig:1e-2scan}, we show $n_\chi=1,2$ results as we increase $m_N/\lambda_{\phi N}$ to be 10 times larger than that in Fig.~2. This results in a 10 times smaller $\lambda_{\phi\nu}$ [cf. Eq.~(5) in the main text], and hence, 100 times smaller $\sigma(\nu\nu\to\phi\to\chi\chi)$, thus a reduced efficiency in $\nu-\chi$ energy transfer. 
%This is reflected in the overall reduction in the light yellow region where neutrino cooling is maximal.
Therefore, $\nu\nu\leftrightarrow\phi\leftrightarrow\chi\chi$ energy transfer is significantly slowed down, which results in an insufficient cooling of neutrinos. This is reflected in the overall reduction in the light yellow region (where neutrino cooling is maximal) compared to that of Fig.~\ref{fig:1e-1scan}.
As we increase $m_N/\lambda_{\phi N}$ even further, this cooling mechanism becomes completely ineffective for $m_N/\lambda_{\phi N}\gtrsim 10$ GeV.

\subsubsection{Normal Ordering Benchmarks}
\begin{figure*}
    \centering
    \includegraphics[width=.8\linewidth]{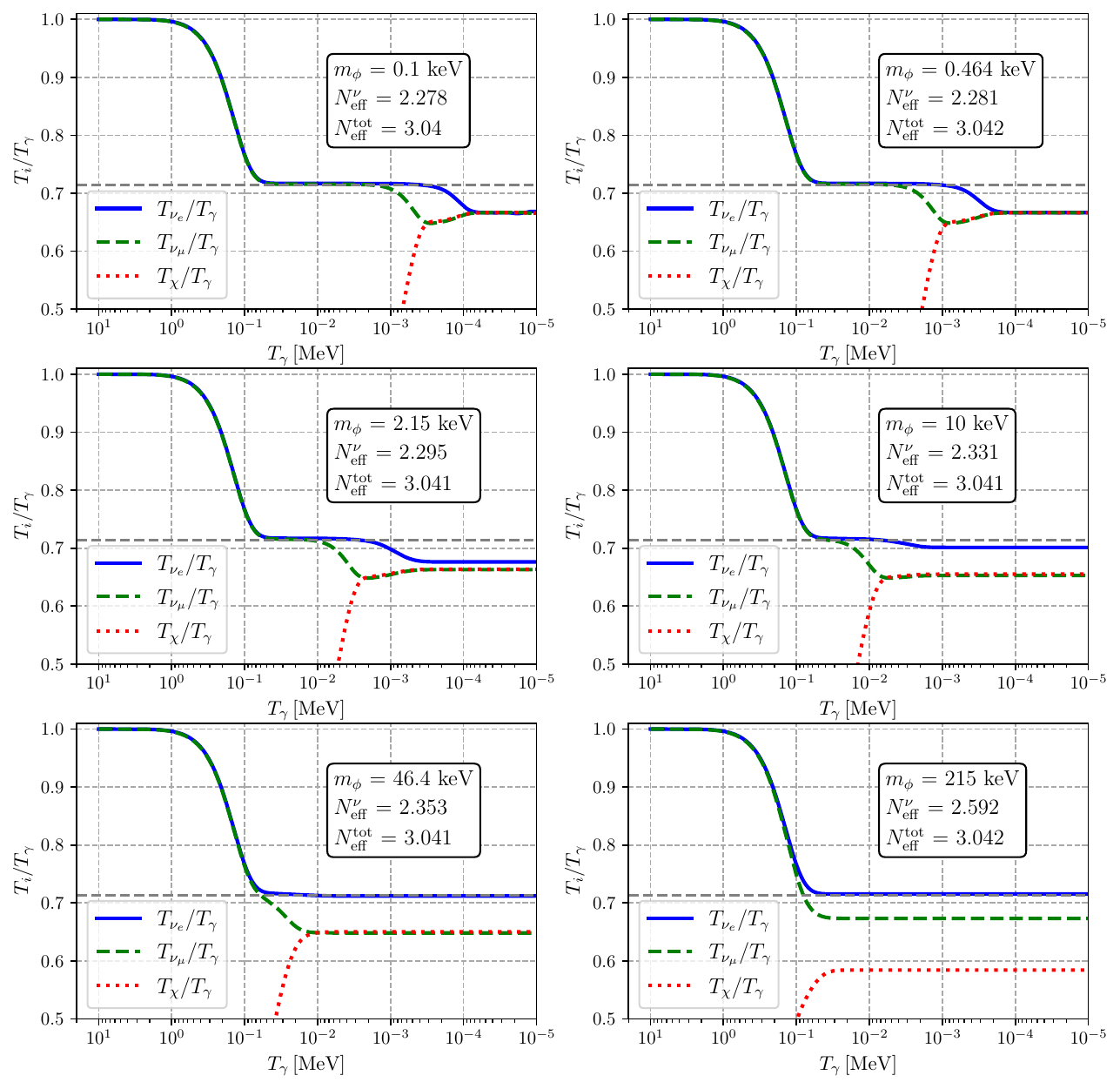}
    \captionsetup{justification=Justified}
    \caption{Temperature evolution of $\nu$ and $\chi$  when varying only $m_{\phi}$. Normal ordering is assumed with the lightest neutrino mass $m_{\nu}^{\rm min}=1$ meV and with $n_\chi=1$, while the other parameters are the same as in Fig.~\ref{fig:1e-1scan}.
    }
    \label{fig:NO}
\end{figure*}

In the analyses above it is assumed that the neutrinos have degenerate masses. In Fig.~\ref{fig:NO} we show benchmarks with normal ordering with the lightest neutrino mass $m_{\nu}^{\rm min}=1$ meV. Compared to Fig.~\ref{fig:temp1}, the overall cooling of neutrinos is not affected for $m_\phi \lesssim 100$ keV. However, in this case, different neutrino species may not be in equilibrium after cooling, which is reflected in different evolution trajectories undergone by $T_{\nu_\mu}$ and $T_{\nu_e}$. The reason is that $\nu_e$ comprises the first two lightest mass eigenstates mostly, thus contributing less to $\nu-\chi$ thermalization, since the neutrino-mediator coupling is directly proportional to neutrino masses. This is especially prominent when $m_\phi$ becomes larger, where $\nu_e$ does not contribute at all to $\nu-\chi$ thermalization and thus has the same temperature as that of the standard cosmology case. 

\subsubsection{Increasing DR Flavors}
\begin{figure*}
    %\centering
    \includegraphics[width=.44\linewidth]{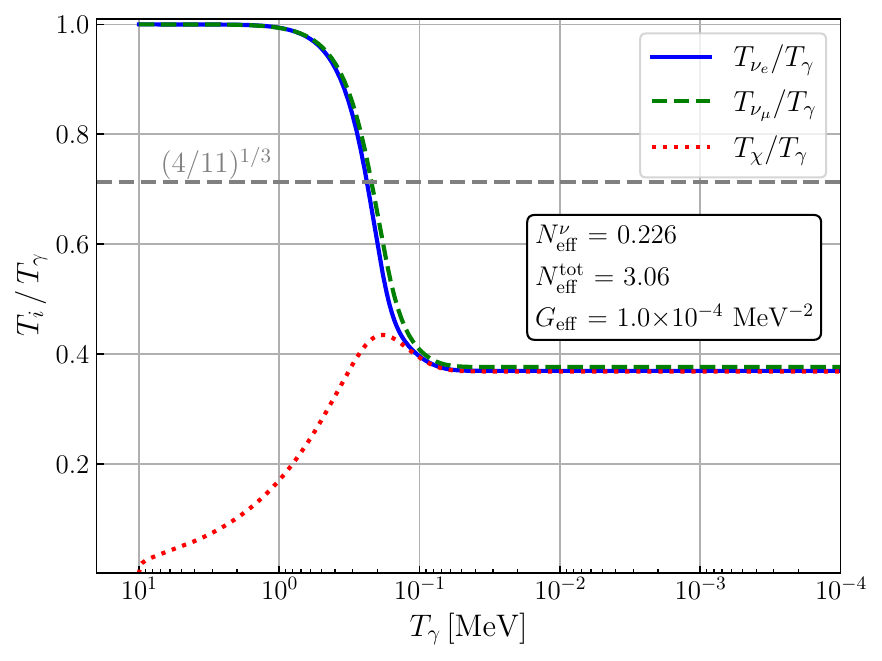}
    \includegraphics[width=.44\linewidth]{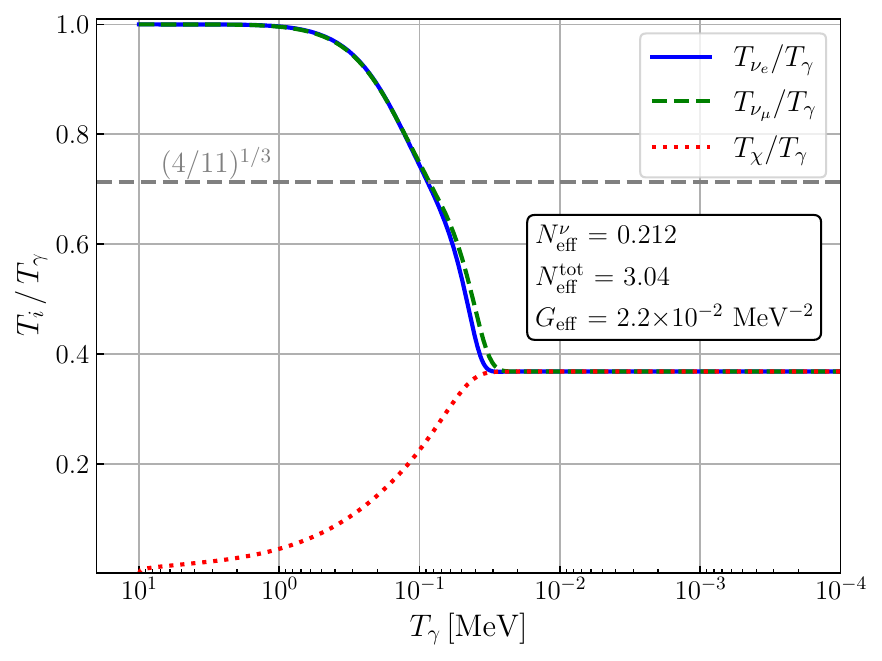}
    \captionsetup{justification=Justified}
    \caption{Temperature evolution of $\nu$ and $\chi$ for $n_\chi=40$. Here we have taken $m_N/\lambda_{\phi N}=100$ MeV and $\sum m_\nu =0.09$ eV assuming degenerate neutrino masses. The left (right) panel is for the MI (SI) mode with $G_{\rm eff}=1.0\times 10^{-4}~{\rm MeV}^{-2}$ ($2.2\times 10^{-2}~{\rm MeV}^{-2}$). Note the sharp drop in the neutrino temperature. }
    \label{fig:nchi40}
\end{figure*}

The previous subsections focused on scenarios with $n_\chi=1$ or 2, mimicking a flavor specific self-interacting neutrino cosmology where only fractions of neutrinos are free-streaming after decoupling. In this subsection, we take the liberty to choose as many DR flavors as we wish. In Fig.~\ref{fig:nchi40} we show the neutrino coupling for a scenario with $n_\chi= 40$ in the MI (left panel) and SI (right panel) modes. It is clear that the thermalization is efficient for both modes, so that the neutrino contribution to $N_{\rm eff}$ is suppressed: $N_{\rm eff}^\nu=3(1+n_\chi/3)^{-1}\approx 0.2$ in this case. The large $n_\chi$ enables an almost complete depletion of neutrino energy, hence free-streaming neutrinos are almost entirely replaced by self-interacting DR, mimicking a flavor-universal self-interacting neutrino in cosmology, while evading the cosmological constraint on $\sum m_\nu$, as well as the laboratory constraints on neutrino self-interaction (see Section B above).

\section{IV. Cosmological Analysis}
\label{app:cosmo}
In this section, we provide more details for our cosmological analysis. Massive free-streaming neutrinos follow the standard Boltzmann hierarchy as given in Ref.~\cite{Ma:1995ey}. The Boltzmann moments of the perturbation to the DR distribution function is defined as follows~\cite{Ma:1995ey}:
\begin{align}
    f_\chi(\mathbf{q},\mathbf{k}, \tau) &= \overline{f}_\chi(q,\tau)(1+\Psi(\mathbf{q},\mathbf{k}, \tau))\;,
\end{align}
where $\overline{f}_\chi(q,\tau)$ is the background Fermi-Dirac distribution function for DR, $\Psi$ is the perturbation on it, $\tau$ is the conformal time, $\mathbf{k}$ is the wave vector of Fourier mode, $\mathbf{q}=a\mathbf{p}$ is the comoving momentum, with $a$ being the scale factor and $\mathbf{p}$ the proper momentum. The momentum-averaged perturbation is given by 
\begin{align}
F_\chi(\hat{\mathbf{q}},\mathbf{k}, \tau) &\equiv \dfrac{\int dq\, q^3\, \overline{f}_\chi\Psi}{\int dq\, q^3\, \overline{f}_\chi}
     \equiv \sum_{\ell = 0}^\infty(-i)^\ell(2\ell +1 )F_{\chi\ell}(k,\tau)P_\ell(\hat{\mathbf{q}}\cdot\hat{\mathbf{k}})\;, 
\end{align}
where $P_\ell$ are the Legendre polynomials, and $F_{\chi\ell}$ are the momentum-averaged $\ell$-th multipoles of the perturbation.

The Boltzmann hierarchy for moments are given by
\begin{align}
    \dot{\delta}_\chi &= -{4 \over 3}\theta_\chi - {2 \over 3}\dot{h}\;,\\
    \dot{\theta}_\chi &= k^2 \left({1 \over 4}\delta_\chi - \sigma_\chi\right) \;,\\
    \dot{F}_{\chi2} &= {8 \over 15} \theta_{\chi} - {3 \over 5} k F_{\chi 3} + {4 \over 15} \dot{h} + {8 \over 5}\dot{\eta} - a  G_{\rm eff}^2 T_\chi^5 \alpha_2 F_{\chi2}\;,\\
    \dot{F}_{\chi\ell} &= {k \over 2\ell +1} \left(\ell F_{\chi(\ell - 1)} - (\ell + 1)F_{\chi(\ell + 1)}\right) 
    - a  G_{\rm eff}^2 T_\chi^5 \alpha_\ell F_{\chi\ell}, \quad \ell \geq 3\;. 
\end{align}
Here, dot represents derivative w.r.t. the conformal time $\tau$; $h$ and $\eta$  are the metric perturbations in the synchronous gauge; $\delta_\chi = F_{\chi0}$ is the density fluctuation, $\theta_\chi = (3/4)kF_{\chi 1}$ is the divergence of fluid velocity, $\sigma_\chi=F_{\chi 2}/2$ is the shear stress; $G_{\rm eff}$ is defined as in Eq.~\eqref{eq:Geff}
. In this analysis, we set $\alpha_\ell \approx \alpha_2 = 0.424 $~\cite{Brinckmann:2022ajr}. Note that, although different $\alpha_\ell$ values differ slightly from $\alpha_2$, the differences have a negligible impact on the solution of the Boltzmann hierarchy. We implemented these modifications in the Boltzmann solver Cosmic Linear Anisotropy Solving System (\texttt{CLASS}), which we used for our Markov Chain Monte Carlo analysis~\cite{Lesgourgues:2011re,Blas:2011rf}.

Since we expect the posterior to be bi-modal, we used Nested sampling for our primary analysis with the Planck and DESI data using \texttt{Multinest}~\cite{Feroz:2007kg,Feroz:2008xx,Feroz:2013hea,Buchner:2014nha} and \texttt{Montepython}~\cite{Audren:2012wb,Brinckmann:2018cvx}. We used the priors on the baryon density $\omega_b\equiv \Omega_b h^2$, cold dark matter density $\omega_c\equiv \Omega_c h^2$, Hubble constant today $H_0\equiv 100h$ km/s/Mpc, the initial super-horizon amplitude of the curvature perturbations $A_s$ at $k=0.05~{\rm Mpc}^{-1}$, the primordial spectral index $n_s$, reionization optical depth $\tau_{\rm reio}$, the effective self-interaction strength $G_{\rm eff}$, and the sum of neutrino masses $\sum m_\nu$ as given in Table~\ref{tab:prior} for the primary runs.

\begin{table}[!htb]
    \centering
    \begin{tabular}{|c|c|}
    \hline
    Parameters & Prior \\
    \hline
         $10^2\omega_b$& $[1.0,  4.0]$\\
    \hline
    $\omega_c$ & $[0.08, 0.16]$\\
    \hline
    $H_0~[{\rm km/s/Mpc}]$ & $[55.0, 85.0]$ \\
    \hline
    $\log (10^{10}A_s)$ & $[2.0,  4.0]$\\
    \hline
    $n_s$ & $[0.8,  1.1]$ \\
    \hline
    $\tau_{\rm reio}$ & $[0.004, 0.25]$\\
    \hline
    $\log_{10}(G_{\rm eff}{\rm MeV}^2)$  & $ [-5.0,1.0]$\\
    \hline
    $\sum m_\nu$ & $[0, 1.5]$\\
    \hline
    \end{tabular}
    \caption{Flat prior ranges for the Nested sampling analysis.}
    \label{tab:prior}
\end{table}
\begin{table}[!htb]
    \centering
    \begin{tabular}{|c|c|}
    \hline
        Mode & $\log_{10}(G_{\rm eff}{\rm MeV}^2)$ Prior \\
        \hline
        MI & $[-5.0, -2.5]$\\
        \hline
        SI & $[-2.5, 0.0]$\\
        \hline
    \end{tabular}
    \caption{Priors on $\log_{10}(G_{\rm eff}{\rm MeV}^2)$ parameter for the separate MI and SI mode analyses.}
    \label{tab:MISI_prior}
\end{table}
\renewcommand{\arraystretch}{1.2}
\begin{table*}[!htb]
    \centering
    \begin{tabular}{|c|c|c|c|c|}
    \hline
    &\multicolumn{4}{c|}{Planck}\\
    \hline
    & \multicolumn{2}{c|}{$0.75N_\chi+2.25N_\nu$} & \multicolumn{2}{c|}{$1.2N_\chi+1.8N_\nu$}\\
    \hline
     Parameters& MI & SI & MI & SI \\
     \hline\hline
%%%%%%%%%%%%
$10^2 \omega_{b}$ & $ 2.23\pm 0.02$ & $ 2.24\pm 0.02$ & $ 2.23\pm 0.02$ & $ 2.24\pm 0.02$\\
\hline
$\omega{}_{c }$ & $ 0.120\pm 0.001$ & $ 0.120\pm 0.001$ & $ 0.120\pm 0.001$ & $ 0.121\pm 0.001$\\
\hline
$H_0 ~[{\rm km/s/Mpc}]$ & $ 67.1^{+1}_{-0.6}$ & $ 67.5^{+1}_{-0.7}$ & $ 67.0^{+1}_{-0.6}$ & $ 67.6^{+1}_{-0.6}$\\
\hline
$\log (10^{10} A_s)$ & $ 3.05\pm 0.02$ & $ 3.03\pm 0.02$ & $ 3.05\pm 0.02$ & $ 3.02\pm 0.02$\\
\hline
$n_{s }$ & $ 0.963^{+0.005}_{-0.004}$ & $ 0.957^{+0.005}_{-0.005}$ & $ 0.962^{+0.005}_{-0.005}$ & $ 0.953\pm 0.005$\\
\hline
$\tau{}_{\rm reio }$ & $ 0.055\pm 0.008$ & $ 0.055^{+0.007}_{-0.008}$ & $ 0.055\pm 0.008$ & $ 0.055^{+0.007}_{-0.008}$\\
\hline
$\sigma_8$ & $ 0.809^{+0.02}_{-0.007}$ & $ 0.809^{+0.02}_{-0.008}$ & $ 0.809^{+0.02}_{-0.007}$ & $ 0.810^{+0.02}_{-0.008}$\\
\hline
$\log_{10}(G_{\rm eff} {\rm MeV}^2)$ & $< -3.34$ & $< -1.23$ & $< -3.48$ & $ -1.7^{+0.3}_{-0.2}$\\
\hline
$\sum m_\nu~[{\rm eV}]$ & $< 0.126$ & $< 0.121$ & $< 0.159$ & $< 0.157$\\
\hline
$\Omega{}_{m }$ & $ 0.319^{+0.008}_{-0.02}$ & $ 0.316^{+0.008}_{-0.01}$ & $ 0.320^{+0.008}_{-0.02}$ & $ 0.315^{+0.008}_{-0.02}$\\
\hline
%%%%%%%%%
    \end{tabular}
    \caption{$1\sigma$ parameter constraints for the separate MI and SI mode analyses with Planck dataset.}
    \label{tab:planck_table}
\end{table*}
\begin{table*}[!htb]
    \centering
    \begin{tabular}{|c|c|c|c|c|}
    \hline
    &\multicolumn{4}{c|}{Planck + DESI}\\
    \hline
    & \multicolumn{2}{c|}{$0.75N_\chi+2.25N_\nu$} & \multicolumn{2}{c|}{$1.2N_\chi+1.8N_\nu$}\\
    \hline
     Parameters& MI & SI & MI & SI \\
    \hline\hline
    $10^2 \omega_{b}$ & $ 2.25\pm 0.01$ & $ 2.25\pm 0.01$ & $ 2.25\pm 0.01$ & $ 2.25\pm 0.01$\\
\hline
$\omega{}_{c }$ & $ 0.1185\pm 0.0009$ & $ 0.1191\pm 0.0009$ & $ 0.1185\pm 0.0009$ & $ 0.119\pm 0.001$\\
\hline
$H_0 ~[{\rm km/s/Mpc}]$ & $ 68.4^{+0.5}_{-0.4}$ & $ 68.6\pm 0.5$ & $ 68.4\pm 0.4$ & $ 68.6\pm 0.5$\\
\hline
$\log(10^{10}A_{s })$ & $ 3.05\pm 0.02$ & $ 3.03\pm 0.02$ & $ 3.05\pm 0.02$ & $ 3.03\pm 0.02$\\
\hline
$n_{s }$ & $ 0.967\pm 0.004$ & $ 0.959^{+0.006}_{-0.004}$ & $ 0.967^{+0.005}_{-0.004}$ & $ 0.957^{+0.005}_{-0.004}$\\
\hline
$\tau{}_{\rm reio }$ & $ 0.058\pm 0.008$ & $ 0.057\pm 0.007$ & $ 0.058^{+0.007}_{-0.008}$ & $ 0.057^{+0.007}_{-0.008}$\\
\hline
$\sigma_8$ & $ 0.817^{+0.008}_{-0.007}$ & $ 0.815^{+0.01}_{-0.008}$ & $ 0.818^{+0.008}_{-0.007}$ & $ 0.819^{+0.01}_{-0.007}$\\
\hline
$\log_{10}(G_{\rm eff} {\rm MeV}^2)$ & $< -3.28$ & $ -1.3^{+0.3}_{-0.9}$ & $< -3.44$ & $ -1.7^{+0.3}_{-0.2}$\\
\hline
$\sum m_\nu~[{\rm eV}]$ & $< 0.0471$ & $< 0.0518$ & $< 0.0575$ & $< 0.0667$\\
\hline
$\Omega{}_{m }$ & $ 0.302\pm 0.006$ & $ 0.302^{+0.005}_{-0.006}$ & $ 0.302^{+0.005}_{-0.006}$ & $ 0.302\pm 0.006$\\
\hline
    \end{tabular}
    \caption{$1\sigma$ parameter constraints for the separate MI and SI mode analyses with Planck + DESI dataset.}
    \label{tab:planck_desi_table}
\end{table*}
\begin{figure*}
    \centering
    \includegraphics[width=\linewidth]{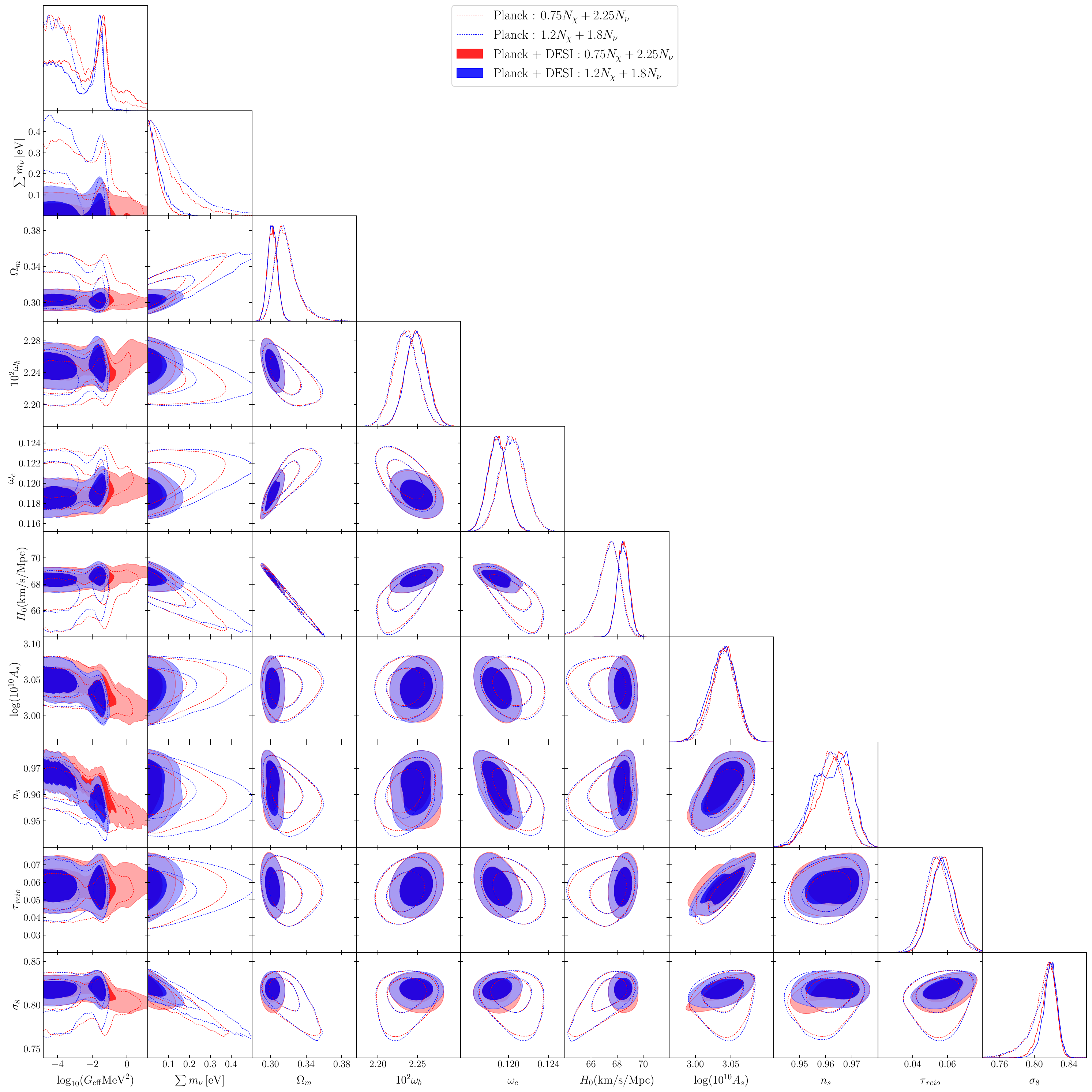}
    \captionsetup{justification=Justified}
    \caption{Triangle plot of all the relevant parameters for our cosmological analysis with Planck and Planck + DESI datasets.}
    \label{fig:tri_all}
\end{figure*}

Since Multinest is very inefficient with a large number of parameters, we used the nuisance marginalized Planck likelihood (\texttt{Plik\_lite}). The `Planck' dataset is comprised of low-$\ell$ TT $( \ell < 30)$,
low-$\ell$ EE $( \ell < 30)$, high-$\ell$ TTTEEE $ \texttt{plik\_lite}( \ell \geq 30)$ and lensing likelihood~\cite{Planck:2019nip,Planck:2018lbu}. We also use the BAO scale measurements from Dark Energy Spectroscopic Instrument (DESI) Data Release 1 (DR1)~\cite{DESI:2024mwx,Herold:2024nvk}. In our analysis, we have set the total $N_{\rm eff} = 3.044$~\cite{Bennett:2020zkv} since we focus on resonant conversion of neutrinos to DR \emph{after} BBN. 
We set $N_{\rm eff}^\chi = 0.75 (1.2) $ and $N_{\rm eff}^\nu = 2.294 (1.244) $ for $n_\chi = 1(2)$. For notational brevity we use $N_{\rm eff}^\nu = 2.25 (1.2) $ in the text.
We also choose the neutrino flavors to have degenerate mass for the cosmological analysis.

The triangle plot for the posteriors from the analysis with Planck and Planck + DESI is shown in Fig.~\ref{fig:tri_all}. The SI mode significance is increased by the addition of the DESI BAO dataset. To study the individual modes, we also performed a separate analysis where we choose different priors on $\log_{10}(G_{\rm eff}{\rm MeV}^2)$ values to study MI and SI mode separately as shown in Table~\ref{tab:MISI_prior}. Having separate priors guarantees that the distributions in the MI and SI modes are unimodal. Therefore, we used the Metropolis–Hastings algorithm~\cite{Metropolis:1953am, Hastings:1970aa} for this analysis. We also used the full high-$\ell$ TTTEEE $ \texttt{plik}( \ell \geq 30)$ with all the nuisance parameters instead of the \texttt{plik\_lite} likelihood used in the previous nested sampling analysis. For the rest of the parameters, we kept the prior ranges the same as those in Table~\ref{tab:prior}. We show the parameter values for the mode-specific analysis in Table~\ref{tab:planck_table} and Table~\ref{tab:planck_desi_table}.

 Our model (equivalently neutrino self-interaction model) is able to partially mitigate the tension in the $\Omega_m$ measurement between the CMB and the DESI BAO data~\cite{DESI:2024mwx,DESI:2025zgx,DESI:2025ejh}. The DESI data show a preference for smaller $\Omega_m$ compared to the Planck measurement, which is one of the major drivers of the `negative' neutrino mass tension~\cite{Craig:2024tky,Green:2024xbb,Naredo-Tuero:2024sgf,DESI:2024mwx,Loverde:2024nfi,DESI:2025ejh}. The presence of strong self-interaction (SI mode) lowers the value of $\Omega_m$ compared to the MI mode (which is equivalent to $\Lambda$CDM) as can be seen from Table~\ref{tab:planck_table} with Planck data alone. This is also the reason for the presence of SI mode over MI mode (as can be seen from Fig.~\ref{fig:tri_all}) when DESI data is added, which prefers a smaller $\Omega_m$. Thus, in addition to relaxing the neutrino mass bound, this model also partially addresses the negative `neutrino mass' tension. A detailed quantitative study of this model for `negative' neutrino mass tension will require either extending the (effective) neutrino mass prior to negative values or some profile-likelihood analysis, which we leave for future work. 
% The  reason is that the SI mode value of the total matter density $\Omega_m$ is slightly smaller than the MI mode and the DESI likelihood prefers a smaller value of $\Omega_m$.
The qualitative features of our analysis are expected to remain unchanged while considering DESI 3-year (DESI-DR2) data release~\cite{DESI:2025zgx,DESI:2025ejh}, since DESI-DR1 and DESI-DR2 are largely consistent with each other.

With respect to other cosmological tensions, the SI mode prefers a slightly larger $H_0$ due to the effects of neutrino-induced phase shift~\cite{Kreisch:2019yzn,Das:2020xke,Das:2023npl} which goes in the right direction for addressing the $H_0$ tension. We have not considered extra radiation in this model, which assists in better addressing the $H_0$ tension in neutrino self-interaction cosmology~\cite{Cyr-Racine:2013jua,Kreisch:2019yzn,RoyChoudhury:2020dmd,Das:2020xke,Poudou:2025qcx}. Finally, strong self-interaction does not produce an appreciable shift in the value of $\sigma_8$ for the datasets used here. However, note that the significance of the $\sigma_8$ tension has gone down, especially with the latest KiDS data~\cite{Wright:2025xka,Pantos:2026koc}, although the latest DES data still show a modest $2.4\sigma$ tension~\cite{DES:2026fyc}. 

In conclusion, the neutrino (or DR) self-interaction cosmology is highly relevant for addressing several cosmological tensions and our model shows that it can be realized free from terrestrial constraints. 

%\bibliography{main}
%\bibliographystyle{JHEP}
\end{document}